\documentclass{CSML}
\pdfoutput=1 

\usepackage{lastpage}
\newcommand{\dOi}{13(2:6)2017}
\lmcsheading{}{1--\pageref{LastPage}}{}{}{Dec.~20, 2015}{May~17, 2017}{}

\newif\ifignore 
\ignorefalse

\newcommand{\auxproof}[1]{
\ifignore\mbox{}\newline
\textbf{PROOF:} \dotfill\newline
{\it #1}\mbox{}\newline
\textbf{ENDPROOF}\dotfill
\fi}

\usepackage{amsmath}
\usepackage{amssymb}
\usepackage{amstext}
\usepackage{mathrsfs}
\usepackage{stmaryrd}
\usepackage{xspace}
\usepackage{cmll}
\usepackage{hyperref}
\usepackage{url}
\usepackage{fancybox}

  \hypersetup{
      colorlinks=true,
      linkcolor=black,
      citecolor=black,
      filecolor=black,
      urlcolor=black,
  }

\usepackage{xypic}
\usepackage[all]{xy}
\xyoption{all}
\xyoption{2cell}
\UseTwocells
\xyoption{tips}
\SelectTips{cm}{}  

\newdimen\proofrulebreadth \proofrulebreadth=.05em
\newdimen\proofdotseparation \proofdotseparation=1.25ex
\newdimen\proofrulebaseline \proofrulebaseline=2ex
\newcount\proofdotnumber \proofdotnumber=3
\let\then\relax
\def\hfi{\hskip0pt plus.0001fil}
\mathchardef\squigto="3A3B
\newif\ifinsideprooftree\insideprooftreefalse
\newif\ifonleftofproofrule\onleftofproofrulefalse
\newif\ifproofdots\proofdotsfalse
\newif\ifdoubleproof\doubleprooffalse
\let\wereinproofbit\relax
\newdimen\shortenproofleft
\newdimen\shortenproofright
\newdimen\proofbelowshift
\newbox\proofabove
\newbox\proofbelow
\newbox\proofrulename
\def\shiftproofbelow{\let\next\relax\afterassignment\setshiftproofbelow\dimen0 }
\def\shiftproofbelowneg{\def\next{\multiply\dimen0 by-1 }%
\afterassignment\setshiftproofbelow\dimen0 }
\def\setshiftproofbelow{\next\proofbelowshift=\dimen0 }
\def\setproofrulebreadth{\proofrulebreadth}

\def\prooftree{
\ifnum  \lastpenalty=1
\then   \unpenalty
\else   \onleftofproofrulefalse
\fi
\ifonleftofproofrule
\else   \ifinsideprooftree
        \then   \hskip.5em plus1fil
        \fi
\fi
\bgroup
\setbox\proofbelow=\hbox{}\setbox\proofrulename=\hbox{}%
\let\justifies\proofover\let\leadsto\proofoverdots\let\Justifies\proofoverdbl
\let\using\proofusing\let\[\prooftree
\ifinsideprooftree\let\]\endprooftree\fi
\proofdotsfalse\doubleprooffalse
\let\thickness\setproofrulebreadth
\let\shiftright\shiftproofbelow \let\shift\shiftproofbelow
\let\shiftleft\shiftproofbelowneg
\let\ifwasinsideprooftree\ifinsideprooftree
\insideprooftreetrue
\setbox\proofabove=\hbox\bgroup$\displaystyle 
\let\wereinproofbit\prooftree
\shortenproofleft=0pt \shortenproofright=0pt \proofbelowshift=0pt
\onleftofproofruletrue\penalty1
}

\def\eproofbit{
\ifx    \wereinproofbit\prooftree
\then   \ifcase \lastpenalty
        \then   \shortenproofright=0pt  
        \or     \unpenalty\hfil         
        \or     \unpenalty\unskip       
        \else   \shortenproofright=0pt  
        \fi
\fi
\global\dimen0=\shortenproofleft
\global\dimen1=\shortenproofright
\global\dimen2=\proofrulebreadth
\global\dimen3=\proofbelowshift
\global\dimen4=\proofdotseparation
\global\count255=\proofdotnumber
$\egroup  
\shortenproofleft=\dimen0
\shortenproofright=\dimen1
\proofrulebreadth=\dimen2
\proofbelowshift=\dimen3
\proofdotseparation=\dimen4
\proofdotnumber=\count255
}

\def\proofover{
\eproofbit 
\setbox\proofbelow=\hbox\bgroup 
\let\wereinproofbit\proofover
$\displaystyle
}%
\def\proofoverdbl{
\eproofbit 
\doubleprooftrue
\setbox\proofbelow=\hbox\bgroup 
\let\wereinproofbit\proofoverdbl
$\displaystyle
}%
\def\proofoverdots{
\eproofbit 
\proofdotstrue
\setbox\proofbelow=\hbox\bgroup 
\let\wereinproofbit\proofoverdots
$\displaystyle
}%
\def\proofusing{
\eproofbit 
\setbox\proofrulename=\hbox\bgroup 
\let\wereinproofbit\proofusing
\kern0.3em$
}

\def\endprooftree{
\eproofbit 
  \dimen5 =0pt
\dimen0=\wd\proofabove \advance\dimen0-\shortenproofleft
\advance\dimen0-\shortenproofright
\dimen1=.5\dimen0 \advance\dimen1-.5\wd\proofbelow
\dimen4=\dimen1
\advance\dimen1\proofbelowshift \advance\dimen4-\proofbelowshift
\ifdim  \dimen1<0pt
\then   \advance\shortenproofleft\dimen1
        \advance\dimen0-\dimen1
        \dimen1=0pt
        \ifdim  \shortenproofleft<0pt
        \then   \setbox\proofabove=\hbox{%
                        \kern-\shortenproofleft\unhbox\proofabove}%
                \shortenproofleft=0pt
        \fi
\fi
\ifdim  \dimen4<0pt
\then   \advance\shortenproofright\dimen4
        \advance\dimen0-\dimen4
        \dimen4=0pt
\fi
\ifdim  \shortenproofright<\wd\proofrulename
\then   \shortenproofright=\wd\proofrulename
\fi
\dimen2=\shortenproofleft \advance\dimen2 by\dimen1
\dimen3=\shortenproofright\advance\dimen3 by\dimen4
\ifproofdots
\then
        \dimen6=\shortenproofleft \advance\dimen6 .5\dimen0
        \setbox1=\vbox to\proofdotseparation{\vss\hbox{$\cdot$}\vss}%
        \setbox0=\hbox{%
                \advance\dimen6-.5\wd1
                \kern\dimen6
                $\vcenter to\proofdotnumber\proofdotseparation
                        {\leaders\box1\vfill}$%
                \unhbox\proofrulename}%
\else   \dimen6=\fontdimen22\the\textfont2 
        \dimen7=\dimen6
        \advance\dimen6by.5\proofrulebreadth
        \advance\dimen7by-.5\proofrulebreadth
        \setbox0=\hbox{%
                \kern\shortenproofleft
                \ifdoubleproof
                \then   \hbox to\dimen0{%
                        $\mathsurround0pt\mathord=\mkern-6mu%
                        \cleaders\hbox{$\mkern-2mu=\mkern-2mu$}\hfill
                        \mkern-6mu\mathord=$}%
                \else   \vrule height\dimen6 depth-\dimen7 width\dimen0
                \fi
                \unhbox\proofrulename}%
        \ht0=\dimen6 \dp0=-\dimen7
\fi
\let\doll\relax
\ifwasinsideprooftree
\then   \let\VBOX\vbox
\else   \ifmmode\else$\let\doll=$\fi
        \let\VBOX\vcenter
\fi
\VBOX   {\baselineskip\proofrulebaseline \lineskip.2ex
        \expandafter\lineskiplimit\ifproofdots0ex\else-0.6ex\fi
        \hbox   spread\dimen5   {\hfi\unhbox\proofabove\hfi}%
        \hbox{\box0}%
        \hbox   {\kern\dimen2 \box\proofbelow}}\doll%
\global\dimen2=\dimen2
\global\dimen3=\dimen3
\egroup 
\ifonleftofproofrule
\then   \shortenproofleft=\dimen2
\fi
\shortenproofright=\dimen3
\onleftofproofrulefalse
\ifinsideprooftree
\then   \hskip.5em plus 1fil \penalty2
\fi
}

\newtheorem{theorem}{Theorem}
\newtheorem{lemma}[theorem]{Lemma}
\newtheorem{proposition}[theorem]{Proposition}
\newtheorem{corollary}[theorem]{Corollary}
\newtheorem{definition}[theorem]{Definition}
\newtheorem{example}[theorem]{Example}

\newenvironment{myproof}[1][Proof.]%
   { \begin{trivlist}%
     \item[\hskip \labelsep {\it #1}]%
   }%
   { \end{trivlist}%
   }

\newcommand{\QEDbox}{\square}
\newcommand{\QED}{\hspace*{\fill}$\QEDbox$}

\newcommand{\klafter}{\mathrel{\raisebox{.15em}{$\scriptscriptstyle\circledcirc$}}}
\newcommand{\idmap}[1][]{\ensuremath{\mathrm{id}_{#1}}}
\newcommand{\after}{\mathrel{\circ}}
\newcommand{\evmap}{\mathrm{ev}}
\newcommand{\inl}{\ensuremath{\mathrm{in}_1}}
\newcommand{\inr}{\ensuremath{\mathrm{in}_2}}
\newcommand{\Bowtie}{\mathord{\bowtie}}
\newcommand{\three}{3}
\newcommand{\orthogonal}{\mathrel{\bot}}
\newcommand{\set}[2]{\{#1\;|\;#2\}}
\newcommand{\setin}[3]{\{#1\in#2\;|\;#3\}}
\newcommand{\supp}{\mathrm{supp}}
\newcommand{\op}[1]{#1^{\mathrm{op}}}
\newcommand{\all}[2]{\forall#1.\,#2}
\newcommand{\allin}[3]{\forall#1\in#2.\,#3}
\newcommand{\ex}[2]{\exists#1.\,#2}
\newcommand{\exin}[3]{\exists#1\in#2.\,#3}
\newcommand{\lam}[2]{\lambda#1.\,#2}
\newcommand{\lamin}[3]{\lambda#1\in#2.\,#3}

\newcommand{\powersetsymbol}{\mathcal{P}}

\newcommand{\distributionsymbol}{\mathcal{D}}

\newcommand{\Pow}{\powersetsymbol}

\newcommand{\Upsets}{\ensuremath{\mathrm{Up}}}
\newcommand{\Dwnsets}{\ensuremath{\mathrm{Dwn}}}

\newcommand{\downset}{\mathop{\downarrow\!}}

\newcommand{\Dst}{\distributionsymbol}
\newcommand{\sDst}{\distributionsymbol_{\leq 1}}

\newcommand{\Giry}{\mathcal{G}}

\newcommand{\Exp}{\mathcal{E}}
\newcommand{\Rad}{\mathcal{R}}
\newcommand{\UF}{\ensuremath{\mathcal{U}{\kern-.75ex}\mathcal{F}}}

\newcommand{\Cat}[1]{\ensuremath{\mathbf{#1}}\xspace}
\newcommand{\cat}[1]{\Cat{#1}}
\newcommand{\Kl}{\mathcal{K}{\kern-.4ex}\ell}
\newcommand{\EM}{\mathcal{E}{\kern-.4ex}\mathcal{M}}

\newcommand{\Sets}{\Cat{Sets}}

\newcommand{\Dcpo}{\Cat{Dcpo}}
\newcommand{\DcPA}{\Cat{DcPA}}
\newcommand{\PoSets}{\Cat{PoSets}}
\newcommand{\MSL}{\Cat{MSL}}
\newcommand{\CH}{\Cat{CH}}   
\newcommand{\CCH}{\Cat{CCH}} 
\newcommand{\CCHsep}{\CCH_{\mathrm{sep}}} 
\newcommand{\Stone}{\Cat{Stone}}   
\newcommand{\BA}{\Cat{BA}} 
\newcommand{\CBA}{\Cat{CBA}} 
\newcommand{\CABA}{\Cat{CABA}} 
\newcommand{\CL}{\Cat{CL}}
\newcommand{\CLJ}{\ensuremath{\CL_{\textstyle\vee}}}
\newcommand{\CLM}{\ensuremath{\CL_{\textstyle\wedge}}}
\newcommand{\CLJO}{\ensuremath{\CL_{{\textstyle\vee},1}}}
\newcommand{\CDL}{\Cat{CDL}}
\newcommand{\CCL}{\Cat{CCL}}

\newcommand{\Meas}{\Cat{Meas}}
\newcommand{\PreFrm}{\Cat{PreFrm}}
\newcommand{\PreFrmZ}{\ensuremath{\Cat{PreFrm}_{0}}}
\newcommand{\Frm}{\Cat{Frm}}

\newcommand{\NNO}{\mathbb{N}}

\newcommand{\R}{\mathbb{R}}

\newcommand{\EA}{\Cat{EA}}
\newcommand{\MVA}{\Cat{MVA}}
\newcommand{\wEA}{\ensuremath{\omega\text{-}\EA}}

\newcommand{\EMod}{\Cat{EMod}}
\newcommand{\MVMod}{\Cat{MVMod}}
\newcommand{\wEMod}{\ensuremath{\omega\text{-}\EMod}}
\newcommand{\DcEMod}{\Cat{DcEMod}}
\newcommand{\Conv}{\Cat{Conv}}
\newcommand{\CstarMap}[1]{\ensuremath{\Cat{Cstar}_{#1}}}

\newcommand{\CstarPU}{\CstarMap{\mathrm{PU}}}

\newcommand{\CCstarMap}[1]{\ensuremath{\Cat{CCstar}_{#1}}}
\newcommand{\CCstarMIU}{\CCstarMap{\mathrm{MIU}}}
\newcommand{\CCstarPU}{\CCstarMap{\mathrm{PU}}}

\newcommand{\WstarMap}[1]{\ensuremath{\Cat{Wstar}_{#1}}}
\newcommand{\WstarNPU}{\WstarMap{\mathrm{NPU}}}
\newcommand{\CWstarMap}[1]{\ensuremath{\Cat{CWstar}_{#1}}}
\newcommand{\CWstarNPU}{\CWstarMap{\mathrm{NPU}}}

\newcommand{\Hom}{\ensuremath{\mathrm{Hom}}}
\newcommand{\Pred}{\ensuremath{\mathrm{Pred}}}
\newcommand{\Stat}{\ensuremath{\mathrm{Stat}}}

\newcommand{\NStat}{\ensuremath{\mathrm{NStat}}}
\newcommand{\Ef}{\ensuremath{\mathcal{E}{\kern-.5ex}f}}
\newcommand{\intd}{{\kern.2em}\mathrm{d}{\kern.03em}}
\newcommand{\indic}[1]{\mathbf{1}_{#1}}
\newcommand{\Cont}{\ensuremath{\mathrm{C}}}
\newcommand{\bigovee}{\mathop{\vphantom{\sum}\mathchoice%
        {\vcenter{\hbox{\huge $\ovee$}}}%
        {\vcenter{\hbox{\Large $\ovee$}}}%
        {\ovee}{\ovee}}\displaylimits}
\newcommand{\Open}{\ensuremath{\mathcal{O}}}
\newcommand{\BiOpen}{\ensuremath{\mathcal{B}{\kern-.15em}\mathcal{O}}}
\newcommand{\OF}{\ensuremath{\mathcal{O}{\kern-.1em}\mathcal{F}}}
\newcommand{\Closed}{\ensuremath{\mathcal{C}{\kern-.05em}\ell}}

\newcommand{\conglongrightarrow}{\mathrel{\smash{\stackrel{
           \raisebox{.5ex}{$\scriptstyle\cong$}}{
           \raisebox{0ex}[0ex][0ex]{$\longrightarrow$}}}}}

\newcommand{\ie}{\textit{i.e.}\xspace}
\newcommand{\eg}{\textit{e.g.}\xspace}

\title{A Recipe for State-and-Effect Triangles}

\author{Bart Jacobs}

\address{
Institute for Computing and Information Sciences, 
Radboud University Nijmegen, The Netherlands. 
}
\urladdr{\url{www.cs.ru.nl/B.Jacobs}}
\email{bart@cs.ru.nl}



\begin{document}

\maketitle

\begin{abstract}
In the semantics of programming languages one can view programs as
state transformers, or as predicate transformers. Recently the author
has introduced `state-and-effect' triangles which capture this
situation categorically, involving an adjunction between state- and
predicate-transformers. The current paper exploits a classical result
in category theory, part of Jon Beck's monadicity theorem, to
systematically construct such a state-and-effect triangle from an
adjunction. The power of this construction is illustrated in many
examples, covering many monads occurring in program semantics,
including (probabilistic) power domains.
\end{abstract}

\section{Introduction}\label{IntroSec}

In program semantics three approaches can be distinguished.
\begin{itemize}
\item Interpreting programs themselves as morphisms in certain
  categories.  Composition in the category then corresponds to
  sequential composition. Parallel composition may be modeled via
  tensors $\otimes$. Since~\cite{Moggi91a} the categories involved are
  often Kleisli categories $\Kl(T)$ of a monad $T$, where the monad
  $T$ captures a specific form of computation: deterministic,
  non-deterministic, probabilistic, \textit{etc}.

\item Interpreting programs via their actions on states, as
  \emph{state transformers}. For instance, in probabilistic
  programming the states may be probabilistic distributions over
  certain valuations (mapping variables to values). Execution of a
  program changes the state, by adapting the probabilities of
  valuations. The state spaces often have algebraic structure, and
  take the form of Eilenberg-Moore categories $\EM(T)$ of a monad $T$.

\item Interpreting programs via their actions on predicates, as
  \emph{predicate transformers}. The predicates involved describe what
  holds at a specific point. This validity may also be quantitative
  (or `fuzzy'), describing that a predicate holds with a certain
  probability in the unit interval $[0,1]$. Execution of a program may
  then adapt the validity of predicates. A particular form of
  semantics of this sort is weakest precondition
  computation~\cite{DijkstraS90}.  In the context of (coalgebraic)
  modal logic, these predicate transformers appear as modal operators.
\end{itemize}

A systematic picture of these three approaches has emerged in
categorical language, using triangles of the form described below,
see~\cite{Jacobs15d}, and also~\cite{Jacobs13a,Jacobs15a,ChoJWW15b}.
\begin{equation}
\label{ComputationTriangle}
\qquad\vcenter{\xymatrix@C-1.2pc@R-1.5pc{
\ovalbox{\textbf{Heisenberg}} & & \ovalbox{\textbf{Schr\"odinger}} \\
\llap{$\op{\Cat{Log}}=$}{\left(\begin{array}{c} \text{predicate} \\[-.3em]
      \text{transformers} \end{array}\right)}\ar@/^1em/[rr] 
& \top &
{\left(\begin{array}{c} \text{state} \\[-.3em]
      \text{transformers} \end{array}\right)}\ar@/^1em/[ll]  \\
\\
& \Big(\text{computations}\Big)\ar[uul]^(0.45){\Pred}\ar[uur]_(0.45){\Stat} &
}}
\end{equation}

The three nodes in this diagram represent categories of which only the
morphisms are described. The arrows between these nodes are functors,
where the two arrows $\rightleftarrows$ at the top form an
adjunction. The two triangles involved should commute. In the case
where two up-going `predicate' and `state' functors $\Pred$ and
$\Stat$ in~\eqref{ComputationTriangle} are full and faithful, we have
three equivalent ways of describing computations.  

On morphisms, the predicate functor $\Pred$
in~\eqref{ComputationTriangle} yields what is called substitution in
categorical logic, but what amounts to a weakest precondition
operation in program semantics, or a modal operator in programming
logic. The upper category on the left is of the form $\op{\Cat{Log}}$,
where $\Cat{Log}$ is some category of logical structures. The opposite
category $\op{(-)}$ is needed because predicate transformers operate
in the reverse direction, taking a postcondition to a precondition.


In a setting of quantum computation this translation back-and-forth
$\rightleftarrows$ in~\eqref{ComputationTriangle} is associated with
the different approaches of Heisenberg (logic-based, working
backwards) and Schr\"odinger (state-based, working forwards), see
\textit{e.g.}~\cite{HeinosaariZ12}. In quantum foundations one speaks
of the duality between states and effects (predicates). Since the
above triangles first emerged in the context of semantics of quantum
computation~\cite{Jacobs15d}, they are sometimes referred to as
`state-and-effect' triangles.

In certain cases the adjunction $\rightleftarrows$
in~\eqref{ComputationTriangle} forms --- or may be restricted to ---
an equivalence of categories, yielding a duality situation. It shows
the importance of duality theory in program semantics and logic; this
topic has a long history, going back to~\cite{Abramsky91}.

In~\cite{Jacobs15d} it is shown that in the presence of relatively
weak structure in a category $\cat{B}$, a diagram of the
form~\eqref{ComputationTriangle} can be formed, with $\cat{B}$ as base
category of computations, with predicates forming effect modules (see
below) and with states forming convex sets. A category with this
relatively weak structure is called an \emph{effectus},
see~\cite{ChoJWW15b}.

The main contribution of this paper is a ``new'' way of generating
state-and-effect triangles, namely from adjunctions. We write the word
`new' between quotes, because the underlying category theory uses a
famous of result of Jon Beck, and is not new at all. What the paper
contributes is mainly a new perspective: it reorganises the work of
Beck in such a way that an appropriate triangle appears, see
Section~\ref{MonadSec}.  The rest of the paper is devoted to
illustrations of this recipe for triangles. These include Boolean and
probabilistic examples, see Sections~\ref{TwoSec} and~\ref{UnitSec}
respectively. The Boolean examples are all obtained from an adjunction
using ``homming into $2 =\{0,1\}$'', whereas the probabilistic
(quantitative) examples all arise from ``homming into $[0,1]$'', where
$[0,1]$ is the unit interval of probabilities.  In between we consider
Plotkin-style constructions via ``homming into 3'', where $3 = \{0,
\Bowtie, 1\}$ is a three-element ordered algebra.

The series of examples in this paper involves many mathematical
structures, ranging from Boolean algebras to compact Hausdorff spaces
and $C^*$-algebras. It is impossible to explain all these notions in
detail here. Hence the reader is assumed to be reasonably familiar
with these structures. It does not matter so much if some of the
examples involve unfamiliar mathematical notions. The structure of
these sections~\ref{TwoSec}, \ref{ThreeSec} and~\ref{UnitSec} is clear
enough --- using 2, $\three$ and $[0,1]$ as dualising object,
respectively --- and it does not matter if some of the examples are
skipped.

An exception is made for the notions of effect algebra and effect
module.  They are explicitly explained (briefly) in the beginning of
Section~\ref{UnitSec} because they play such a prominent role in
quantitative logic.

The examples involve many adjunctions that are known in the
literature. Here they are displayed in triangle form. In several cases
monads arise that are familiar in coalgebraic research, like the
neighbourhood monad $\mathcal{N}$ in Subsection~\ref{SetsSetsSubsec},
the monotone neighbourhood monad $\mathcal{M}$ in
Subsection~\ref{SetsPosetsSubsec}, the Hoare power domain monad
$\mathcal{H}$ in Subsection~\ref{DcpoCLSubsec}, the Smyth power domain
monad $\mathcal{S}$ in Subsection~\ref{DcpoPreFrmSubsec}, the infinite
distribution monad $\Dst_{\infty}$ in
Subsection~\ref{SetsDcEModSubsec}, the Giry monad $\Giry$ in
Subsection~\ref{MeaswEModSubsec}, and the valuation monad
$\mathcal{V}$ in Subsection~\ref{DcpoDcEModSubsec}. Also we will see
several examples where we have pushed the recipe to a limit, and where
the monad involved is simply the identity.

This paper extends the earlier conference version~\cite{Jacobs15b}
with several order-theoretic examples, notably using complete lattices
and directed complete partial orders (for various power domains).

\section{A basic result about monads}\label{MonadSec}

We assume that the reader is familiar with the categorical concept of
a monad $T$, and with its double role, describing a form of
computation, via the associated Kleisli category $\Kl(T)$, and
describing algebraic structure, via the category $\EM(T)$ of
Eilenberg-Moore algebras.

The following result is a basic part of the theory of monads, see
\eg~\cite[Prop.~3.15 and Exercise (KEM)]{BarrW85} or~\cite[Prop.~6.5
  and~6.7]{LambekS86} or~\cite[Thm.~20.42]{AdamekHS90}, and describes
the initiality and finality of the Kleisli category and
Eilenberg-Moore category as `adjunction resolutions' giving rise to a
monad.

\begin{theorem}
\label{AdjTriangleThm}
Consider an adjunction $F \dashv G$ with induced monad $T = GF$. Then
there are `comparison' functors $\Kl(T) \rightarrow \cat{A} \rightarrow
\EM(T)$ in a diagram:
\begin{equation}
\label{ResolutionDiag}
\vcenter{\xymatrix@C+1pc@R-0pc{
\Kl(T)\ar@/^1.3ex/[rr]^-{L}\ar@/^1ex/[drr] & &
   \cat{A}\ar@/^1.5ex/[rr]^-{K}\ar@/^1.1ex/[d]_-{\dashv}^-{G} & \top &
   \EM\rlap{$(T)$}\ar@/^1ex/[dll]_(0.42)[@]{\top}\ar@{.>}@/^1.2ex/[ll]^(0.7){M}
\\
& & \cat{B}\ar@/^1ex/[urr]\ar@/^1.1ex/[u]^-{F}\ar@(dl,dr)_{T=GF}
   \ar@/^1ex/[ull]_(0.45)[@]{\raisebox{-3em}{$\scriptstyle\bot$}} &
}}
\end{equation}

\noindent where the functor $L \colon \Kl(T) \rightarrow \cat{A}$ is
full and faithful.

In case the category $\Cat{A}$ has coequalisers (of reflexive pairs),
then $K$ has a left adjoint $M$, as indicated via the dotted arrow,
satisfying $MKL \cong L$.



\end{theorem}

The famous monadicity theorem of Jon Beck gives conditions that
guarantee that the functor $K \colon \cat{A} \rightarrow \EM(T)$ is an
equivalence of categories, so that objects of $\cat{A}$ are algebras.
The existence of the left adjoint $M$ is the part of this theorem that
we use in the current setting. Other (unused) parts of Beck's theorem
require that the functor $G$ preserves and reflects coequalisers of
reflexive pairs. For convenience we include a proof sketch.

\begin{myproof}
We write $\eta,\varepsilon$ for the unit and counit of the adjunction
$F \dashv G$, so that $\eta$ is also the unit of the induced monad $T
= GF$, with multiplication $\mu = G(\varepsilon F)$.  Define $L(X) =
F(X)$ and $\smash{L\big(X \stackrel{f}{\rightarrow} GF(Y)\big)} =
\varepsilon_{F(Y)} \after F(f) \colon F(X) \rightarrow F(Y)$. This
functor $L$ is full and faithful because there is a bijective adjoint
correspondence:
$$\begin{prooftree}
\xymatrix{ F(X)\ar[r] & F(Y)}
\Justifies
\xymatrix{ X\ar[r] & GF(Y) \rlap{$\;=T(Y)$}}
\end{prooftree}$$

\auxproof{
We check that this is a functor indeed:
$$\begin{array}{rcl}
L(\idmap[X])
& = &
L(\eta_{X}) \\
& = &
\varepsilon_{F(X)} \after F(\eta_{X}) \\
& = &
\idmap[F(X)] \\
L(g \klafter f)
& = &
\varepsilon \after F(\mu \after T(g) \after f) \\
& = &
\varepsilon \after FG(\epsilon) \after FGF(g) \after F(f) \\
& = &
\varepsilon \after F(g) \after \varepsilon \after F(f) \\
& = &
L(g) \after L(f).
\end{array}$$
}

The functor $K \colon \cat{A} \rightarrow \EM(T)$ is defined as:
$$\begin{array}{rclcrcl}
K(A)
& = &
\ensuremath{\left(\xy
(0,4)*{GFG(A)};
(0,-4)*{G(A)};
{\ar^{G(\varepsilon_{A})\!\!\!} (0,2); (0,-2)};
\endxy\right)} 
& \qquad\mbox{and}\qquad &
K\big(A\stackrel{f}{\rightarrow} B\big)
& = &
G(f).
\end{array}$$

\auxproof{
$$\begin{array}{rcl}
G(\varepsilon_{A}) \after \eta_{G(A)}
& = &
\idmap \\
G(\varepsilon_{A}) \after \mu_{G(A)}
& = &
G(\varepsilon_{A}) \after G(\varepsilon_{FG(A)}) \\
& = &
G(\varepsilon_{A}) \after GFG(\varepsilon_{A}) \\
& = &
G(\varepsilon_{A}) \after TG(\varepsilon_{A}).
\end{array}$$

\noindent For $f\colon A \rightarrow B$ we have that $G(f)$ is a
map of algebras, by naturality:
$$\begin{array}{rcccccl}
G(\varepsilon_{B}) \after TG(f)
& = &
G(\varepsilon_{B} \after FG(f))
& = &
G(f \after \varepsilon_{A}) 
& = &
G(f) \after G(\varepsilon_{A}).
\end{array}$$
}

\noindent We leave it to the reader to see that $K$ is well-defined.
On an object $X\in\Kl(T)$, that is, on $X\in\cat{B}$, the result
$KL(X)$ is the multiplication $\mu_{X} = G(\varepsilon_{FX})$ of the
monad $T = GF$.  For a Kleisli map $f\colon X \rightarrow T(Y)$ the
map $KL(f)$ is Kleisli extension:
$$\begin{array}{rcccl}
KL(f)
& = &
G(\varepsilon_{F(Y)} \after F(f))
& = &
\mu_{Y} \after T(f) \,\colon\, T(X) \longrightarrow T(Y).
\end{array}$$

Assume now that the category $\Cat{A}$ has coequalisers. For an
algebra $a\colon T(X) \rightarrow X$ let $M(X,a)$ be the (codomain of
the) coequaliser in:
$$\vcenter{\xymatrix{
\llap{$FG$}F(X)\ar@/^1ex/[rr]^-{F(a)}\ar@/_1ex/[rr]_-{\varepsilon_{F(X)}} 
   & & F(X)\ar@{->>}[r]^-{c} & M(X,a)
}}
$$

\noindent It is not hard to see that there is a bijective correspondence:
$$\begin{prooftree}
\xymatrix{M(X,a)\ar[r]^-{f} & A \rlap{\hspace*{8.4em}in $\Cat{A}$}}
\Justifies
\xymatrix{\ensuremath{\left(\xy
(0,4)*{T(X)};
(0,-4)*{X};
{\ar^{a} (0,2); (0,-2)};
\endxy\right)}
\ar[r]_-{g} &
\ensuremath{\left(\xy
(0,4)*{TG(A)};
(0,-4)*{G(A)};
{\ar^{G(\varepsilon_{A})\!\!\!} (0,2); (0,-2)};
\endxy\right)}\rlap{$=K(A)$}
\rlap{\hspace*{6em}in $\EM(T)$}}
\end{prooftree}\hspace*{5em}$$

\auxproof{
\noindent This is done as follows.
\begin{itemize}
\item Given $f\colon M(X,a) \rightarrow A$ as above, take
  $\overline{f} = G(f \after c) \after \eta_{X} \colon X \rightarrow
  G(A)$. We prove that $\overline{f}$ is an algebra homomorphism by
  using the triangular identities and the above
  coequaliser~\eqref{AdjTriangleCoeq}:
$$\begin{array}{rcl}
G(\varepsilon_{A}) \after T(\overline{f})
& = &
G\big(\varepsilon_{A} \after FG(f \after c) \after F(\eta_{X})\big) \\
& = &
G\big(f \after c \after \varepsilon_{F(X)} \after F(\eta_{X})\big) \\
& = &
G\big(f \after c\big) \\
& = &
G\big(f \after c \after \varepsilon_{F(X)}\big) \after \eta_{GF(X)} \\
& = &
G\big(f \after c \after F(a)\big) \after \eta_{GF(X)} \\
& = &
G(f \after c) \after \eta_{X} \after a \\
& = &
\overline{f} \after a.
\end{array}$$

\item Conversely, given an algebra map $g\colon X \rightarrow G(A)$,
  its transpose $\varepsilon_{A} \after F(g)$ coequalises the two
  parallel maps in~\eqref{AdjTriangleCoeq}:
$$\begin{array}{rcl}
\big(\varepsilon_{A} \after F(g)\big) \after F(a)
& = &
\varepsilon_{A} \after F(g \after a) \\
& = &
\varepsilon_{A} \after F(G(\varepsilon_{A}) \after T(g)) \\
& = &
\big(\varepsilon_{A} \after F(g)\big) \after \varepsilon_{F(X)}.
\end{array}$$

\noindent Hence there is a unique map $\overline{g} \colon M(X,a)
\rightarrow X$, with $\overline{g} \after c = \varepsilon_{A} \after
F(g)$.
\end{itemize}

Clearly, $\smash{\overline{\overline{f}} = f}$ and
  $\smash{\overline{\overline{g}} = g}$.
$$\begin{array}{rcl}
\overline{\overline{f}} \after c
& = &
\varepsilon_{A} \after F(\overline{f}) \\
& = &
\varepsilon_{A} \after FG(f \after c) \after F(\eta_{X}) \\
& = &
f \after c \after \varepsilon_{F(X)} \after F(\eta_{X}) \\
& = &
f \after c \\
\overline{\overline{g}}
& = &
G(\overline{g} \after c) \after \eta_{X} \\
& = &
G(\varepsilon_{A} \after F(g)) \after \eta_{X} \\
& = &
G(\varepsilon_{A}) \after \eta_{G(A)} \after g \\
& = &
g.
\end{array}$$
}

\noindent What remains is to show $MKL \cong L$. This follows because
for each $X\in\cat{B}$, the following diagram is a coequaliser in
$\Cat{A}$.
$$\vcenter{\xymatrix@C+1pc{
\llap{$FG$}FGF(X)\ar@/^1ex/[rr]^-{F(\mu_{X}) = FG(\varepsilon_{F(X)})}
   \ar@/_1ex/[rr]_-{\varepsilon_{FGF(X)}} 
   & & FGF(X)\ar@{->>}[r]^-{\varepsilon_{F(X)}} & F(X)
}}$$

\auxproof{
We have:
$$\begin{array}{rcccl}
\varepsilon_{F(X)} \after F(\mu_{X})
& = &
\varepsilon_{F(X)} \after FG(\varepsilon_{F(X)}) 
& = &
\varepsilon_{F(X)} \after \varepsilon_{FGF(X)}.
\end{array}$$

\noindent And if $f\colon FGF(X) \rightarrow A$ satisfies $f \after
F(\mu_{X}) = f \after \varepsilon_{FGF(X)}$, then $\overline{f} = f
\after F(\eta_{X}) \colon F(X) \rightarrow A$ satisfies:
$$\begin{array}{rcl}
\overline{f} \after \varepsilon_{F(X)}
& = &
f \after F(\eta_{X}) \after \varepsilon_{F(X)} \\
& = &
f \after \varepsilon_{FGF(X)} \after FGF(\eta_{X}) \\
& = &
f \after F(\mu_{X}) \after FT(\eta_{X}) \\
& = &
f.
\end{array}$$

\noindent If also $g\colon F(X) \rightarrow A$ satisfies $g \after 
\varepsilon_{F(X)} = f$, then:
$$\begin{array}{rcccccl}
\overline{f}
& = &
f \after F(\eta_{X})
& = &
g \after \varepsilon_{F(X)} \after F(\eta_{X})
& = &
g.
\end{array}$$
}

\noindent Hence the codomain $MKL(X)$ of the coequaliser of $FKL(X) =
FG(\varepsilon_{F(X)})$ and the counit map $\varepsilon_{FGF(X)}$ is
isomorphic to $F(X) = L(X)$. Proving naturality of $MKL \cong L$
(w.r.t.\ Kleisli maps) is a bit of work, but is essentially
straightforward. \QED

\auxproof{
For $f\colon X \rightarrow GF(Y)$, write $f_{*} = \mu_{Y} \after GF(f) =
G(\varepsilon_{F(Y)} \after F(f)) \colon GF(X) \rightarrow GF(Y)$ for the
Kleisli extension. We get $MK(f) = M(f_{*}) = P(f)$ because the
rectangle on the right below commutes.
$$\xymatrix@C+1pc{
\llap{$FG$}FGF(X)\ar@/^1ex/[rr]^-{F(\mu_{X}) = FG(\varepsilon_{F(X)})}
   \ar@/_1ex/[rr]_-{\varepsilon_{FGF(X)}}\ar[d]_{FGF(f_{*})}
   & & FGF(X)\ar@{->>}[r]^-{\varepsilon_{F(X)}}\ar[d]^{F(f_{*})} & 
   F(X)\ar[d]^{P(f) = \varepsilon_{F(Y)} \after F(f)}
\\
\llap{$FG$}FGF(Y)\ar@/^1ex/[rr]^-{F(\mu_{Y}) = FG(\varepsilon_{F(Y)})}
   \ar@/_1ex/[rr]_-{\varepsilon_{FGF(Y)}} 
   & & FGF(Y)\ar@{->>}[r]^-{\varepsilon_{F(Y)}} & F(Y)
}$$

\noindent Indeed,
$$\begin{array}{rcl}
P(f) \after \varepsilon_{F(X)}
& = &
\varepsilon_{F(Y)} \after F(f) \after \varepsilon_{F(X)} \\
& = &
\varepsilon_{F(Y)} \after FG(\varepsilon_{F(Y)} \after F(f)) \\
& = &
\varepsilon_{F(Y)} \after F(f_{*}).
\end{array}$$
}
\end{myproof}

An essential `aha moment' underlying this paper is that the above
result can be massaged into triangle form. This is what happens in the
next result, to which we will refer as the `triangle corollary'. It is
the `recipe' that occurs in the title of this paper.

\begin{corollary}
\label{TriangleCor}
Consider an adjunction $F\dashv G$, where $F$ is a functor $\cat{B}
\rightarrow \cat{A}$, the category $\cat{A}$ has coequalisers, and the
induced monad on $\cat{B}$ is written as $T =
GF$. Diagram~\eqref{ResolutionDiag} then gives rise to a triangle as
below, where both up-going functors are full and faithful.
\begin{equation}
\label{TriangleCorDiag}
\vcenter{\xymatrix@R-.5pc@C-1pc{
\cat{A}\ar@/^0.7em/[rr]^-{K} & \top & \EM\rlap{$(T)$}\ar@/^0.6em/[ll]^-{M}
\\
& \Kl(T)\ar[ul]^{\Pred = L}\ar[ur]_{KL = \Stat} &
}}
\end{equation}

\noindent This triangle commutes, trivially from left to right, and
up-to-isomorphism from right to left, since $MKL \cong L$. In this
context we refer to the functor $L$ as the `predicate' functor
$\Pred$, and to the functor $KL$ as the `states' functor $\Stat$. \QED
\end{corollary}

The remainder of the paper is devoted to instances of this triangle
corollary. In each of these examples the category $\cat{A}$ will be of
the form $\op{\cat{P}}$, where $\cat{P}$ is a category of predicates
(with equalisers).  The full and faithfulness of the functors $\Pred
\colon \Kl(T) \rightarrow \op{\cat{P}}$ and $\Stat \colon \Kl(T)
\rightarrow \EM(T)$ means that there are bijective correspondences
between:
\begin{equation}
\label{ComputationTransformers}
\begin{prooftree}
\xymatrix{X\ar[rr]^-{\text{computations}} & & T(Y)}
\Justifies
\xymatrix{\Pred(Y)\ar[rrr]_-{\text{predicate transformers}} & & & \Pred(X)}
\end{prooftree}
\qquad
\begin{prooftree}
\xymatrix{X\ar[rr]^-{\text{computations}} & & T(Y)}
\Justifies
\xymatrix{\Stat(X)\ar[rrr]_-{\text{state transformers}} & & & \Stat(Y)}
\end{prooftree}
\end{equation}

\noindent Since $\Stat(X) = T(X)$, the correspondence on the right is
given by Kleisli extension, sending a map $f\colon X \rightarrow T(Y)$
to $\mu \after T(f) \colon T(X) \rightarrow T(Y)$. This bijective
correspondence on the right is a categorical formality. But the
correspondence on the left is much more interesting, since it
precisely describes to which kind of predicate transformers
(preserving which structure) computations correspond. Such a
correspondence is often referred to as `healthiness' of the
semantics. It is built into our triangle recipe, as will be
illustrated below.

Before looking at triangle examples, we make the following points.
\begin{itemize}
\item As discussed in~\cite{Jacobs15d}, the predicate functor $\Pred
  \colon \Kl(T) \rightarrow \cat{A}$ is in some cases an
  \emph{enriched} functor, preserving additional structure that is of
  semantical/logical relevance. For instance, operations on programs,
  like $\cup$ for non-deterministic sum, may be expressed as structure
  on Kleisli homsets.  Preservation of this structure by the functor
  $\Pred$ gives the logical rules for dealing with such structure in
  weakest precondition computations. These enriched aspects will not
  be elaborated in the current context.

\item The triangle picture that we use here is refined
  in~\cite{HinoKHJ16}.  In all our examples, the adjunction $F \dashv
  G$ arises by homming into a dualising object $\Omega$. The induced
  monad $T$ is then of the `double dual' form
  $\Omega^{\Omega^{(-)}}$. The approach of~\cite{HinoKHJ16} uses
  monads $S$ having a map of monads $S \Rightarrow T = GF$; this monad
  map corresponds bijectively to an Eilenberg-Moore algebra $S(\Omega)
  \rightarrow \Omega$, which is understood as a logical modality.
\end{itemize}

\section{Dualising with 2}\label{TwoSec}

We split our series of examples in three parts, determined by the
dualising object: $2$, $3$, or $[0,1]$. The first series of Boolean
examples is obtained via adjunctions that involve `homming into $2$',
where $2 = \{0,1\}$ is the 2-element set of Booleans.

\subsection{Sets and sets}\label{SetsSetsSubsec}

We will present examples in the following manner, in three stages.
$$\vcenter{\xymatrix@R-2pc{
\op{\Sets}\ar@/^2ex/[dd]^{\Pow = \Hom(-,2)} \\
\dashv \\
\Sets\ar@/^2ex/[uu]^{\Pow = \Hom(-,2)}\ar@(dl,dr)_{\mathcal{N}=\Pow\Pow}
}}
\qquad
\begin{prooftree}
\begin{prooftree}
\xymatrix{\Pow(X)\ar[r]^-{\op{\Sets}} & Y}
\Justifies
\xymatrix{Y\ar[r]^-{\Sets} & \Pow(X)}
\end{prooftree}
\Justifies
\xymatrix{X\ar[r]_-{\Sets} & \Pow(Y)}
\end{prooftree}%
\qquad
\vcenter{\xymatrix@R-.5pc@C-2pc{
\op{\Sets}\ar@/^0.7em/[rr] & \top & \EM\rlap{$(\mathcal{N}) = \CABA$}\ar@/^0.6em/[ll]
\\
& \Kl(\mathcal{N})\ar[ul]^{\Pred}\ar[ur]_{\Stat} &
}}\hspace*{7em}$$

\noindent On the left we describe the adjunction that forms the basis
for the example at hand, together with the induced monad. In this case
we have the familiar fact that the contravariant powerset functor
$\Pow \colon \Sets \rightarrow \op{\Sets}$ is adjoint to itself, as
indicated. The induced double-powerset monad $\Pow\Pow$ on $\Sets$ is
known in the coalgebra/modal logic community as the neighbourhood
monad $\mathcal{N}$, because its coalgebras are related to
neighbourhood frames in modal logic.

In the middle, the bijective correspondence is described that forms
the basis of the adjunction. In this case there is the obvious
correspondence between functions $Y \rightarrow \Pow(X)$ and functions
$X \rightarrow \Pow(Y)$ --- which are all relations on $X\times Y$.

On the right the result is shown of applying the triangle
corollary~\ref{TriangleCor} to the adjunction on the left. The full
and faithfulness of the predicate functor $\Pred \colon
\Kl(\mathcal{N}) \rightarrow \op{\Sets}$ plays an important role in
the approach to coalgebraic dynamic logic in~\cite{HansenKL14},
relating coalgebras $X \rightarrow \mathcal{N}(X)$ to predicate
transformer functions $\Pow(X) \rightarrow \Pow(X)$, going in the
opposite direction. The category $\EM(\mathcal{N})$ of Eilenberg-Moore
algebras of the neighbourhood monad $\mathcal{N}$ is the category
$\CABA$ of complete atomic Boolean algebras (see
\eg~\cite{Taylor02}). The adjunction $\op{\Sets} \rightleftarrows
\EM(\mathcal{N})$ is thus an equivalence.


\subsection{Sets and posets}\label{SetsPosetsSubsec}

We now restrict the adjunction in the previous subsection to posets.
$$\vcenter{\xymatrix@R-2pc{
\op{\PoSets}\ar@/^2ex/[dd]^{\Upsets = \Hom(-,2)} \\
\dashv \\
\Sets\ar@/^2ex/[uu]^{\Pow = \Hom(-,2)}\ar@(dl,dr)_{\mathcal{M}=\Upsets\Pow}
}}
\qquad
\begin{prooftree}
\xymatrix@C+.5pc{Y\ar[r]^-{\PoSets} & \Pow(X)}
\Justifies
\xymatrix{X\ar[r]_-{\Sets} & \Upsets(Y)}
\end{prooftree}
\qquad
\vcenter{\xymatrix@R-.5pc@C-2pc{
\op{\PoSets}\ar@/^0.7em/[rr] & \top & \EM(\mathcal{M})\rlap{$ = \CDL$}\ar@/^0.6em/[ll]
\\
& \Kl(\mathcal{M})\ar[ul]^{\Pred}\ar[ur]_{\Stat} &
}}\hspace*{6em}$$

\auxproof{
We check the bijective correspondence, for $Y\in\PoSets$a, and $X\in\Sets$,
$$\begin{prooftree}
\xymatrix@C+.5pc{Y\ar[r]^-{f} & \Pow(X)}
\Justifies
\xymatrix{X\ar[r]_-{g} & \Upsets(Y)}
\end{prooftree}$$

\begin{itemize}
\item Given $f\colon Y \rightarrow \Pow(X)$ in $\PoSets$, define
  $\overline{f} \colon X \rightarrow \Upsets(Y)$ as $\overline{f}(x) =
  \setin{y}{Y}{x\in f(y)}$. Each $\overline{f}(x)$ is an upset, since
  if $z \geq y \in \overline{f}(x)$, then $f(z) \supseteq f(y)$, so
  $x\in f(y)$ implies $x\in f(z)$, and thus $z\in \overline{f}(x)$.

\item For $g\colon X \rightarrow \Upsets(Y)$ define $\overline{g}
  \colon Y \rightarrow \Pow(X)$ as $\overline{g}(y) =
  \setin{x}{X}{y\in g(x)}$. This is a monotone function, since if $y
  \leq z$ in $Y$, and $x\in \overline{g}(y)$, then $y \in g(x)$, and
  thus $z\in g(x)$, since $g(x)$ is an upset. Hence $x\in
  \overline{g}(z)$.
\end{itemize}
}

\noindent The functor $\Upsets \colon \op{\PoSets} \rightarrow \Sets$
sends a poset $Y$ to the collection of upsets $U\subseteq Y$,
satisfying $y \geq x \in U$ implies $y\in U$. These upsets can be
identified with monotone maps $p\colon Y \rightarrow 2$, namely as
$p^{-1}(1)$.

Notice that this time there is a bijective correspondence between
computations $X \rightarrow \mathcal{M}(Y) = \Upsets\Pow(Y)$ and
\emph{monotone} predicate transformers $\Pow(Y) \rightarrow \Pow(X)$.
This fact is used in~\cite{HansenKL14}. The algebras of the monad
$\mathcal{M}$ are completely distributive lattices,
see~\cite{Markowsky79} and~\cite[I, Prop.~3.8]{Johnstone82}.


\subsection{Sets and meet-semilattices}\label{SetsMSLSubsec}

We now restrict the adjunction further to meet semilattices, that is,
to posets with finite meets $\wedge, \top$.
$$\vcenter{\xymatrix@R-2pc{
\op{\MSL}\ar@/^2ex/[dd]^{\Hom(-,2)} \\
\dashv \\
\Sets\ar@/^2ex/[uu]^{\Pow = \Hom(-,2)}\ar@(dl,dr)_{\mathcal{F}=\MSL(\Pow(-), 2)}
}}
\qquad
\begin{prooftree}
\xymatrix@C+.5pc{Y\ar[r]^-{\MSL} & \Pow(X)}
\Justifies
\xymatrix{X\ar[r]_-{\Sets} & \MSL(Y, 2)}
\end{prooftree}
\qquad
\vcenter{\xymatrix@R-.5pc@C-2pc{
\op{\MSL}\ar@/^0.7em/[rr] & \top & \EM(\mathcal{F})\rlap{$ = \CCL$}\ar@/^0.6em/[ll]
\\
& \Kl(\mathcal{F})\ar[ul]^{\Pred}\ar[ur]_{\Stat} &
}}\hspace*{6em}$$

\noindent Morphisms in the category $\MSL$ of meet semilattices
preserve the meet $\wedge$ and the top element $\top$ (and hence the
order too).  For $Y\in\MSL$ one can identify a map $Y\rightarrow 2$
with a \emph{filter} of $Y$, that is, with an upset $U\subseteq Y$
closed under $\wedge, \top$.

\auxproof{
\begin{itemize}
\item A filter $U\subseteq Y$ yields a function $p_{U} \colon Y
  \rightarrow 2$ by $p_{U}(y) = 1$ iff $y\in U$.
\begin{itemize}
\item If $x \leq y$ and $p_{U}(x) = 1$, then $x\in U$, so $y\in U$, and
thus $p_{U}(y) = 1$. Hence $p_{U}(x) \leq p_{U}(y)$.

\item $p_{U}(\top) = 1$, since $\top \in U$.

\item In order to get $p_{U}(x \wedge y) = p_{U}(x) \wedge p_{U}(y)$,
  the non-trivial part is $(\geq)$. So assume $p_{U}(x) \wedge
  p_{U}(y) = 1$.  Then $x\in U$ and $y\in U$, so that $x\wedge y\in
  U$. Hence $p_{U}(x\wedge y) = 1$.
\end{itemize}

\item A map $p \colon Y \rightarrow 2$ yields a filter $U_{p} =
  p^{-1}(1) = \setin{y}{Y}{p(y) = 1}$. This is a filter:
\begin{itemize}
\item $y \geq x \in U_{p}$ means $p(x) = 1$, so that $p(y) \geq p(x) =
  1$, and thus $y\in U_{p}$.

\item $\top \in U_{p}$ since $p(\top) = 1$.

\item If $x, y\in U_{p}$, then $p(x\wedge y) = p(x) \wedge p(y) = 1
  \wedge 1 = 1$, so that $x\wedge y \in U_{p}$.
\end{itemize}
\end{itemize}

\noindent Moreover,
$$\begin{array}{rcl}
U_{p_U}
& = &
\set{y}{p_{U}(y) = 1} \\
& = &
\set{y}{y\in U} \\
& = &
U \\
p_{U_p}(y) = 1
& \Longleftrightarrow &
y \in U_{p} \\
& \Longleftrightarrow &
p(y) = 1.
\end{array}$$
}

The resulting monad $\mathcal{F}(X) = \MSL(\Pow(X), 2)$ gives the
filters in $\Pow(X)$. This monad is thus called the \emph{filter
  monad}. In~\cite{Wyler81} it is shown that its category of algebras
$\EM(\mathcal{F})$ is the category $\CCL$ of continuous complete
lattices, that is, of complete lattices in which each element $x$ is
the (directed) join $x = \bigvee\set{y}{y \ll x}$ of the elements way
below it.

\subsection{Sets and complete lattices}\label{SetsCLSubsec}

A poset is called a complete lattice if each subset has a join, or
equivalently, if each subset has a meet. Since these complete lattices
will be used in several examples, we elaborate some basic properties
first. We shall consider two categories with complete lattices as
objects, namely:
\begin{itemize}
\item $\CLJ$ whose morphisms preserve all joins $\bigvee$;

\item $\CLM$ whose morphisms preserve all meets $\bigwedge$.
\end{itemize}

\noindent We write $\op{L}$ for the complete lattice obtained from $L$
by reversing the order. Thus, $f\colon L \rightarrow K$ in
$\CLJ$ gives a map $f \colon \op{L} \rightarrow \op{K}$ in
$\CLM$. Hence we have an isomorphism $\CLJ \cong
\CLM$. Notice that we have:
$$\begin{array}{rcl}
\CLM(L, K)
& \cong &
\CLJ(\op{L}, \op{K})\quad\mbox{as sets}
\end{array}$$

\noindent But:
$$\begin{array}{rcl}
\op{\CLM(L, K)}
& \cong &
\CLJ(\op{L}, \op{K})\quad\mbox{as posets}
\end{array}$$

There is another isomorphism between these two categories of complete
lattices. A basic fact in order theory is that each map $f\colon L
\rightarrow K$ in $\CLJ$ has a right adjoint $f^{\#} \colon K
\rightarrow L$ in $\CLM$, given by:
\begin{equation}
\label{CLmapAdjointEqn}
\begin{array}{rcl}
f^{\#}(b)
& = &
\bigvee\setin{x}{L}{f(x) \leq b}.
\end{array}
\end{equation}

\noindent Clearly, $f(a) \leq b$ implies $a \leq f^{\#}(b)$. For the
reverse direction we apply $f$ to an inequality $a \leq f^{\#}(b)$ and
obtain:
$$\begin{array}{rcccccccl}
f(a)
& \,\leq\, &
f\big(f^{\#}(b)\big)
& \,=\, &
f\big(\bigvee\set{x}{f(x) \leq b}\big)
& \,=\, &
\bigvee\set{f(x)}{f(x)\leq b}
& \,\leq\, &
b.
\end{array}$$

\noindent This gives an isomorphism of categories $\CLJ \cong
\op{\big(\CLM\big)}$. Via a combination with the above
isomorphism $\CLJ \cong \CLM$ we see that the two
categories $\CLJ$ and $\CLM$ are self-dual.

\begin{lemma}
\label{CLTwoLem}
For a complete lattice $L$ there are isomomorphisms of posets:
\begin{equation}
\label{CLJTwoEqn}
\xymatrix{
\CLJ\big(L, 2\big)\ar[r]^-{\cong} & \op{L} &
   \CLJ\big(L, \op{2}\big)\ar[l]_-{\cong}
}
\end{equation}

\noindent Similarly there are isomorphisms:
\begin{equation}
\label{CLMTwoEqn}
\xymatrix{
\CLM\big(L, 2\big)\ar[r]^-{\cong} & \op{L} &
   \CLM\big(L, \op{2}\big)\ar[l]_-{\cong}
}
\end{equation}
\end{lemma}

\begin{myproof}
We restrict ourselves to describing the four isomorphisms.  The
isomorphism on the left in~\eqref{CLJTwoEqn} sends a join-preserving
map $\varphi \colon L \rightarrow 2$ and an element $a\in L$ to:
$$\begin{array}{rclcrcl}
\widehat{\varphi}
& = &
\bigvee\setin{x}{L}{\varphi(x) = 0}
& \qquad\mbox{and}\qquad
\widehat{a}(x)
& = &
\left\{\begin{array}{ll}
0 \quad & \mbox{if } x \leq a \\
1 & \mbox{otherwise.}
\end{array}\right.
\end{array}$$

\noindent The isomorphism on the right in~\eqref{CLJTwoEqn} maps a
$\varphi \colon L \rightarrow \op{2}$ and $a\in L$ to:
$$\begin{array}{rclcrcl}
\widetilde{\varphi}
& = &
\bigvee\set{x}{\varphi(x)=1}
& \qquad\mbox{and}\qquad &
\qquad \widetilde{a}(x) = 1
& \Longleftrightarrow &
x \leq a.
\end{array}$$

\auxproof{
The map from left to right in~\eqref{CLJTwoEqn} sends a
join-preserving map $\varphi \colon L \rightarrow 2$ to the element:
$$\begin{array}{rcl}
\widehat{\varphi}
& = &
\bigvee\setin{x}{L}{\varphi(x) = 0}.
\end{array}$$

\noindent If $\varphi \leq \psi$ in $\CLJ\big(L, 2\big)$,
then $\varphi(x) \leq \psi(x)$ for each $x\in L$, so that:
$$\begin{array}{rcl}
\set{x}{\varphi(x) = 0}
& \supseteq &
\set{x}{\psi(x) = 0},
\end{array}$$

\noindent and thus:
$$\begin{array}{rcccccl}
\widehat{\varphi}
& = &
\bigvee\set{x}{\varphi(x) = 0}
& \geq &
\bigvee\set{x}{\psi(x) = 0}
& = &
\widehat{\psi}.
\end{array}$$

\noindent In the other direction one sends an element $a\in L$ to the
function $\widehat{a} \colon L \rightarrow 2$ given by:
$$\begin{array}{rcl}
\widehat{a}(x)
& = &
\left\{\begin{array}{ll}
0 \quad & \mbox{if } x \leq a \\
1 & \mbox{otherwise.}
\end{array}\right.
\end{array}$$

\noindent Clearly, $\widehat{a}$ preserves joins:
$$\begin{array}{rcl}
\widehat{a}(\bigvee_{i}x_{i}) = 0
& \Longleftrightarrow &
\bigvee_{i}x_{i} \leq a \\
& \Longleftrightarrow &
\all{i}{x_{i} \leq a} \\
& \Longleftrightarrow &
\all{i}{\widehat{a}(x_{i}) = 0} \\
& \Longleftrightarrow &
\bigvee_{i}\widehat{a}(x_{i}) = 0.
\end{array}$$

\noindent Also, if $a \leq b$ then $\widehat{b} \leq \widehat{a}$
since: from $\widehat{a}(x) = 0$ we get $x\leq a$ and thus $x\leq b$
so that $\widehat{b}(x) = 0$.

We have an isomorphism on the left in~\eqref{CLJTwoEqn} since:
$$\begin{array}{rcccccl}
\widehat{\widehat{a}}
& = &
\bigvee\set{x}{\widehat{a}(x) = 0} 
& = &
\bigvee\set{x}{x \leq a} 
& = &
a.
\end{array}$$

\noindent And:
$$\begin{array}{rcccl}
\widehat{\widehat{\varphi}}(y) = 0
& \Longleftrightarrow &
y \leq \widehat{\varphi} = \bigvee\set{x}{\varphi(x)=0}
& \smash{\stackrel{(*)}{\Longleftrightarrow}} &
\varphi(y) = 0.
\end{array}$$

\noindent The direction $(\Leftarrow)$ of the marked equivalence is
obvious. For $(\Rightarrow)$ assume $y \leq \widehat{\varphi}$. Then:
$$\begin{array}{rcccccccl}
\varphi(y)
& \leq &
\varphi(\widehat{\varphi})
& = &
\varphi(\bigvee\set{x}{\varphi(x)=0})
& = &
\bigvee\set{\varphi(x)}{\varphi(x)=0} 
& = &
0.
\end{array}$$

We turn to the isomorphism on the right in~\eqref{CLJTwoEqn}. It is
given by:
$$\begin{array}{rclcrcl}
\widetilde{\varphi}
& = &
\bigvee\set{x}{\varphi(x)=1}
& \qquad\mbox{and}\qquad &
\qquad \widetilde{a}(x) = 1
& \Longleftrightarrow &
x \leq a.
\end{array}$$

\noindent These mappings are monotone. If $\varphi \leq \psi$, then
$\varphi(x) \geq \psi(x)$ for all $x\in L$, so that
$\set{x}{\varphi(x)=1} \supseteq \set{x}{\psi(x)=1}$, and thus
$\widetilde{\varphi} = \bigvee\set{x}{\varphi(x)=1} \geq
\bigvee\set{x}{\psi(x)=1} = \widetilde{\psi}$.

The map $\widetilde{a} \colon L \rightarrow \op{2}$ preserves joins,
since it sends joins to meets:
$$\begin{array}{rcl}
\widetilde{a}(\bigvee_{i}x_{i})=1
& \Longleftrightarrow &
\bigvee_{i}x_{i} \leq a \\
& \Longleftrightarrow &
\all{i}{x_{i} \leq a} \\
& \Longleftrightarrow &
\all{i}{\widetilde{a}(x_{i}) = 1} \\
& \Longleftrightarrow &
\bigwedge_{i}\widetilde{a}(x_{i}) = 1.
\end{array}$$

\noindent We have:
$$\begin{array}{rcccccl}
\widetilde{\widetilde{a}}
& = &
\bigvee\set{x}{\widetilde{a}(x)=1}
& = &
\bigvee\set{x}{x \leq a}
& = &
a.
\end{array}$$

\noindent And:
$$\begin{array}{rcccl}
\widetilde{\widetilde{\varphi}}(x) = 1
& \Longleftrightarrow &
x \leq \widetilde{\varphi} = \bigvee\set{y}{\varphi(y)=1}
& \Longleftrightarrow &
\varphi(x) = 1.
\end{array}$$

\noindent For the direction $(\Rightarrow)$ of the last equivalence
we use that $\varphi$ reverses the order and sends joins to meets:
$$\begin{array}{rcccccccl}
\varphi(x)
& \geq &
\varphi(\widetilde{\varphi})
& = &
\varphi(\bigvee\set{y}{\varphi(y)=1})
& = &
\bigwedge\set{\varphi(y)}{\varphi(y)=1}
& = &
1.
\end{array}$$
}

We turn to the isomorphisms in~\eqref{CLMTwoEqn}. They are a consequence
of~\eqref{CLJTwoEqn} since:
$$\begin{array}{rcccccl}
\CLM(L, 2) 
& \cong &
\op{\CLJ(\op{L}, \op{2})}
& \cong &
\op{(\op{(\op{L})})}
& \cong &
\op{L}.
\end{array}$$

\noindent And similarly:
$$\begin{array}{rcccccl}
\CLM(L, \op{2}) 
& \cong &
\op{\CLJ(\op{L}, 2)}
& \cong &
\op{(\op{(\op{L})})}
& \cong &
\op{L}.
\end{array}$$

\noindent The isomorphism on the left in~\eqref{CLMTwoEqn} is
described explicitly by:
$$\begin{array}{rclcrcl}
\widehat{\varphi}
& = &
\bigwedge\setin{x}{L}{\varphi(x) = 1}
& \qquad\mbox{and}\qquad &
\qquad\widehat{a}(x) = 1
& \Longleftrightarrow &
a \leq x.
\end{array}$$

\auxproof{
\noindent These operations are monotone. If $\varphi \leq \psi$ in
$\CLM(L,2)$ then $\varphi(x) \leq \psi(x)$ for all $x\in L$. Hence
$\set{x}{\varphi(x)=1} \subseteq \set{x}{\psi(x)=1}$, and so
$\widehat{\varphi} = \bigwedge\set{x}{\varphi(x)=1} \geq
\bigwedge\set{x}{\psi(x)=1} = \widehat{\psi}$. In the other direction,
if $a \leq b$, then $\widehat{a} \geq \widehat{b}$, since if
$\widehat{a}(x) \leq \widehat{b}(x)$: if $\widehat{b}(x)= 1$, then $b
\leq x$, so $a\leq x$ and thus $\widehat{a}(x) = 1$.

The map $\widehat{a}$ preserves meets:
$$\begin{array}{rcl}
\widehat{a}(\bigwedge_{i}x_{i}) = 1
& \Longleftrightarrow &
a \leq \bigwedge_{i}x_{i} \\
& \Longleftrightarrow &
\all{i}{a \leq x_{i}} \\
& \Longleftrightarrow &
\all{i}{\widehat{a}(x_{i}) = 1} \\
& \Longleftrightarrow &
\bigwedge_{i}\widehat{a}(x_{i}) = 1.
\end{array}$$

\noindent We have $\widehat{\widehat{a}} =
\bigwedge\set{x}{\widehat{a}(x) = 1} = \bigwedge\set{x}{a \leq x} =
a$, and:
$$\begin{array}{rcccl}
\widehat{\widehat{\varphi}}(x)=1
& \Longleftrightarrow &
\widehat{\varphi} = \bigwedge\set{y}{\varphi(y)=1} \leq x
& \Longleftrightarrow &
\varphi(x)=1.
\end{array}$$

\noindent For the last $(\Rightarrow)$ we use:
$$\begin{array}{rcccccccl}
\varphi(x)
& \geq &
\varphi(\widehat{\varphi})
& = &
\varphi(\bigwedge\set{y}{\varphi(y)=1})
& = &
\bigwedge\set{\varphi(y)}{\varphi(y)=1}
& = &
1.
\end{array}$$
}

\noindent The isomorphism on the right in~\eqref{CLMTwoEqn} is
described explicitly by:
$$\begin{array}{rclcrcl}
\widetilde{\varphi}
& = &
\bigwedge\setin{x}{L}{\varphi(x) = 0}
& \qquad\mbox{and}\qquad &
\qquad\widetilde{a}(x) = 0
& \Longleftrightarrow &
a \leq x.
\end{array}\eqno{\QEDbox}$$

\auxproof{
\noindent We check that these operations are monotone. If $\varphi
\leq \psi$ in $\CLM(L, \op{2})$, then $\widehat{\varphi} \geq
\beta(\psi)$, since $\varphi \leq \psi$ implies $\varphi(x) \geq
\psi(x)$ for all $x$. Hence $\set{x}{\varphi(x) = 0} \subseteq
\set{x}{\psi(x) = 0}$, and thus $\widehat{\varphi} =
\bigwedge\set{x}{\varphi(x) = 0} \geq \bigwedge\set{x}{\psi(x) = 0} =
\widehat{\psi}$. Next, if $a \leq b$, then $\widetilde{a} \geq
\widetilde{b}$ since $\widetilde{a}(x) \leq \widetilde{b}(x)$ for all
$x$: if $\widetilde{b}(x) = 0$, then $b \leq x$, so that $a\leq x$,
and thus $\widetilde{a}(x) = 0$.

The function $\widetilde{a}$ sends meets to joins:
$$\begin{array}{rcl}
\widetilde{a}(\bigwedge_{i}x_{i}) = 0
& \Longleftrightarrow &
a \leq \bigwedge x_{i} \\
& \Longleftrightarrow &
\all{i}{a \leq x_{i}} \\
& \Longleftrightarrow &
\all{i}{\widetilde{a}(x_{i}) = 0} \\
& \Longleftrightarrow &
\bigvee_{i} \widetilde{a}(x_{i}) = 0.
\end{array}$$

\noindent We have a bijective correspondence since:
$$\begin{array}{rcccccl}
\widetilde{\widetilde{a}}
& = &
\bigwedge\set{x}{\widetilde{a}(x) = 0} 
& = &
\bigwedge\set{x}{a \leq x} 
& = &
a.
\end{array}$$

\noindent And:
$$\begin{array}{rcccl}
\widetilde{\widetilde{\varphi}}(x) = 0
& \Longleftrightarrow &
\bigwedge\set{y}{\varphi(y)=0} = \widetilde{\varphi} \leq x 
& \Longleftrightarrow &
\varphi(x) = 0.
\end{array}$$

\noindent For the last $(\Rightarrow)$ we use that $\varphi$ sends
meets to joins and thus reverses the order:
$$\begin{array}{rcccccccl}
\varphi(x)
& \leq &
\varphi(\widetilde{\varphi})
& = &
\varphi(\bigwedge\set{y}{\varphi(y)=0})
& = &
\bigvee\set{\varphi(y)}{\varphi(y)=0}
& = &
0.
\end{array}\eqno{\QEDbox}$$
}
\end{myproof}

We note that the composite isomorphisms $\CLJ(L, 2) \cong \CLJ(L,
\op{2})$ in~\eqref{CLJTwoEqn} and $\CLM(L, 2) \cong \CLM(L, \op{2})$
in~\eqref{CLMTwoEqn} are given by $\varphi\mapsto\neg\varphi$, where
$\neg\varphi(x) = 1$ iff $\varphi(x)=0$.

\auxproof{
If $\varphi$ sends joins to joins, then $\neg\varphi$ sends joins to
meets:
$$\begin{array}{rcl}
\neg\varphi(\bigvee_{i}x_{i}) = 1
& \Longleftrightarrow &
\bigvee_{i}\varphi(x_{i}) = \varphi(\bigvee_{i}x_{i}) = 0 \\
& \Longleftrightarrow &
\all{i}{\varphi(x_{i}) = 0} \\
& \Longleftrightarrow &
\all{i}{\neg\varphi(x_{i}) = 1} \\
& \Longleftrightarrow &
\bigwedge_{i}\neg\varphi(x_{i}) = 1.
\end{array}$$

\noindent In the opposite direction, if $\varphi$ sends joins to meets,
then $\neg\varphi$ sends joins to joins:
$$\begin{array}{rcl}
\neg\varphi(\bigvee_{i}x_{i}) = 0
& \Longleftrightarrow &
\bigwedge_{i}\varphi(x_{i}) = \varphi(\bigvee_{i}x_{i}) = 1 \\
& \Longleftrightarrow &
\all{i}{\varphi(x_{i}) = 1} \\
& \Longleftrightarrow &
\all{i}{\neg\varphi(x_{i}) = 0} \\
& \Longleftrightarrow &
\bigvee_{i}\neg\varphi(x_{i}) = 0.
\end{array}$$

\noindent For the isomorphism $\CLM(L, 2) \cong \CLM(L, \op{2})$
we use that if $\varphi$ sends meets to meets, then $\neg\varphi$
sends meets to joins:
$$\begin{array}{rcl}
\neg\varphi(\bigwedge_{i}x_{i}) = 0
& \Longleftrightarrow &
\bigwedge_{i}\varphi(x_{i}) = \varphi(\bigwedge_{i}x_{i}) = 1 \\
& \Longleftrightarrow &
\all{i}{\varphi(x_{i}) = 1} \\
& \Longleftrightarrow &
\all{i}{\neg\varphi(x_{i}) = 0} \\
& \Longleftrightarrow &
\bigvee_{i}\neg\varphi(x_{i}) = 0.
\end{array}$$

\noindent In the other direction, if $\varphi$ sends meets to joins,
then $\neg\varphi$ sends meets to meets:
$$\begin{array}{rcl}
\neg\varphi(\bigwedge_{i}x_{i}) = 1
& \Longleftrightarrow &
\bigvee_{i}\varphi(x_{i}) = \varphi(\bigwedge_{i}x_{i}) = 0 \\
& \Longleftrightarrow &
\all{i}{\varphi(x_{i}) = 0} \\
& \Longleftrightarrow &
\all{i}{\neg\varphi(x_{i}) = 1} \\
& \Longleftrightarrow &
\bigwedge_{i}\neg\varphi(x_{i}) = 1.
\end{array}$$
}

The state-and-effect triangle of this subsection is given by
the following situation.
$$\vcenter{\xymatrix@R-2pc{
\op{\big(\CLM\big)}\ar@/^2ex/[dd]^{\Hom(-,\op{2}) \cong \op{(-)}} \\
\dashv \\
\Sets\ar@/^2ex/[uu]^{\Pow = \Hom(-,\op{2})}\ar@(dl,dr)_{\Pow(-)}
}}
\quad
\begin{prooftree}
\xymatrix@C+.5pc{L\ar[r]^-{\CLM} & \Pow(X)}
\Justifies
\xymatrix{X\ar[r]_-{\Sets} & L}
\end{prooftree}
\quad
\vcenter{\xymatrix@R-.5pc@C-2pc{
\op{\big(\CLM\big)}\ar@/^0.7em/[rr] & \top & 
   \EM(\Pow)\rlap{$\; = \CLJ$}\ar@/^0.6em/[ll]
\\
& \Kl(\Pow)\ar[ul]^{\Pred}\ar[ur]_{\Stat} &
}}\hspace*{6em}$$

\noindent The upgoing functor on the left $\Pow = \Hom(-,\op{2})$ is
the contravariant powerset functor. In the other direction, the
functor $L \mapsto \Hom(L,\op{2}) \cong \op{L}$, by~\eqref{CLMTwoEqn},
maps a complete lattice $L$ to its underlying set. It sends a
$\bigwedge$-preserving map $L \rightarrow K$ to the associated
($\bigvee$-preserving) map $K \rightarrow L$.

\auxproof{
The functor $\Pow = \Hom(-,\op{2}) \colon \Sets \rightarrow
\op{\big(\CLM\big)}$ is the contravariant powerset functor, since
$\varphi,\psi \colon X \rightarrow \op{2}$ are ordered by $\varphi
\sqsubseteq \psi$ iff $\varphi(x) \geq \psi(x)$ for all $x\in X$. In
particular, for subsets $U,V \subseteq X$ the associated indicator
functions satisfy:
$$\begin{array}{rcccccl}
\indic{U} \sqsubseteq \indic{V}
& \Longleftrightarrow &
\all{x}{\indic{U}(x) \geq \indic{V}(x)}
& \Longleftrightarrow &
\all{x}{x \in V \Rightarrow x \in U}
& \Longleftrightarrow &
V \subseteq U.
\end{array}$$
}

The adjoint correspondence in the middle sends a meet-preserving map
$f\colon L \rightarrow \Pow(X)$ and a function $g\colon X \rightarrow
L$ to the transposes:
$$\begin{array}{rclcrcl}
\overline{f}(x)
& = &
\bigwedge\setin{a}{L}{x\in f(a)}
& \qquad\mbox{and} &
\qquad \overline{g}(a)
& = &
\set{x}{g(x) \leq a}.
\end{array}$$

\auxproof{
\noindent Clearly, $\overline{g}$ preserves meets, and
$\overline{\overline{g}} = g$. We also have:
$$\begin{array}{rcccl}
x\in \overline{\overline{f}}(a)
& \Longleftrightarrow &
\overline{f}(x) = \bigwedge\set{b}{x\in f(b)} \leq a
& \Longleftrightarrow &
x\in f(a).
\end{array}$$

\noindent The direction $(\Leftarrow)$ of the last equivalence is
obvious, and the direction $(\Rightarrow)$ is obtained from:
$$\begin{array}{rcccccccl}
x
& \in &
\bigcap\set{f(b)}{x\in f(b)}
& = &
f(\bigvee\set{b}{x\in f(b)})
& = &
f(\overline{f}(x))
& \subseteq &
f(a).
\end{array}$$
}

\noindent By taking $L = \Pow(Y)$ we get the classical healthiness of
the $\Box$-predicate transformer semantics for non-deterministic
computation~\cite{DijkstraS90}, with a bijective correspondence
between Kleisli maps $X \rightarrow \Pow(Y)$ and meet-preserving maps
$\Pow(Y)\rightarrow \Pow(X)$.

The adjunction $\rightleftarrows$ in the state-and-effect triangle on
the right is an isomorphism of categories, as discussed before
Lemma~\ref{CLTwoLem}. This triangle captures the essence of
non-deterministic program semantics from~\cite{DijkstraS90}, involving
computations, predicate transformation and state transformation.

There is also an adjunction that gives rise to $\Diamond$-predicate
transformer semantics, as join preserving maps. In order to describe
it properly, with opposite orders, we need to use posets instead of
sets, see Subsection~\ref{PosetCLSubsec} below.

\auxproof{
There is a similar adjunction with join-preserving maps:
$$\vcenter{\xymatrix@R-2pc{
\op{\big(\CLJ\big)}\ar@/^2ex/[dd]^{\Hom(-,2) \cong (-)} \\
\dashv \\
\Sets\ar@/^2ex/[uu]^{\Pow = \Hom(-,2)}\ar@(dl,dr)_{\Pow(-)}
}}
\qquad
\begin{prooftree}
\xymatrix@C+.5pc{L\ar[r]^-{f} & \Pow(X)}
\Justifies
\xymatrix{X\ar[r]_-{g} & L}
\end{prooftree}
\qquad
\vcenter{\xymatrix@R-.5pc@C-2pc{
\op{\big(\CLJ\big)}\ar@/^0.7em/[rr] & \top & 
   \EM\rlap{$(\Pow) = \CLJ$}\ar@/^0.6em/[ll]
\\
& \Kl(\Pow)\ar[ul]^{\Pred}\ar[ur]_{\Stat} &
}}\hspace*{6em}$$

\noindent The bijective correspondence works as follows.
\begin{itemize}
\item Given a join preserving map $f\colon L \rightarrow \Pow(X)$ we
  define $\overline{f} \colon X \rightarrow L$ in $\Sets$ as
  $\overline{f}(x) = \bigvee\setin{a}{L}{x \not\in f(a)}$.

\item In the other direction, given a function $g\colon X \rightarrow
  L$ we take $\overline{g} \colon L \rightarrow \Pow(X)$ to be
  $\overline{g}(a) = \setin{x}{X}{a \not\leq g(x)}$. This map
  $\overline{g}$ preserves joins since:
$$\begin{array}{rcccccccl}
x \not\in \overline{g}(\bigvee_{i}a_{i})
& \Longleftrightarrow &
\bigvee_{i}a_{i} \leq g(x) \\
& \Longleftrightarrow &
\all{i}{a_{i} \leq g(x)} \\
& \Longleftrightarrow &
\all{i}{x \not\in \overline{g}(a_{i})} \\
& \Longleftrightarrow &
x \not\in \bigcup_{i}\overline{g}(a_{i}).
\end{array}$$
\end{itemize}

\noindent The operations are each other's inverse:
$$\begin{array}{rcccccl}
\overline{\overline{g}}(x)
& = &
\bigvee\set{a}{x\not\in \overline{g}(a)} 
& = &
\bigvee\set{a}{a \leq g(x)} 
& = &
g(x).
\end{array}$$

\noindent And:
$$\begin{array}{rcccl}
x\not\in\overline{\overline{f}}(a)
& \Longleftrightarrow &
a \leq \overline{f}(x) = \bigvee\set{b}{x\not\in f(b)} 
& \smash{\stackrel{(*)}{\Longleftrightarrow}} &
x \not\in f(a).
\end{array}$$

\noindent The direction $(\Leftarrow)$ is obvious, and for
$(\Rightarrow)$ we reason as follows. Let $a \leq \overline{f}(x) =
\bigvee\set{b}{x\not\in f(b)}$. Then:
$$\begin{array}{rcccl}
f(a)
& \subseteq &
f\big(\bigvee\set{b}{x\not\in f(b)}\big)
& = &
\bigcup\set{f(b)}{x\not\in f(b)}.
\end{array}$$

\noindent Hence if $x\in f(a)$, then $x\in f(b)$ for some $b\in L$
with $x\not\in f(b)$. Clearly, this is impossible.

In the induced triangle we have the self-duality of $\CLJ$ as
isomorphism between categories of predicates and of states. For a
Kleisli map $g\colon X \rightarrow \Pow(Y)$ the induced substitution
functor is written as $\Pred(g) \colon \Pow(Y) \rightarrow \Pow(X)$.
It is not the familiar $\diamond$-formula, but:
$$\begin{array}{rcccccl}
\Pred(g)(V)
& = &
\overline{g}(V)
& = &
\setin{x}{X}{V \not\subseteq g(x)}
& = &
\setin{x}{X}{V \cap \neg g(x) \neq \emptyset}.
\end{array}$$

\noindent This is curious!

If we restrict ourselves to powerset complete lattices, then we do have
the familiar $\diamond$-formula in the bijective correspondence:
$$\begin{prooftree}
\xymatrix{\Pow(Y)\ar[r]^-{f} & \Pow(X) \mbox{ $\bigvee$-preserving}}
\Justifies
\xymatrix{X\ar[r]_-{g} & \Pow(Y)}
\end{prooftree}$$

\noindent via:
$$\begin{array}{rclcrcl}
\overline{f}(x)
& = &
\set{y}{x\in f(\{y\})}
& \qquad\mbox{and} &
\qquad \overline{g}(V)
& = &
\set{x}{g(x) \cap V \neq \emptyset}.
\end{array}$$

\noindent Then indeed:
$$\begin{array}{rcl}
\overline{\overline{g}}(x)
& = &
\set{y}{x\in \overline{g}(\{y\})} \\
& = &
\set{y}{g(x) \cap \{y\} \neq \emptyset} \\
& = &
g(x)
\\
\overline{\overline{f}}(V)
& = &
\set{x}{\overline{f}(x) \cap V \neq \emptyset} \\
& = &
\set{x}{\exin{y}{V}{x\in f(\{y\})}} \\
& = &
\set{x}{x \in \bigcup_{y\in V} f(\{y\})} \\
& = &
\set{x}{x \in  f(\bigcup_{y\in V}\{y\})} \\
& = &
\set{x}{x \in f(V)} \\
& = &
f(V).
\end{array}$$

\noindent The problem is that this correspondence for powerset
lattices cannot be formulated for arbitrary lattices, since it uses
singleton sets.

\auxproof{
The functor $\op{(\CLM)} \rightarrow \Sets$ can be described in this
setting as $L \mapsto \CLM(L, \op{2})$, see~\eqref{CLMTwoEqn}. Let's
write the induced double dual monad on $\Sets$ as $T(X) = \CLM(2^{X},
\op{2})$. We have maps:
$$\xymatrix@R-2pc{
\Pow(X)\ar[rr]^-{\sigma^{\Box}} & & T(X)=\CLM(2^{X}, \op{2}) \\
U\ar@{|->}[rr] & & {\lamin{p}{2^{X}}{\left\{\begin{array}{ll}
   0 \quad & \mbox{if } \allin{x}{U}{p(x)=1}
   \rlap{\quad (written as $\indic{U} \leq p$)} \\
   1 & \mbox{otherwise} \end{array}\right.}}
}\hspace*{5em}$$

\noindent For each $U\subseteq X$ the map $\sigma^{\Box}(U)$ sends
meets to joins since:
$$\begin{array}{rcl}
\sigma^{\Box}(U)(\bigwedge_{i}p_{i}) = 0
& \Longleftrightarrow &
\indic{U} \leq \bigwedge_{i}p_{i} \\
& \Longleftrightarrow &
\all{i}{\indic{U} \leq p_{i}} \\
& \Longleftrightarrow &
\all{i}{\sigma^{\Box}(U)(p_{i}) = 0} \\
& \Longleftrightarrow &
\bigvee_{i}\sigma^{\Box}(U)(p_{i}) = 0.
\end{array}$$

\noindent Moreover, $\sigma^{\Box}$ is an isomorphism, with inverse:
$$\begin{array}{rcl}
(\sigma^{\Box})^{-1}(h)
& = &
\bigcap\setin{V}{\Pow(X)}{h(\indic{V}) = 0}.
\end{array}$$

\noindent Indeed:
$$\begin{array}{rcl}
\big((\sigma^{\Box})^{-1} \after \sigma^{\Box}\big)(U)
& = &
\bigcap\set{V}{\sigma^{\Box}(U)(\indic{V}) = 0} \\
& = &
\bigcap\set{V}{\indic{U} \leq \indic{V}} \\
& = &
\bigcap\set{V}{U \subseteq V} \\
& = &
U
\\
\big(\sigma^{\Box} \after (\sigma^{\Box})^{-1}\big)(h)(p) = 0
& \Longleftrightarrow &
\sigma^{\Box}\big((\sigma^{\Box})^{-1}(h)\big)(p) = 0 \\
& \Longleftrightarrow &
\indic{(\sigma^{\Box})^{-1}(h)} = \bigwedge\set{\indic{V}}{h(\indic{V}) = 0}
   \leq p \\
& \Longleftrightarrow &
h(p) = 0
\end{array}$$

\noindent For $(\Leftarrow)$ write $p = \indic{U}$, so that
$h(\indic{U}) = 0$. Hence $\bigwedge\set{\indic{V}}{h(\indic{V}) = 0}
\leq \indic{U} = p$. For $(\Rightarrow)$, we use that $h$ maps
meets to joins and reverses the order. Hence:
$$\begin{array}{rcccccl}
h(p)
& \leq &
h(\bigwedge\set{\indic{V}}{h(\indic{V}) = 0})
& = &
\bigvee\set{h(\indic{V})}{h(\indic{V}) = 0}
& = &
0.
\end{array}$$

The modality $\tau^{\Box} \colon \Pow(2) \rightarrow \op{2}$ corresponding
to $\sigma^{\Box}$is obtained as:
$$\begin{array}{rcccccl}
\tau^{\Box}(U)
& = &
\sigma^{\Box}(U)(\idmap[2])
& = &
\left\{\begin{array}{ll}
   0 \quad & \mbox{if } \allin{x}{U}{x=1} \\
   1 & \mbox{otherwise} \end{array}\right\}
& = &
\neg(U\subseteq \{1\}).
\end{array}$$

There is also a map:
$$\xymatrix@R-2pc{
\Pow(X)\ar[rr]^-{\sigma^{\Diamond}} & & T(X)=\CLM(2^{X}, \op{2}) \\
U\ar@{|->}[rr] & & {\lamin{p}{2^{X}}{\left\{\begin{array}{ll}
   1 \quad & \mbox{if } \allin{x}{U}{p(x)=0} \\
   0 & \mbox{otherwise} \end{array}\right.}}
}\hspace*{5em}$$

We prove that $\sigma^{\Diamond}(U)$ sends meets to join:
$$\begin{array}{rcl}
\sigma^{\Diamond}(U)(\bigwedge_{i}p_{i}) = 0
& \Longleftrightarrow &
\exin{x}{U}{(\bigwedge_{i}p_{i})(x) = 1} \\
& \Longleftrightarrow &
\exin{x}{U}{\all{i}{p_{i}(x) = 1}} \\
& \Longleftrightarrow &

\end{array}$$

\noindent The induced modality $\tau^{\Diamond} \colon \Pow(2)
\rightarrow \op{2}$ is given by:
$$\begin{array}{rcccccl}
\tau^{\Diamond}(U)
& = &
\sigma^{\Diamond}(U)(\idmap[2])
& = &
\left\{\begin{array}{ll}
   0 \quad & \mbox{if } \allin{x}{\neg U}{x=1} \\
   1 & \mbox{otherwise} \end{array}\right\}
& = &
\neg(1\in U).
\end{array}$$

There is, by-the-way, a slightly different adjunction $\Sets
\leftrightarrows \op{\big(\CLJ\big)}$ but it also does not
give the $\diamond$-formula. This adjunction sends a set $X$ to the
complete lattice $\op{\Pow(X)}$, and involves:
$$\begin{prooftree}
\xymatrix{L\ar[r]^-{f} & \op{\Pow(X)} \mbox{ $\bigvee$-preserving}}
\Justifies
\xymatrix{X\ar[r]_-{g} & L}
\end{prooftree}$$

\noindent via:
$$\begin{array}{rclcrcl}
\overline{f}(x)
& = &
\bigvee\setin{a}{L}{x\in f(a)}
& \qquad\mbox{and} &
\qquad \overline{g}(a)
& = &
\set{x}{a \leq g(x)}.
\end{array}$$

\noindent This works since:
$$\begin{array}{rcl}
\overline{g}(\bigvee_{i}a_{i})
& = &
\set{x}{\bigvee_{i}a_{i} \leq g(x)} \\
& = &
\set{x}{\all{i}{a_{i} \leq g(x)}} \\
& = &
\bigcap_{i}\set{x}{a_{i} \leq g(x)} \\
& = &
\bigcap_{i} \overline{g}(a_{i}).
\end{array}$$

\noindent And:
$$\begin{array}{rcl}
\overline{\overline{g}}(x)
& = &
\bigvee\set{a}{x\in\overline{g}(a)} \\
& = &
\bigvee\set{a}{a \leq g(x)} \\
& = &
g(x)
\\
x \in \overline{\overline{f}}(a)
& \Leftrightarrow &
a \leq \overline{f}(x) = \bigvee\set{b}{x\in f(b)} \\
& \Leftrightarrow &
x \in f(a).
\end{array}$$

\noindent In the last step the direction $(\Leftarrow)$ is obvious.
For $(\Rightarrow)$ let $a \leq \overline{f}(x)$, then we are done by:
$$\begin{array}{rcccccl}
f(a)
& \supseteq &
f(\overline{f}(x))
& = &
\bigcap\set{f(b)}{x\in f(b)}
& \ni &
x.
\end{array}$$

\noindent In this case we get for $g\colon X \rightarrow \Pow(Y)$,
$$\begin{array}{rcccl}
\Pred(g)(V)
& = &
\overline{g}(V)
& = &
\set{x}{V \subseteq g(x)}.
\end{array}$$
}
}

\subsection{Sets and Boolean algebras}\label{SetsBASubsec}

We further restrict the adjunction $\op{\MSL} \leftrightarrows \Sets$
from Subsection~\ref{SetsMSLSubsec} to the category $\BA$ of Boolean
algebras.
$$\vcenter{\xymatrix@R-2pc{
\op{\BA}\ar@/^2ex/[dd]^{\Hom(-,2)} \\
\dashv \\
\Sets\ar@/^2ex/[uu]^{\Pow = \Hom(-,2)}\ar@(dl,dr)_{\mathcal{U}=\BA(\Pow(-), 2)}
}}
\qquad
\begin{prooftree}
\xymatrix@C+.5pc{Y\ar[r]^-{\BA} & \Pow(X)}
\Justifies
\xymatrix{X\ar[r]_-{\Sets} & \BA(Y, 2)}
\end{prooftree}
\qquad
\vcenter{\xymatrix@R-.5pc@C-2pc{
\op{\BA}\ar@/^0.7em/[rr] & \top & \EM\rlap{$(\mathcal{U}) = \CH$}\ar@/^0.6em/[ll]
\\
& \Kl(\mathcal{U})\ar[ul]^{\Pred}\ar[ur]_{\Stat} &
}}\hspace*{6em}$$

\noindent The functor $\Hom(-, 2) \colon \op{\BA} \rightarrow \Sets$
sends a Boolean algebra $Y$ to the set $\BA(Y,2)$ of Boolean algebra
maps $Y\rightarrow 2$. They can be identified with \emph{ultrafilters}
of $Y$. The resulting monad $\mathcal{U} = \BA(\Pow(-), 2)$ is the
\emph{ultrafilter monad}, sending a set $X$ to the BA-maps $\Pow(X)
\rightarrow 2$, or equivalently, the ultrafilters of $\Pow(X)$.

An important result of Manes (see~\cite{Manes69}, and also~\cite[III,
  2.4]{Johnstone82}) says that the category of Eilenberg-Moore
algebras of the ultrafilter monad $\mathcal{U}$ is the category $\CH$
of compact Hausdorff spaces. This adjunction $\op{\BA}
\rightleftarrows \CH$ restricts to an equivalence $\op{\BA} \simeq
\Stone$ called Stone duality, where $\Stone \hookrightarrow \CH$ is
the full subcategory of Stone spaces --- in which each open subset is
the union of the clopens contained in it.

\subsection{Sets and complete Boolean algebras}\label{SetsCBASubsec}

We can restrict the adjunction $\op{\BA} \rightleftarrows \Sets$ from
the previous subsection to an adjunction $\op{\CBA} \rightleftarrows
\Sets$ between \emph{complete} Boolean algebras and sets. The
resulting monad on $\Sets$ is of the form $X \mapsto \CBA(\Pow(X),
2)$. But here we hit a wall, since this monad is the identity.

\begin{lemma}
\label{CBALem}
For each set $X$ the unit map $\eta \colon X \rightarrow \CBA(\Pow(X),
2)$, given by $\eta(x)(U) = 1$ iff $x\in U$, is an isomorphism.
\end{lemma}

\begin{myproof}
Let $h\colon \Pow(X) \rightarrow 2$ be a map of complete Boolean
algebras, preserving the BA-structure and all joins (unions). Since
each $U\in\Pow(X)$ can be described as union of singletons, the
function $h$ is determined by its values $h(\{x\})$ for $x\in X$. We
have $1 = h(X) = \bigcup_{x\in X} h(\{x\})$. Hence $h(\{x\}) = 1$ for
some $x\in X$.  But then $h(X - \{x\}) = h(\neg \{x\}) = \neg h(\{x\})
= \neg 1 = 0$. This implies $h(\{x'\}) = 0$ for each $x'\neq x$. Hence
$h = \eta(x)$. \QED
\end{myproof}

\subsection{Posets and complete lattices}\label{PosetCLSubsec}

We return to complete lattices, from Subsection~\ref{SetsCLSubsec},
but now consider them with join-preserving maps:
$$\vcenter{\xymatrix@R-2pc{
\op{\big(\CLJ\big)}\ar@/^2ex/[dd]^{\Hom(-,2)\cong\op{(-)}} \\
\dashv \\
\PoSets\ar@/^2ex/[uu]^{\Upsets = \Hom(-,2)}\ar@(dl,dr)_{\Dwnsets}
}}
\quad
\begin{prooftree}
\xymatrix@C+.5pc{L\ar[r]^-{\CLJ} & \Upsets(X)}
\Justifies
\xymatrix{X\ar[r]_-{\PoSets} & \op{L}}
\end{prooftree}
\quad
\vcenter{\xymatrix@R-.5pc@C-2.5pc{
\op{\big(\CLJ\big)}\ar@/^0.7em/[rr] & \top & \EM(\Dwnsets)\rlap{$ =\! \CLJ$}\ar@/^0.6em/[ll]
\\
& \Kl(\Dwnsets)\ar[ul]^{\Pred}\ar[ur]_{\Stat} &
}}\hspace*{10em}$$

\noindent Recall from Subsection~\ref{SetsPosetsSubsec} that we write
$\Upsets(X)$ for the poset of upsets in a poset $X$, ordered by
inclusion.  This poset is a complete lattice via unions. For a
monotone function $f\colon X \rightarrow Y$ between posets, the
inverse image map $f^{-1}$ restricts to $\Upsets(Y) \rightarrow
\Upsets(X)$ and preserves unions. This gives the functor $\Upsets \colon
\PoSets \rightarrow \op{(\CLJ)}$, which is isomorphic to $\Hom(-,2)$,
as already noted in Subsection~\ref{SetsPosetsSubsec}.

\auxproof{
If $V \subseteq Y$ is a downset, and $x' \geq x \in f^{-1}(V)$, then
$f(x') \geq f(x) \in V$, so that $f(x') \in V$, and $x'\in f^{-1}(V)$.

For an upset $U\subseteq X$ define $\widehat{U}\colon X \rightarrow 2$
by $\widehat{U}(x) = 1$ iff $x\in U$. This map is monotone: if $x \leq
x'$, and $\widehat{U}(x) = 1$, then $x\in U$, so $x'\in U$, and thus
$\widehat{U}(x') = 1$.

For a monotone map $\varphi \colon X \rightarrow 2$ define
$\widehat{\varphi} = \set{x}{\varphi(x) = 1}$. This is an upset: if
$x' \geq x\in \widehat{\varphi}$, then $\varphi(x') \geq \varphi(x) =
1$, so $x'\in \widehat{\varphi}$.

We have:
$$\begin{array}{rcccccl}
\widehat{\widehat{U}}
& = &
\set{x}{\widehat{U}(x) = 1}
& = &
\set{x}{x\in U}
& = &
U.
\end{array}$$

\noindent And:
$$\begin{array}{rcccl}
\widehat{\widehat{\varphi}}(x) = 1
& \Longleftrightarrow &
x \in \widehat{\varphi}
& \Longleftrightarrow &
\varphi(x) = 1.
\end{array}$$
}

The downgoing functor $\Hom(-,2) \colon \op{(\CLJ)} \rightarrow
\PoSets$ is isomorphic to taking the opposite order $\op{(-)}$, see
Lemma~\ref{CLTwoLem}. A map $f\colon L \rightarrow K$ in $\CLJ$ is
mapped to the monotone adjoint function $f^{\#} \colon \op{K}
\rightarrow \op{L}$, as in~\eqref{CLmapAdjointEqn}, given by
$f^{\#}(a) = \bigvee\set{b}{f(b) \leq a}$.

\auxproof{
We use the isomorphism on the left in~\eqref{CLJTwoEqn} twice:
$$\begin{array}{rcccccl}
\op{f}(a)
& = &
\widehat{\widehat{a} \after f}
& = &
\bigvee\set{b}{\widehat{a}(f(b)) = 0}
& = &
\bigvee\set{b}{f(b) \leq a}.
\end{array}$$
}

We elaborate the bijective correspondence in the middle in detail.
\begin{itemize}
\item Given a join preserving map $f\colon L \rightarrow \Upsets(X)$ we
  define $\overline{f} \colon X \rightarrow \op{L}$ in $\PoSets$ as
  $\overline{f}(x) = \bigvee\setin{a}{L}{x \not\in f(a)}$. It is easy
to see that $\overline{f}$ is monotone.

\auxproof{
Let $x \leq y$. Since each $f(a)$ is an upset, we have:
$$\begin{array}{rcl}
\set{a}{x\not\in f(a)}
& \supseteq &
\set{a}{y\not\in f(a)}
\end{array}$$

\noindent and thus:
$$\begin{array}{rcccccl}
\overline{f}(x)
& = &
\bigvee\set{a}{x\not\in f(a)}
& \geq &
\bigvee\set{a}{y\not\in f(a)}
& = &
\overline{f}(y).
\end{array}$$
}

\item In the other direction, given a monotone function $g\colon X
  \rightarrow \op{L}$ we take $\overline{g} \colon L \rightarrow
  \Upsets(X)$ to be $\overline{g}(a) = \setin{x}{X}{a \not\leq
    g(x)}$. This yields an upset: if $x' \geq x \in \overline{g}(a)$,
  then $a \not\leq g(x')$. If $a \leq g(x')$ then $a \leq g(x)$ since
  $g(x') \leq g(x)$ because $g$ reverses the order. This map
  $\overline{g}$ preserves joins since:
$$\begin{array}{rcl}
x \not\in \overline{g}(\bigvee_{i}a_{i})
\hspace*{\arraycolsep}\Longleftrightarrow\hspace*{\arraycolsep}
\bigvee_{i}a_{i} \leq g(x) 
& \Longleftrightarrow &
\all{i}{a_{i} \leq g(x)} \\
& \Longleftrightarrow &
\all{i}{x \not\in \overline{g}(a_{i})} 
\hspace*{\arraycolsep}\Longleftrightarrow\hspace*{\arraycolsep}
x \not\in \bigcup_{i}\overline{g}(a_{i}).
\end{array}$$
\end{itemize}

\noindent The transformations are each other's inverse:
$$\begin{array}{rcccccl}
\overline{\overline{g}}(x)
& = &
\bigvee\set{a}{x\not\in \overline{g}(a)} 
& = &
\bigvee\set{a}{a \leq g(x)} 
& = &
g(x).
\end{array}$$

\noindent And:
$$\begin{array}{rcccl}
x\not\in\overline{\overline{f}}(a)
& \Longleftrightarrow &
a \leq \overline{f}(x) = \bigvee\set{b}{x\not\in f(b)} 
& \smash{\stackrel{(*)}{\Longleftrightarrow}} &
x \not\in f(a).
\end{array}$$

\noindent The direction $(\Leftarrow)$ of the marked equivalence is
obvious, and for $(\Rightarrow)$ we reason as follows. Let $a \leq
\overline{f}(x) = \bigvee\set{b}{x\not\in f(b)}$. Then, using that
$f$ preserves joins:
$$\begin{array}{rcccl}
f(a)
& \subseteq &
f\big(\bigvee\set{b}{x\not\in f(b)}\big)
& = &
\bigcup\set{f(b)}{x\not\in f(b)}.
\end{array}$$

\noindent Hence if $x\in f(a)$, then $x\in f(b)$ for some $b\in L$
with $x\not\in f(b)$. Clearly, this is impossible.

We notice that the induced monad on $\PoSets$ is given by taking
downsets $\Dwnsets(-)$, since the reversed poset $\op{\Upsets(X)}$ is
the poset $\Dwnsets(X)$ of downsets of $X$, ordered by inclusion.  The
isomorphism $\op{\Upsets(X)} \cong \Dwnsets(X)$ is given by
complements. For a monotone map $f\colon X \rightarrow Y$ the function
$\Dwnsets(f) \colon \Dwnsets(X) \rightarrow \Dwnsets(Y)$ sends a
downset $U\subseteq X$ to the downclosure of the image: $\downset f(U)
= \setin{y}{Y}{\exin{x}{U}{y\leq f(x)}}$. This function $\Dwnsets(f)$ is
clearly monotone.

\auxproof{
For an upset $U \subseteq X$ the complement $\neg U$ is a downset: let
$x' \leq x \in \neg U$; we need to prove $x'\in\neg U$. If not, then
$x'\in U$, but then also $x\in U$ since $U$ is an upset. Similarly, if
$V \subseteq X$ is a downset, then $\neg V$ is an upset: if $x' \geq x
\in \neg V$, then $x'\in\neg V$, because if $x'\in V$, then $x\in V$
since $V$ is a downset.
}

If we incorporate this isomorphism $\op{\Upsets(X)} \cong \Dwnsets(X)$,
then the adjoint correspondence specialises to:
\begin{equation}
\label{PosetCLMuDiamondCorr}
\begin{prooftree}
\xymatrix@C+.5pc{\Upsets(Y)\ar[r]^-{f} & \Upsets(X)}
\Justifies
\xymatrix{X\ar[r]_-{g} & \Dwnsets(Y)}
\end{prooftree}
\qquad\mbox{given by}\qquad
\left\{\begin{array}{rcl}
\overline{f}(x)
& = &
\bigcap\setin{U}{\Dwnsets(X)}{x\not\in f(\neg U)} \\
\overline{g}(V)
& = &
\setin{x}{X}{g(x) \cap V \neq \emptyset}
\end{array}\right.
\end{equation}

\noindent We see that in this adjunction $\op{(\CLJ)} \leftrightarrows
\PoSets$ gives rise to the $\Diamond$-predicate transformer. Again,
healthiness is built into the construction.

This correspondence gives a handle on the downsets monad $\Dwnsets$ on
$\PoSets$. The unit $\eta\colon X \rightarrow \Dwnsets(X)$ is obtained
by transposing the identity on $\Upsets(X)$, so that:
$$\begin{array}{rcccccccl}
\eta(x)
& = &
\overline{\idmap}(x)
& = &
\bigcap\setin{U}{\Dwnsets(X)}{x\not\in \neg U}
& = &
\bigcap\setin{U}{\Dwnsets(X)}{x\in U}
& = &
\downset x.
\end{array}$$

\auxproof{
With this unit we can also derive what $\Dwnsets(f)$, for $f\colon X
\rightarrow Y$ should be, namely the $\op{(-)}$ of the transpose
$\Upsets(Y) \rightarrow \Upsets(X)$ of $\eta \after f\colon X
\rightarrow \Dwnsets(Y)$, pre- and post-composed with complement:
$$\begin{array}{rcl}
\Dwnsets(f)(U)
& = &
\neg\op{(\overline{\eta \after f})}(\neg U) \\
& = &
\neg\bigcup\setin{V}{\Upsets(Y)}
   {\overline{\eta \after f}(V) \subseteq \neg U} \\
& = &
\bigcap\setin{V}{\Dwnsets(Y)}
   {\overline{\eta \after f}(\neg V) \subseteq \neg U} \\
& = &
\bigcap\setin{V}{\Dwnsets(Y)}
   {\set{x}{\downset f(x) \cap \neg V \neq \emptyset} \subseteq \neg U} \\
& = &
\bigcap\setin{V}{\Dwnsets(Y)}
   {U \subseteq \set{x}{\downset f(x) \cap \neg V = \emptyset}} \\
& = &
\bigcap\setin{V}{\Dwnsets(Y)}{\allin{x}{U}{\downset f(x) \subseteq V}} \\
& = &
\bigcap\setin{V}{\Dwnsets(Y)}{\allin{x}{U}{f(x) \in V}} \\
& = &
\bigcap\setin{V}{\Dwnsets(Y)}{f(U) \subseteq V} \\
& = &
\overline{f(U)}.
\end{array}$$
}

\noindent The multiplication $\mu \colon \Dwnsets^{2}(X) \rightarrow
\Dwnsets(X)$ is given by union. To see this, we first transpose the
identity map on $\Dwnsets(X)$ upwards, giving a map $\varepsilon\colon
\Upsets(X) \rightarrow \Upsets(\Dwnsets(X))$ described by:
$$\begin{array}{rcl}
\varepsilon(V)
& = &
\setin{U}{\Dwnsets(X)}{U \cap V \neq \emptyset}.
\end{array}$$

\noindent We then obtain the multiplication map $\mu$ of the downset
monad by applying the $\op{(-)}$ functor to $\varepsilon$, and using
complement on both sides:
\begin{equation}
\label{PosetCLMuEqn}
\begin{array}{rcl}
\mu(B)
\hspace*{\arraycolsep}=\hspace*{\arraycolsep}
\neg\op{\varepsilon}(\neg B) 
& = &
\neg\bigcup\setin{V}{\Upsets(X)}{\varepsilon(V) \subseteq \neg B} \\
& = &
\bigcap\setin{V}{\Dwnsets(X)}{B \subseteq \neg\varepsilon(\neg V)} \\
& = &
\bigcap\setin{V}{\Dwnsets(X)}{\allin{U}{B}{U\cap \neg V = \emptyset}} \\
& = &
\bigcap\setin{V}{\Dwnsets(X)}{\bigcup B\subseteq V} \\
& = &
\bigcup B.
\end{array}
\end{equation}

\noindent This last equation holds because the union of downclosed sets is
downclosed.

\auxproof{
We check that $\eta$ and $\mu$ are natural transformations:
$$\begin{array}{rcccccccl}
\Dwnsets(f)\big(\eta(x)\big)
& = &
\set{y}{\exin{x'}{\downset x}{y \leq f(x')}} 
& = &
\set{y}{y \leq f(x)} 
& = &
\downset f(x)
& = &
\eta(f(x)).
\end{array}$$

\noindent Similarly:
$$\begin{array}{rcl}
\big(\mu \after \Dwnsets^{2}(f)\big)(A)
& = &
\mu\big(\set{V}{\exin{U}{A}{V \subseteq \Dwnsets(f)(U)}}\big) \\
& = &
\bigcup\set{V}{\exin{U}{A}{V \subseteq \Dwnsets(f)(U)}} \\
& = &
\bigcup\set{\Dwnsets(f)(U)}{U\in A} \\
& = &
\set{y}{\exin{U}{A}{\exin{x}{U}{y \leq f(x)}}} \\
& = &
\set{y}{\ex{x\in\bigcup A}{y \leq f(x)}} \\
& = &
\set{y}{\ex{x\in\mu(A)}{y \leq f(x)}} \\
& = &
\big(\Dwnsets(f) \after \mu\big)(A)
\end{array}$$

We also check the monad equations. First, for $U\in\Dwnsets(X)$,
$$\begin{array}{rcl}
\big(\mu \after \eta\big)(U)
& = &
\mu(\downset U) \\
& = &
\bigcup \downset U \\
& = &
U
\\
\big(\mu \after \Dwnsets(\eta)\big)(U)
& = &
\mu\big(\downset\set{\downset x}{x\in U}\big) \\
& = &
\bigcup \downset\set{\downset x}{x\in U} \\
& = &
\bigcup \set{\downset x}{x\in U} \\
& = &
\downset U \\
& = &
U.
\end{array}$$

\noindent Next, for $A\in\Dwnsets^{3}(X)$,
$$\begin{array}{rcl}
\big(\mu \after \Dwnsets(\mu)\big)(A)
& = &
\mu\big(\downset\set{\mu(B)}{B\in A}\big) \\
& = &
\bigcup\downset\set{\mu(B)}{B\in A} \\
& = &
\set{x}{\exin{U}{\downset\set{\mu(B)}{B\in A}}{x\in U}} \\
& = &
\set{x}{\ex{U}{\exin{B}{A}{U\subseteq \mu(B) \mbox{ and } x\in U}}} \\
& = &
\set{x}{\exin{B}{A}{x\in \mu(B)}} \\
& = &
\set{x}{\exin{B}{A}{\exin{U}{B}{x\in B}}} \\
& = &
\set{x}{\exin{U}{\bigcup A}{x\in U}} \\
& = &
\set{x}{x \in \bigcup\bigcup A} \\
& = &
\bigcup\bigcup A \\
& = &
\mu\big(\bigcup A\big) \\
& = &
\big(\mu \after \mu\big)(A)
\end{array}$$
}

The category $\EM(\Dwnsets)$ of Eilenberg-Moore algebras of this
downset monad $\Dwnsets$ is the category $\CLJ$ of complete lattices
and join-preserving maps. Hence the adjunction $\rightleftarrows$
above on the right is an isomorphism of categories.

\auxproof{
If $X$ is a complete lattice, then then its join map forms an
$\Dwnsets$-algebra $\bigvee \colon \Dwnsets(X) \rightarrow X$, since:
\begin{itemize}
\item $(\bigvee \after \eta)(x) = \bigvee \downset x = x$.

\item For a downset of downsets $A\subseteq \Dwnsets(X)$,
$$\begin{array}{rcl}
\big(\bigvee \after \Dwnsets(\bigvee)\big)(A)
& = &
\bigvee\downset\set{\bigvee U}{U\in A} \\
& = &
\bigvee\set{\bigvee U}{U\in A} \\
& = &
\bigvee\bigcup A \\
& = &
\big(\bigvee \after \mu\big)(A).
\end{array}$$
\end{itemize}

\noindent A join-preserving map is obviously a map of algebras.

In the other direction, if a dcpo $X$ carries an algebra $\alpha
\colon \Dwnsets(X) \rightarrow X$ in $\Dcpo$, then each subset
$U\subseteq X$ has a join $\bigvee U = \alpha(\downset U)$, since:
\begin{itemize}
\item if $x\in U$, then $\downset x \subseteq \downset U$, so that $x
  = \alpha(\downset x) \leq \alpha(\downset U) = \bigvee U$.

\item If $x\leq y$ for each $x\in U$, then $U \subseteq \downset y$,
  so that $\downset U \subseteq \downset y$. Hence $\bigvee U =
  \alpha(\downset U) \leq \alpha(\downset y) = y$.
\end{itemize}
}

\subsection{Dcpo's and complete lattices}\label{DcpoCLSubsec}

We write $\Dcpo$ for the category with directed complete partial
orders (dcpos) as objects, and (Scott) continuous functions
(preserving directed joins) as morphisms between them. A subset
$U\subseteq X$ of a dcpo $X$ is (Scott) open if $U$ is an upset
satisfying for each directed collection $(x_{i})$, if
$\bigvee_{i}x_{i}\in U$, then $x_{i}\in U$ for some index $i$. The
(Scott) closed sets are then the downsets that are closed under
directed joins. We write $\Open(X)$ and $\Closed(X)$ for the sets of
open and closed subsets of $X$.

\begin{lemma}
\label{DcpoTwoLem}
For each dcpo $X$ there are isomorphisms:
\begin{equation}
\label{DcpoTwoEqn}
\begin{array}{rclcrcl}
\Open(X)
& \cong &
\Dcpo(X, 2)
& \qquad\mbox{and}\qquad & \qquad
\Closed(X)
& \cong &
\Dcpo(X, \op{2}).
\end{array}
\end{equation}

\noindent Moreover, via complements we have an isomorphism of complete
lattices $\op{\Open(X)} \cong \Closed(X)$. In combination
with~\eqref{CLJTwoEqn} we get $\Closed(X) \cong \CLJ(\Open(X), 2)$.
\end{lemma}

\begin{myproof}
The first isomorphism in~\eqref{DcpoTwoEqn} sends an open subset
$U\subseteq X$ to the function $\widehat{U} \colon X \rightarrow 2$
given by $\widehat{U}(x) = 1$ iff $x\in U$. In the other direction,
for a continuous function $\varphi\colon X \rightarrow 2$ we take the
open subset $\widehat{\varphi} = \setin{x}{L}{\varphi(x) = 1}$.
Similarly, the second isomorphism sends a closed subset $V$ to the
function $\widetilde{V} \colon X \rightarrow \op{2}$ with
$\widetilde{V}(x) = 1$ iff $x\not\in V$, and conversely sends $\psi
\colon X \rightarrow \op{2}$ to $\widetilde{\psi} =
\setin{x}{X}{\psi(x) = 0}$. \QED

\auxproof{
Then:
$$\begin{array}{rcl}
\widehat{U}(\bigvee_{i}x_{i}) = 1
& \Longleftrightarrow &
\bigvee_{i}x_{i}\in U \\
& \Longleftrightarrow &
\exin{i}{x_{i}\in U} \\
& \Longleftrightarrow &
\exin{i}{\widehat{U}(x_{i}) = 1} \\
& \Longleftrightarrow &
\bigvee_{i} \widehat{U}(x_{i}) = 1.
\end{array}$$

\noindent These operations are monotone. If $U\subseteq V$, then
$\widehat{U} \leq \widehat{V}$, since $\widehat{U}(x) \leq
\widehat{V}(x)$ for each $x$: if $\widehat{U}(x) = 1$, then $x\in U$,
so $x\in V$ and thus $\widehat{V}(x) = 1$. Similarly, if $\varphi \leq
\psi$, then $\widehat{\varphi} = \set{x}{\varphi(x) = 1} \subseteq
\set{x}{\psi(x) = 1} = \widehat{\psi}$.

These operations are each other's inverse:
$$\begin{array}{rcccccl}
\widehat{\widehat{U}}
& = &
\set{x}{\widehat{U}(x) = 1} 
& = &
\set{x}{x\in U} 
& = &
U.
\end{array}$$

\noindent And:
$$\begin{array}{rcccl}
\widehat{\widehat{\varphi}}(x) = 1
& \Longleftrightarrow &
x\in\widehat{\varphi} 
& \Longleftrightarrow &
\varphi(x) = 1.
\end{array}$$

\noindent For a closed subset $V\subseteq X$ we define $\widetilde{V}
\colon X \rightarrow \op{2}$ as $\widetilde{V}(x) = 1$ iff $x\not\in
V$.  And for a continuous map $\varphi\colon X \rightarrow \op{2}$ we
take $\widetilde{\varphi} = \setin{x}{X}{\varphi(x) = 0}$. We check:
\begin{itemize}
\item The function $\widetilde{V}$ is continuous:
$$\begin{array}{rcl}
\widetilde{V}(\bigvee_{i}x_{i}) = 1
& \Longleftrightarrow &
\bigvee_{i}x_{i} \in \neg V \\
& \Longleftrightarrow &
\ex{i}{x_{i}\in \neg V} \\
& \Longleftrightarrow &
\ex{i}{\widetilde{V}(x_{i}) = 1} \\
& \Longleftrightarrow &
\bigvee_{i}\widetilde{V}(x_{i}) = 1.
\end{array}$$

\item The set $\widetilde{\varphi}$ is closed, since its complement
$\neg\widetilde{\varphi} = \set{x}{\varphi(x) = 1}$ is open.

\item If $U\subseteq V$, then $\widetilde{U} \leq \widetilde{V}$ since
  $\widetilde{U}(x) \geq \widetilde{V}(x)$ for each $x\in X$: if
  $\widetilde{V}(x) = 1$, then $x\not\in V$, so $x\not\in U$, and thus
  $\widetilde{U}(x) = 1$.

\item If $\varphi \leq \psi$, then $\widetilde{\varphi} \subseteq
  \widetilde{\psi}$. Indeed, if $\varphi(x) \geq \psi(x)$, then
  $\widetilde{\varphi} = \set{x}{\varphi(x) = 0} \subseteq
  \set{x}{\psi(x) = 0} = \widetilde{\psi}$.

\item We have:
$$\begin{array}{rcccccl}
\widetilde{\widetilde{V}}
& = &
\set{x}{\widetilde{V}(x) = 0}
& = &
\set{x}{x \in V}
& = &
V.
\end{array}$$

\item And:
$$\begin{array}{rcccccl}
\widetilde{\widetilde{\varphi}}(x) = 1
& \Longleftrightarrow &
x\not\in\widetilde{\varphi}
& \Longleftrightarrow &
\varphi(x) \neq 0
& \Longleftrightarrow &
\varphi(x) = 1.
\end{array}$$
\end{itemize}
}
\end{myproof}

We shall be using a subcategory $\CLJO \hookrightarrow \CLJ$ of
complete lattices where maps are not only join-preserving but also
preserve the top element $1$. The following is then an easy adaptation
of Lemma~\ref{CLTwoLem} and Lemma~\ref{DcpoTwoLem}.

\begin{lemma}
\label{CLJOTwoLem}
For a complete lattice $L$ and a dcpo $X$ there are isomorphisms:
$$\begin{array}{rclcrcl}
\CLJO(L,2)
& \cong &
\op{\big(L\backslash 1\big)}
& \qquad\mbox{and thus}\qquad &
\CLJO(\Open(X), 2)
& \cong &
\Closed(X)\backslash\emptyset.
\end{array}$$
\end{lemma}

\begin{myproof}
Following the proof of Lemma~\ref{CLTwoLem} one easily shows that
$\varphi\colon L \rightarrow 2$ in $\CLJ$ preserves $1$ iff the
corresponding element $\widehat{\varphi} =
\bigvee\set{x}{\varphi(x)=0}\in L$ is not $1$. This gives the first
isomorphism. The second one then easily follows, see
Lemma~\ref{DcpoTwoLem}. \QED

\auxproof{
\begin{itemize}
\item Let $\varphi(1) = 1$. Then $\widehat{\varphi}=1$ leads to a
contradiction:
$$\begin{array}{rcccccccccl}
1
& = &
\varphi(1)
& = &
\varphi(\widehat{\varphi})
& = &
\varphi(\bigvee\set{x}{\varphi(x)=0})
& = &
\bigvee\set{\varphi(x)}{\varphi(x)=0}
& = &
0.
\end{array}$$

\item Let now $\widehat{\varphi} \neq 1$. Assume, towards a
  contradication, that $\varphi(1) \neq 1$, so $\varphi(1) = 0$. But
  then $1 \leq \bigvee\set{x}{\varphi(x)=0} = \widehat{\varphi}$.
\end{itemize}

The resulting isomorphism $\CLJO(\Open(X), 2) \cong
\Closed(X)\backslash\emptyset$ is given as follows.
\begin{itemize}
\item For $\varphi \colon \Open(X) \rightarrow 2$, take:
$$\begin{array}{rcccl}
\widehat{\varphi}
& = &
\neg \bigcup\setin{U}{\Open(X)}{\varphi(U) = 0}
& = &
\bigcap\setin{V}{\Closed(X)}{\varphi(\neg V) = 0}.
\end{array}$$

\noindent Suppose $\widehat{\varphi} = \emptyset$. Then:
$$\begin{array}{rcccccccl}
1
& = &
\varphi(X)
& = &
\varphi(\neg\widehat{\varphi})
& = &
\bigvee\set{\varphi(U)}{\varphi(U) = 0}
& = &
0.
\end{array}$$

\item Given $V\in\Closed(X)$ non-empty. Take:
$$\begin{array}{rcl}
\varphi(U)
& = &
\left\{\begin{array}{ll}
0 \quad & \mbox{if $U\subseteq \neg V$, i.e. $U\cap V = \emptyset$} \\
1 & \mbox{otherwise}
\end{array}\right.
\end{array}$$

\noindent Then:
$$\begin{array}{rcl}
\varphi(\bigcup_{i}U_{i}) = 0
& \Longleftrightarrow &
\bigcup_{i}U_{i} \subseteq \neg V \\
& \Longleftrightarrow &
\all{i}{U_{i} \subseteq \neg V} \\
& \Longleftrightarrow &
\all{i}{\varphi(U_{i}) = 0} \\
& \Longleftrightarrow &
\bigvee_{i} \varphi(U_{i}) = 0
\end{array}$$

\noindent Since $V\neq \emptyset$, we have $\neg V \neq X$, so $X
\not\subseteq \neg V$, and thus $\varphi(X) = 1$.
\end{itemize}

\noindent Finally, $\widehat{\widehat{V}} =
\bigcap\set{W}{\widehat{V}(\neg W) = 0} = \bigcap\set{W}{\neg W
  \subseteq \neg V} = \bigcap\set{W}{V\subseteq W} = V$. Similarly,
$\widehat{\widehat{\varphi}}(U) = 0$ iff $U\subseteq
\neg\widehat{\varphi} = \bigcup\set{V}{\varphi(V)=0}$ iff $\varphi(U)
= 0$. For the last equivalence, the (if)-part is easy. For (only if),
let $U\subseteq \neg\widehat{\varphi}$. Then $\varphi(U) \leq
\bigvee\set{\varphi(U)}{\varphi(U)=0} = 0$.
}
\end{myproof}

We now restrict the adjunction $\op{(\CLJ)} \leftrightarrows \PoSets$
from Subsection~\ref{PosetCLSubsec} to dcpos.
$$\vcenter{\xymatrix@R-2pc{
\op{\big(\CLJO\big)}\ar@/^2ex/[dd]^{\Hom(-,2)} \\
\dashv \\
\Dcpo\ar@/^2ex/[uu]^{\Open = \Hom(-,2)}\ar@(dl,dr)_{\mathcal{H}}
}}
\qquad
\begin{prooftree}
\xymatrix@C+.5pc{L\ar[r]^-{\CLJO} & \Open(X)}
\Justifies
\xymatrix{X\ar[r]_-{\Dcpo} & \op{(L\backslash 1)}}
\end{prooftree}
\qquad
\vcenter{\xymatrix@R-.5pc@C-2pc{
\op{\big(\CLJO\big)}\ar@/^0.7em/[rr] & \top & \EM(\mathcal{H})\ar@/^0.6em/[ll]
\\
& \Kl(\mathcal{H})\ar[ul]^{\Pred}\ar[ur]_{\Stat} &
}}\hspace*{0em}$$

\noindent In this situation we encounter Smyths~\cite{Smyth83}
topological view on predicate transformers, as maps between complete
lattices of open subsets $\Open(X) \cong \Hom(X, 2)$, see
Lemma~\ref{DcpoTwoLem}. Notice that the poset $\op{(L\backslash 1)}$
is a dcpo, with directed joins given by meets in $L$.

The adjoint transposes for the above adjunction are defined precisely
as in Subsection~\ref{PosetCLSubsec}. We only have to prove some
additional properties.
\begin{itemize}
\item For $f\colon L \rightarrow \Open(X)$ in $\CLJO$ we have
$\overline{f}(x) = \bigvee\set{a}{x\not\in f(a)}$. We check:
\begin{itemize}
\item $\overline{f}(x)\neq 1$ for each $x\in X$. Towards a
  contradiction, let $\overline{f}(x) = 1$. Then, using that $f$
  preserves $1$ and $\bigvee$ we get:
$$\begin{array}{rcccccccl}
x
& \in &
X
& = &
f(1)
& = &
f(\overline{f}(x))
& = &
\bigcup\set{f(a)}{x\not\in f(a)}.
\end{array}$$

\noindent We get $x\in \bigcup\set{f(a)}{x\not\in f(a)}$, which is
impossible.

\item The function $\overline{f} \colon X \rightarrow
  \op{(L\backslash 1)}$ sends directed joins $\bigvee_{i}x_{i}$ to
  meets. By monotonicity of $\overline{f} \colon X \rightarrow \op{L}$
  we have $\overline{f}(\bigvee_{i}x_{i}) \leq \overline{f}(x_{j})$,
  for each $j$, and thus $\overline{f}(\bigvee_{i}x_{i}) \leq
  \bigwedge_{i}\overline{f}(x_{i})$. For the reverse inequality we
  reason as follows.
\begin{itemize}
\item We have $x_{j} \not\in f(\bigwedge_{i}\overline{f}(x_{i}))$, for
  each $j$; otherwise, because $f\colon L \rightarrow \Open(X)$ is
  monontone and preserves joins, we get a contradiction:
$$\qquad\begin{array}{rcl}
x_{j}
\hspace*{\arraycolsep}\in\hspace*{\arraycolsep}
f\big(\bigwedge_{i}\overline{f}(x_{i})\big)
\hspace*{\arraycolsep}\leq\hspace*{\arraycolsep}
f\big(\overline{f}(x_{j})\big) 
& = &
f\big(\bigvee\set{y}{x_{j}\not\in f(y)}\big) \\
& = &
\bigcup\set{f(y)}{x_{j}\not\in f(y)}.
\end{array}$$

\item Since $f(\bigwedge_{i}\overline{f}(x_{i}))$ is open, we get
  $\bigvee_{i}x_{i} \not\in f(\bigwedge_{i}\overline{f}(x_{i}))$.

\item But then $\bigwedge_{i}\overline{f}(x_{i}) \leq
  \bigvee\set{y}{\bigvee_{i}x_{i} \not \in f(y)} =
  \overline{f}(\bigvee_{i}x_{i})$.
\end{itemize}
\end{itemize}

\item We also check that $\overline{g}(a) = \set{x}{a \not\leq g(x)}$
  is open. We already know from Subsection~\ref{PosetCLSubsec} that it
  is an upset. So let $\bigvee_{i}x_{i} \in \overline{g}(a)$. Then $a
  \not\leq g(\bigvee_{i}x_{i})$. Let $a \leq g(x_{i})$ for all
  $i$. Then $a \leq \bigwedge_{i}g(x_{i}) = g(\bigvee_{i}x_{i})$,
  which is impossible.  Hence $a \not\leq g(x_{i})$ for some index
  $i$. But then $x_{i} \in \overline{g}(a)$.

We need to add that $\overline{g}$ preserves the top element $1$, \ie
that $\overline{g}(1) = X$. We thus have to show that
$x\in\overline{g}(1)$ holds for each $x$. But this is clear, since $1
\not\leq g(x)$ \ie $g(x)\neq 1$.  The latter holds because $g$ has
type $X \rightarrow \op{(L\backslash 1)}$.
\end{itemize}

The induced monad on $\Dcpo$ is $X \mapsto \op{\big(\Open(X)\backslash
  X\big)} \cong \Closed(X)\backslash\emptyset$. This is what is called
the Hoare power monad~\cite{AbramskyJ94a}, written as $\mathcal{H}$,
which sends a dcpo to its non-empty closed subsets. For a continuous
map $f\colon X \rightarrow Y$ we have $\mathcal{H}(f) \colon
\mathcal{H}(X) \rightarrow \mathcal{H}(Y)$ given by $\mathcal{H}(f)(U)
= \overline{f(U)}$, that is, by the (topological) closure of the
image.  The unit $\eta\colon X \rightarrow \mathcal{H}(X)$ of the
Hoare monad is determined as $\eta(x) = \downset x$, and the
multiplication $\mu \colon \mathcal{H}^{2}(X) \rightarrow
\mathcal{H}(X)$ as $\mu(A) = \overline{\bigcup A}$. This closure
arises in the last step of~\eqref{PosetCLMuEqn}.

The predicate transformer $\Open(Y) \rightarrow \Open(X)$ that is
bijectively associated with a Kleisli map $g\colon X \rightarrow
\mathcal{H}(Y)$ is the $\Diamond$-version, given by $g^{\Diamond}(V) =
\set{x}{V \cap g(x) \neq \emptyset}$. Like
in~\eqref{PosetCLMuDiamondCorr} the bijective correspondence has to
take the isomorphism $\op{\Open(X)} \cong \Closed(X)$ via complement
$\neg$ into account.

The Eilenberg-Moore algebra of the Hoare monad are the dcpos with a
binary join operation. They are also called affine complete lattices,
see \eg~\cite{Jacobs94a}.

\auxproof{
An Eilenberg-Moore algebra $\alpha \colon
\mathcal{H}(X) \rightarrow X$ provides a dcpo $X$ with finite joins,
and makes it into a complete lattice. Hence $\EM(\mathcal{H}) = \CLJ$.

If $X$ is a complete lattice we can define $\bigvee \colon
\mathcal{H}(X) \rightarrow X$, which is an algebra, as usual.  This
follows since $\bigvee U = \bigvee \overline{U}$. The direction
$(\leq)$ is obvious, and for $(\geq)$ we reason as follows. We have $U
\subseteq \downset\bigvee U$, and because $\downset\bigvee U$ is
closed also $\overline{U} \subseteq \downset\bigvee U$. Hence
$\bigvee\overline{U} \leq \bigvee U$.

Conversely, if $X$ is a dcpo with algebra $\alpha\colon \mathcal{H}(X)
\rightarrow X$, then we define finite joins in $X$ as:
$$\begin{array}{rcl}
x_{1} \vee \cdots \vee x_{n}
& = &
\alpha\big(\downset x_{1} \cup \ldots \cup \downset x_{n}\big).
\end{array}$$

\noindent We show that this forms a join.
\begin{itemize}
\item Clearly $\downset x_{i} \subseteq \downset x_{1} \cup \ldots
  \cup \downset x_{n}$. Hence $x_{i} = \alpha(\downset x_{i}) \leq
  \alpha\big(\downset x_{1} \cup \ldots \cup \downset x_{n}\big) =
  x_{1} \vee \cdots \vee x_{n}$.

\item If $x_{i} \leq y$ for each $i$, then $\downset x_{i} \subseteq
  \downset y$, and so $\downset x_{1} \cup \ldots \cup \downset x_{n}
  \subseteq \downset y$. Hence $x_{1} \vee \cdots \vee x_{n} =
  \alpha\big(\downset x_{1} \cup \ldots \cup \downset x_{n}\big) \leq
  \alpha(\downset y) = y$.
\end{itemize}

$\Dcpo\leftrightarrows\op{(\CLJ)}$ given in:
$$\vcenter{\xymatrix@R-2pc{
\op{(\CLJ)}\ar@/^2ex/[dd]^{\Hom(-,2)\cong\op{(-)}} \\
\dashv \\
\Dcpo\ar@/^2ex/[uu]^{\Open = \Dcpo(-,2)}\ar@(dl,dr)_{\mathcal{H}}
}}
\qquad
\begin{prooftree}
\xymatrix@C+.5pc{Y\ar[r]^-{f} & \Open(X) \cong \Dcpo(X, 2)}
\Justifies
\xymatrix{X\ar[r]_-{g} & \op{Y} \cong \CLJ(Y, 2)}
\end{prooftree}$$

\noindent The bijective correspondence is given as follows.
\begin{itemize}
\item Given a join-preserving function $f\colon Y \rightarrow
  \Open(X)$ we take $\overline{f} \colon X \rightarrow Y$ as:
$$\begin{array}{rcl}
\overline{f}(x)
& = &
\bigvee\setin{y}{Y}{x\not\in f(y)}.
\end{array}$$

\noindent This definition arises through the above isomorphisms
$\Open(X) \cong \Dcpo(X, 2)$ and $\CLJ(Y, 2) \cong \op{Y}$ given by,
as in~\eqref{DcpoMapEqn} and~\eqref{CLJTwoEqn}.

\auxproof{
$$\begin{array}{rcccccl}
\overline{f}(x)
& = &
\widehat{\lam{y}{\widehat{f(y)}(x)}}
& = &
\bigvee\set{y}{\widehat{f(y)}(x) = 0}
& = &
\bigvee\set{y}{x \not\in f(y)}.
\end{array}$$
}

It is easy to see that $\overline{f}$ is a monotone map $X \rightarrow
\op{Y}$. If $x \leq x'$, then $x\in f(y) \Rightarrow x'\in f(y)$ since
$f(y)$ is open and thus an upset, so $\set{y}{x'\not\in f(y)}
\subseteq \set{y}{x\not\in f(y)}$, and thus $\overline{f}(x') =
\bigvee\set{y}{x'\not\in f(y)} \leq \bigvee\set{y}{x\not\in f(y)} =
\overline{f}(x)$.

Next, the map $\overline{f}$ is continuous: it sends directed joins to
meets. This is the non-trivial part of the proof. By monotonicity we
have $\overline{f}(\bigvee_{i}x_{i}) \leq \overline{f}(x_{j})$, for
each $j$, and thus $\overline{f}(\bigvee_{i}x_{i}) \leq
\bigwedge_{i}\overline{f}(x_{i})$. For the reverse inequality we
reason as follows.
\begin{itemize}
\item We have $x_{j} \not\in f(\bigwedge_{i}\overline{f}(x_{i}))$. If not,
then, because $f$ is monontone and preserves joins, we get:
$$\begin{array}{rcccccccl}
x_{j}
& \in &
f\big(\bigwedge_{i}\overline{f}(x_{i})\big)
& \leq &
f\big(\overline{f}(x_{j})\big)
& = &
f\big(\bigvee\set{y}{x_{j}\not\in f(y)}\big)
& = &
\bigcup\set{f(y)}{x_{j}\not\in f(y)}.
\end{array}$$

\noindent This is impossible.

\item Since $f(\bigwedge_{i}\overline{f}(x_{i}))$ is open, we get
  $\bigvee_{i}x_{i} \not\in f(\bigwedge_{i}\overline{f}(x_{i}))$.

\item But then $\bigwedge_{i}\overline{f}(x_{i}) \leq
  \bigvee\set{y}{\bigvee_{i}x_{i} \not \in f(y)} =
  \overline{f}(\bigvee_{i}x_{i})$.
\end{itemize}

\item In the other direction, given a continous map $g\colon X
  \rightarrow \op{Y}$ we define $\overline{g} \colon Y \rightarrow
  \Open(X)$ as:
$$\begin{array}{rcl}
\overline{g}(y)
& = &
\setin{x}{X}{y \not\leq g(x)}.
\end{array}$$

\noindent We first have to check that $\overline{g}(y)$ is open.
\begin{itemize}
\item If $x'\geq x\in\overline{g}(y)$, then $y \not\leq g(x)$.  Since
  $g(x') \leq g(x)$, having $y \leq g(x')$ is impossible. Hence $x'\in
  \overline{g}(y)$.

\item Let $\bigvee_{i}x_{i} \in \overline{g}(y)$, and assume towards a
  contradication that $x_{i}\not\in\overline{g}(y)$ for all $i$. Then
  $y \leq g(x_{i})$, so that $y \leq \bigwedge_{i} g(x_{i}) =
  g(\bigvee_{i}x_{i})$, \textit{quod non}.
\end{itemize}

\noindent Next we have to check that $\overline{g}$ is a
join-preserving function $Y \rightarrow \Open(X)$. Well,
$$\begin{array}{rcl}
x\not\in \overline{g}(\bigvee_{i}y_{i})
& \Longleftrightarrow &
\bigvee_{i}y_{i} \leq g(x) \\
& \Longleftrightarrow &
\all{i}{y_{i} \leq g(x)} \\
& \Longleftrightarrow &
\all{i}{x \not\in \overline{g}(y_{i})} \\
& \Longleftrightarrow &
x\not\in \bigcup_{i}\overline{g}(y_{i}).
\end{array}$$
\end{itemize}

\noindent Finally we have:
$$\begin{array}{rcccccl}
\overline{\overline{g}}(x)
& = &
\bigvee\set{y}{x \not\in \overline{g}(y)}
& = &
\bigvee\set{y}{y \leq g(x)}
& = &
g(x).
\end{array}$$

\noindent And:
$$\begin{array}{rcccl}
x \not\in \overline{\overline{f}}(y)
& \Longleftrightarrow &
y \leq \overline{f}(x) = \bigvee\set{z}{x\not\in f(z)}
& \Longleftrightarrow &
x \not\in f(y)
\end{array}$$

\noindent As usual, $(\Leftarrow)$ is easy, and for $(\Rightarrow)$ we
use that $f$ preserves meets in:
$$\begin{array}{rcccccl}
x 
& \in &
\bigcap\set{f(z)}{x\in f(z)}
& = &
f(\bigwedge\set{z}{x\in f(z)})
& \subseteq &
f(y).
\end{array}$$

Keimel~\cite{Keimel15} describes for dcpo's $X,Y$ the following
correspondence
$$\begin{prooftree}
\xymatrix{\Open(Y)\ar[r]^-{f} & \Open(X) \mbox{ in }\CLJ}
\Justifies
\xymatrix{X\ar[r]_-{g} & \mathcal{H}(Y)}
\end{prooftree}$$

\noindent where $\mathcal{H}(Y) \subseteq \Closed(Y)$ is the Hoare
power domain given by \emph{non-empty} closed subsets.

The correspondence is given as follows.
\begin{itemize}
\item Given a join-preserving map $f\colon \Open(Y) \rightarrow
  \Open(X)$ define $\overline{f} \colon X \rightarrow \mathcal{H}(Y)$
  as:
$$\begin{array}{rcl}
\overline{f}(x)
& = &
\bigcap\set{\neg V}{V\in\Open(Y) \mbox{ with } x\not\in f(V)}.
\end{array}$$ 

\noindent We first check that $\overline{f}(x)$ is non-empty. Towards
a contradiction, let $\overline{f}(x) = \emptyset$. Since $\emptyset$
is open, we get $f(\overline{f}(x)) = f(\emptyset) = \emptyset$. Thus
$x\not\in f(\overline{f}(x))$, so that $\overline{f}(x) \subseteq
\neg\overline{f}(x)$. ??

Next, $\overline{f}$ preserves directed joins:
$$\begin{array}{rcl}
y\in\overline{f}(\bigvee_{i}x_{i})
& \Longleftrightarrow &
\allin{V}{\Open(Y)}{\bigvee_{i}x_{i} \not\in f(V) \Rightarrow y\in\neg V} \\
& \Longleftrightarrow &
\allin{V}{\Open(Y)}{y\in V \Rightarrow \bigvee_{i}x_{i} \in f(V)} \\
& \Longleftrightarrow &
\allin{V}{\Open(Y)}{y\in V \Rightarrow \ex{i}{x_{i} \in f(V)}} \\
& \Longleftrightarrow &
\allin{V}{\Open(Y)}{(\all{i}{x_{i} \not\in f(V)}) \Rightarrow y\in\neg V} \\
& \smash{\stackrel{(*)}{\Longleftrightarrow}} &
\ex{i}{\allin{V}{\Open(Y)}{x_{i} \not\in f(V) \Rightarrow y\in\neg V}} \\
& \Longleftrightarrow &
y \in \bigcup_{i}\overline{f}(x_{i}).
\end{array}$$

\noindent For the marked implication $(\Leftarrow)$ let $i$ be given,
and $V\in\Open(V)$ with $\all{i}{x_{i} \not\in f(V)}$. Then $x_{i}
\not\in f(V)$ in particular, so that $y\in\neg V$ by assumption.  For
$(\Rightarrow)$, suppose towards a contradiction $\all{i}{\ex{V}{
    x_{i} \not\in f(V) \mbox{ and } y\in V}}$. We pick an $i$, using
that a directed collection is non-empty, and a $V\in\Open(Y)$ with
$x_{i}\not\in f(V)$ and $y\in V$. The antecedent then gives
$\neg\all{i}{x_{i} \not\in f(V)}$, so that $x_{j} \in f(V)$ for some
$j$. The assumption gives a $V'\in\Open(Y)$ with $x_{j}\not\in V'$ and
$y\in V'$. Hence $y\in V\cap V'$

\item Given a continuous map $g\colon X \rightarrow \mathcal{H}(Y)$,
define $\overline{g} \colon \Open(Y) \rightarrow \Open(X)$ as:
$$\begin{array}{rcl}
\overline{g}(V)
& = &
\set{x}{g(x) \cap V \neq \emptyset}.
\end{array}$$

\noindent We have to check that $\overline{g}(V)$ is open.
\begin{itemize}
\item If $x' \geq x \in \overline{g}(V)$, say via $y\in g(x) \cap V$,
then $y\in g(x') \cap V$ since $g(x) \subseteq g(x')$.

\item If $\bigvee_{i}x_{i} \in \overline{g}(V)$, say via $y\in
  g(\bigvee_{i}x_{i}) \cap V = (\bigcup_{i} g(x_{i}))\cap V =
  \bigcup_{i} g(x_{i}) \cap V$, then $y\in g(x_{i})\cap V$ for some
  $i$.  But then $x_{i} \in \overline{g}(V)$.
\end{itemize}

\noindent Next, the function $\overline{g} \colon \Open(Y) \rightarrow
\Open(V)$ preserves joins (unions):
$$\begin{array}{rcl}
x \in \overline{g}(\bigcup_{i}U_{i})
& \Longleftrightarrow &
g(x) \cap \bigcup_{i}U_{i} = \bigcup_{i}(g(x)\cap U_{i}) \neq \emptyset \\
& \Longleftrightarrow &
\ex{i}{g(x) \cap U_{i} \neq \emptyset} \\
& \Longleftrightarrow &
\ex{i}{x \in \overline{g}(U_{i})} \\
& \Longleftrightarrow &
x \in \bigcup_{i}\overline{g}(U_{i}).
\end{array}$$
\end{itemize}

\noindent The correspondence is obtained via:
$$\begin{array}{rcl}
y\in\overline{\overline{g}}(x)
& \Longleftrightarrow &
\allin{V}{\Open(Y)}{x\not\in \overline{g}(V) \Rightarrow y\in\neg V} \\
& \Longleftrightarrow &
\allin{V}{\Open(Y)}{g(x) \cap V = \emptyset \Rightarrow y\not\in V} \\
& \Longleftrightarrow &
y \in g(x).
\end{array}$$

\noindent The implication $(\Rightarrow)$ is obtained by taking $V =
\neg g(V)\in\Open(Y)$. For $(\Leftarrow)$, let $y\in g(x)$ and $g(x)
\cap V = \emptyset$. Then evidently $y\not\in V$.

In the other direction:
$$\begin{array}{rcl}
x \in \overline{\overline{f}}(V)
& \Longleftrightarrow &
\overline{f}(x) \cap V \neq \emptyset \\
& \Longleftrightarrow &
\exin{y}{V}{\allin{U}{\Open(Y)}{x\not\in f(U) \Rightarrow y\in\neg U}} \\
& \Longleftrightarrow &
\exin{y}{V}{\allin{U}{\Open(Y)}{y\in U \Rightarrow x\in f(U)}} \\
& \Longleftrightarrow &
x \in f(V).
\end{array}$$

\noindent For the implication $(\Rightarrow)$, let $y\in V$ and take
$U = V$. Then $x\in f(V)$. For $(\Leftarrow)$ let $x\in f(V) =
f(\bigcup_{y\in V}\{y\}) = \bigcup_{y\in V}f(\{y\})$. Hence $x \in
f(\{y\})$ for some $y\in V$. Then, for an arbitrary open $U$, if $y\in
U$, then $\{y\} \subseteq U$, so that $x \in f(\{y\}) \subseteq f(U)$.
}

\subsection{Dcpo's and Preframes}\label{DcpoPreFrmSubsec}

A \emph{preframe} is a dcpo with finite meets, in which the binary
meet operation $\wedge$ is continuous in both variables. We write
$\PreFrm$ for the category of preframes, where maps are both (Scott)
continuous and preserve finite meets $(\wedge, \top)$. The two-element
set $2 = \{0,1\} = \{\bot, \top\}$ is a preframe, with obvious joins
and meets. Each set of opens of a topological space is also a
preframe. 

In fact we shall use a subcategory $\PreFrmZ \hookrightarrow \PreFrm$
of preframes with a bottom element $0$, which is preserved by
(preframe) homomorphisms. We shall use this category as codomain of
the functor $\Open = \Hom(-,2) \colon \Dcpo \rightarrow
\op{(\PreFrmZ)}$.

We obtain a functor in the opposite direction also by homming into
$2$. We note that for a preframe $L$ the preframe-homomorphisms
$f\colon L \rightarrow 2$ correspond to Scott open filters $f^{-1}(1)
\subseteq L$, that is, to filters which are at the same time open
subsets in the Scott topology. If we require that $f$ is a map in
$\PreFrmZ$, additionally preserving $0$, then the Scott open filter
$f^{-1}(1)$ is proper, that is, not the whole of $L$.

We shall write the resulting functor as $\OF = \Hom(-,2) \colon
\op{(\PreFrmZ)} \rightarrow \Dcpo$.  Here we use that these proper
Scott open filters, ordered by inclusion, form a dcpo.

\auxproof{
First we check the correspondence:
$$\begin{prooftree}
{\xymatrix{L\ar[r]^-{f} & 2 \mbox{ in $\PreFrm$}}}
\Justifies
U\subseteq L \mbox{ Scott open filter}
\end{prooftree}$$

\begin{itemize}
\item Given $f$, then $\overline{f} = f^{-1}(1) = \setin{a}{L}{f(a) = 1}
\subseteq L$. We check that this is a Scott open filter.
\begin{itemize}
\item Clearly $\top\in\overline{f}$ since $f(\top) = 1$.

\item If $a,b\in \overline{f}$, then $a\wedge b$ too, since
$f(a\wedge b) = f(a) \wedge f(b) = 1 \wedge 1 = 1$.

\item If $b \geq a\in \overline{f}$, then $f(b) \geq f(a) = 1$, so
  $f(b) = 1$ too. Hence $\overline{f}$ is a filter.

\item It is Scott open since if $\bigvee_{i}a_{i} \in \overline{f}$,
  then $\bigvee_{i}f(a_{i}) = f(\bigvee_{i}a_{i}) = 1$, so that
  $f(a_{i}) = 1$ for some $i$.
\end{itemize}

\item Conversely, given a Scott open filter $U\subseteq L$, we define
  a function $\overline{U} \colon L \rightarrow 2$ by $\overline{U}(a)
  = 1$ iff $a\in U$. We check that $\overline{U}$ is a map of
  preframes.
\begin{itemize}
\item $\overline{U}(\top) = 1$ since $\top\in U$.

\item $\overline{U}$ is monotone since: if $a\leq b$, and
  $\overline{U}(a) = 1$, then $a\in U$, and thus $b\in U$, since $U$
  is an upset, so that $\overline{U}(b) = 1$.

\item $\overline{U}$ preserves $\wedge$. By monotonicity we have
  $\overline{U}(a\wedge b) \leq \overline{U}(a) \wedge
  \overline{U}(b)$. For the other direction, let $\overline{U}(a)
  \wedge \overline{U}(b) = 1$. Then $\overline{U}(a) = \overline{U}(b)
  = 1$, so that $a\in U$ and $b\in U$. But then $a\wedge b \in U$, so
  that $\overline{U}(a\wedge b) = 1$.

\item Again by monotonicity we have $\bigvee_{i}\overline{U}(a_{i})
  \leq \overline{U}(\bigvee_{i}a_{i})$. For the other direction, let
  $\overline{U}(\bigvee_{i}a_{i}) = 1$. Then $\bigvee_{i}a_{i} \in U$,
  so that $a_{i}\in U$ for some $i$. Hence $\overline{U}(a_{i}) = 1$,
  and thus $\bigvee_{i}\overline{U}(a_{i}) = 1$.
\end{itemize}
\end{itemize}

\noindent These operations are each other's inverses:
$$\begin{array}{rcccl}
\overline{\overline{f}}(a) = 1
& \Longleftrightarrow &
a\in \overline{f} 
& \Longleftrightarrow &
f(a) = 1.
\end{array}$$

\noindent And:
$$\begin{array}{rcccccl}
\overline{\overline{U}}
& = &
\set{a}{\overline{U}(a) = 1}
& = &
\set{a}{a\in U}
& = &
U.
\end{array}$$

We check that we have a functor $\OF \colon \op{\PreFrm} \rightarrow
\Dcpo$. Suppose we have directed collection $(U_{i})$ of open filters.
We claim that the union $U = \bigcup_{i}U_{i}$ is again an open filter.
\begin{itemize}
\item We have $\top\in U$ because a directed collection is non-empty.

\item If $a_{i}, a_{2} \in U$, say via $a_{j} \in U_{i_j}$, then there
  is an index $k$ with $U_{i_1} \subseteq U_{k}$ and $U_{i_2}
  \subseteq U_{k}$, using directedness. Hence $a_{1},a_{2} \in U_{k}$,
  and thus also $a_{1} \wedge a_{2} \in U_{k} \subseteq U$.

\item If $b \geq a \in U$, say via $a\in U_{i}$, then $b\in U_{i}
  \subseteq U$.

\item If $\bigvee_{j}a_{j}\in U$, say via $\bigvee_{j}a_{j}\in U_{i}$,
  then $a_{j}\in U_{i} \subseteq U$ for some $j$.
\end{itemize}

\noindent If $f\colon L \rightarrow K$ is a map of preframes, then
$f^{-1}$ restricts to $\OF(K) \rightarrow \OF(L)$. The easiest way to
see this is via the above correspondence $\OF(K) \cong \Hom(K,2)$.
The inverse image map $f^{-1}$ obviously preserves (directed) unions,
and is thus Scott continuous.
}

If we put things together we obtain:
$$\vcenter{\xymatrix@R-2pc{
\op{(\PreFrmZ)}\ar@/^2ex/[dd]^{\Hom(-,2)\cong\OF} \\
\dashv \\
\Dcpo\ar@/^2ex/[uu]^{\Open = \Hom(-,2)}\ar@(dl,dr)_{\mathcal{S}}
}}
\qquad
\begin{prooftree}
\xymatrix@C+1pc{L\ar[r]^-{\PreFrmZ} & \Open(X)}
\Justifies
\xymatrix{X\ar[r]_-{\Dcpo} & \OF(L)}
\end{prooftree}
\qquad
\vcenter{\xymatrix@R-.5pc@C-2pc{
\op{(\PreFrmZ)}\ar@/^0.7em/[rr] & \top & 
   \;\EM(\mathcal{S})\quad\ar@/^0.6em/[ll]
\\
& \Kl(\mathcal{S})\ar[ul]^{\Pred}\ar[ur]_{\Stat} &
}}\hspace*{0em}$$

\auxproof{
We check that correspondence:
$$\begin{prooftree}
\xymatrix@C+.5pc{L\ar[r]^-{f} & \Open(X)}
\Justifies
\xymatrix{X\ar[r]_-{g} & \OF(L)}
\end{prooftree}$$

\begin{itemize}
\item Given $f\colon L \rightarrow \Open(X)$, take $\overline{f}(x) = 
\setin{a}{L}{x\in f(a)}$. This is well-defined.
\begin{itemize}
\item $\top\in\overline{f}(x)$ since $x\in f(\top) = X$.

\item If $b \geq a \in \overline{f}(x)$, then $x \in f(a) \subseteq
  f(b)$, so $b\in \overline{f}(x)$.

\item If $a,b\in \overline{f}(x)$, then $x\in f(a)$ and $x\in f(b)$,
  so that $x\in f(a) \cap f(b) = f(a \wedge b)$, and thus $a\wedge b
  \in \overline{f}(x)$.

\item If $\bigvee_{i}a_{i} \in \overline{f}(x)$, then $x \in
  f(\bigvee_{i}a_{i}) = \bigcup_{i} f(a_{i})$, so that $x\in f(a_{i})$
  for some $i$. But then $a_{i} \in \overline{f}(x)$. These points
  show that $\overline{f}(x)$ is a Scott open filter.

\item $\overline{f}$ is continuous, since for a directed collection
  $(x_{i})$ we have:
$$\begin{array}{rcl}
a \in \overline{f}(\bigvee_{i}x_{i})
& \Longleftrightarrow &
\bigvee_{i}x_{i} \in f(a) \\
& \Longleftrightarrow &
\ex{i}{x_{i} \in f(a)} \\
& \Longleftrightarrow &
\ex{i}{a \in \overline{f}(x_{i})} \\
& \Longleftrightarrow &
a \in \bigcup_{i}\overline{f}(x_{i}).
\end{array}$$
\end{itemize}

\item Conversely, given $g\colon X \rightarrow \OF(L)$, define
  $\overline{g}(a) = \setin{x}{X}{a\in g(x)}$. This is also
  well-defined.
\begin{itemize}
\item $y \geq x \in \overline{g}(a)$ implies $a\in g(x) \subseteq
  g(y)$, so that $y\in \overline{g}(a)$.

\item If $\bigvee_{i}x_{i} \in \overline{g}(a)$, then $a \in
  g(\bigvee_{i}x_{i}) = \bigcup_{i}g(x_{i})$, so that $a\in g(x_{i})$
  for some $i$. But then $x_{i} \in \overline{g}(a)$. Hence
  $\overline{g}(a)$ is a Scott open subset.

\item $\overline{g}(\top) = \set{x}{\top\in g(x)} = X$ since $g(x)$ is
  a filter.

\item $\overline{g}(a\wedge b) = \set{x}{a\wedge b \in g(x)} =
  \set{x}{a \in g(x) \mbox{ and } b \in g(x)} = \set{x}{a \in g(x)}
  \cap \set{x}{b \in g(x)} = \overline{g}(a) \cap \overline{g}(b)$.

\item $\overline{g}(\bigvee_{i}a_{i}) = \set{x}{\bigvee_{i}a_{i} \in
  g(x)} = \set{x}{\ex{i}{a_{i}\in g(x)}} = \bigcup_{i}
  \set{x}{a_{i}\in g(x)}$.
\end{itemize}
\end{itemize}

\noindent Clearly we have:
$$\begin{array}{rcl}
\overline{\overline{f}}(a)
& = &
\set{x}{a \in \overline{f}(x)} \\
& = &
\set{x}{x\in f(a)} \\
& = &
f(a)
\\
\overline{\overline{g}}(x)
& = &
\set{a}{x \in \overline{g}(a)} \\
& = &
\set{a}{a \in g(x)} \\
& = &
g(x).
\end{array}$$
}

\noindent The induced monad $\mathcal{S}(X) = \OF(\Open(X))$ takes the
proper Scott open filters in the preframe $\Open(X)$ of Scott open
subsets of a dcpo $X$. This is the Smyth power domain,
see~\cite{Keimel12}.  We recall the Hofmann-Mislove
theorem~\cite{HofmannM81,KeimelP94}: in a sober topological space $Y$,
the Scott open filters in $\Open(Y)$ correspond to compact saturated
subsets of $Y$. This subset is non-empty if and only if the
corresponding filter is proper. We also recall that if $X$ is a
continuous dcpo, where each element is the directed join of elements
way below it, then its Scott topology is sober, see \eg~\cite[VII,
  Lemma~2.6]{Johnstone82}. This explains why the Smyth power domain is
often defined on continuous dcpos. We shall not follow this route
here, and will continue to work with functions instead of with
subsets.


The induced functor $\Pred \colon \Kl(\mathcal{S}) \rightarrow
\op{(\PreFrmZ)}$ is full and faithful, corresponding to healthiness of
the predicate transformer semantics. Specifically, for a Kleisli map
$g\colon X \rightarrow \mathcal{S}(Y) = \OF(\Open(Y))$ we have
$\Pred(g) \colon \Open(Y) \rightarrow \Open(X)$ given by the preframe
homomorphism:
$$\begin{array}{rcl}
\Pred(g)(V)
& = &
\setin{x}{X}{V \in g(x)}.
\end{array}$$

\noindent The Eilenberg-Moore algebras of the Smyth power domain monad
$\mathcal{S}$ are dcpos with an additional binary meet operation.

\section{Dualising with 3}\label{ThreeSec}

Using a three-element set $3$ as dualising object is unusual. We will
elaborate one example only, leading to a description of the Plotkin
power domain on the category of dcpos. We start from the following
notion, which seems new, but is used implicitly in the theory of
Plotkin power domains, notably in~\cite{Heckmann97}, see
also~\cite{BattenfeldKS14}.

\begin{definition}
\label{PlotkinAlgDef}
A \emph{Plotkin algebra} is a poset $X$ with least and greatest
elements $0,1\in X$, and with a binary operation $\amalg$ and a
special element $\Bowtie\in X$ such that:
\begin{itemize}
\item $\amalg$ is idempotent, commutative, associative, and monotone;
\item $\Bowtie$ is an absorbing element for $\amalg$, so that $x
  \amalg \Bowtie = \Bowtie = \Bowtie \amalg x$.
\end{itemize}

\noindent A Plotkin algebra is called directed complete if the poset
$X$ is a dcpo and the operation $\amalg$ is continuous. We write
$\DcPA$ for the category of directed complete Plotkin algebras. A
morphism in this category is a continuous function that preserves
$\amalg$ and $\Bowtie, 0, 1$.
\end{definition}

Each meet semilattice $(X, \wedge, 1)$ with a least element $0$ is a
Plotkin algebra with $\Bowtie = 0$. Similarly, each join semilattice
$(X, \vee, 0)$ with a greatest element $1$ is a Plotkin algebra with
$\Bowtie = 1$. These observations can be extended to the directed
complete case via functors:
$$\xymatrix{
\CLJO\ar[r] & \DcPA & \PreFrmZ\ar[l]
}$$

\noindent They give a connection with the categories that we have seen
in Subsections~\ref{DcpoCLSubsec} and~\ref{DcpoPreFrmSubsec} for the
Hoare and Smyth power domain. 

A \emph{frame} is complete lattice whose binary meet operation
$\wedge$ preserves all joins on both sides. The morphisms in the
category $\Frm$ of frames preserve both joins $\bigvee$ and finite
meets $(\wedge, 1)$. Hence there are forgetful functors.
$$\xymatrix{
\CLJO & \Frm\ar[l]\ar[r] & \PreFrmZ
}$$

\noindent But there is also another construction to obtain a Plotkin
algebra from a frame.

\begin{definition}
\label{FrmDcPADef}
Each frame $X$ gives rise to a directed complete Plotkin algebra,
written as $X\ltimes X$, via:
$$\begin{array}{rcl}
X\ltimes X
& = &
\setin{(x,y)}{X\times X}{x \geq y}.
\end{array}$$

\noindent It carries the product dcpo structure, and forms a Plotkin
algebra with:
$$\begin{array}{rclcrcl}
(x,y) \amalg (x',y')
& = &
(x \vee x', y \wedge y')
& \qquad\mbox{and}\qquad &
\Bowtie
& = &
(1, 0).
\end{array}$$

\noindent This operation $\amalg$ is continuous since $X$ is a frame.
\end{definition}

Explicitly, the projections form maps of Plotkin algebras in:
\begin{equation}
\label{PiMapDiag}
\vcenter{\xymatrix@C+1pc{
(X, 0, 1, \vee, 1)
& 
(X\ltimes X, (0,0), (1,1), \amalg, \Bowtie)
   \ar[l]_-{\pi_1}\ar[r]^-{\pi_2}
& 
(X, 0, 1, \wedge, 0)
}}
\end{equation}

\noindent We shall also use functions $\inl, \inr \colon X \rightarrow
X\ltimes X$ defined by:
$$\begin{array}{rclcrcl}
\inl(x)
& = &
(x, 0)
& \qquad\mbox{and}\qquad &
\inr(y)
& = &
(1, y).
\end{array}$$

\noindent These are \emph{not} maps of Plotkin algebras, since
$\inl(1) = \Bowtie \neq 1$ and $\inr(0) = \Bowtie \neq 0$. But we do
have $\inl(0) = 0$ and $\inr(1) = 1$, and also the following structure
is preserved.
\begin{equation}
\label{InMapDiag}
\vcenter{\xymatrix@C+1pc{
(X, \vee, 1)\ar[r]^-{\inl}
& 
(X\ltimes X, \amalg, \Bowtie)
& 
(X, \wedge, 0)\ar[l]_-{\inr}
}}
\end{equation}

\auxproof{
$$\begin{array}{rclcrcl}
\inl(0)
& = &
(0,0)
& \hspace*{5em} &
\inr(0)
& = &
(1, 0)
\\
& = &
0
& & 
& = &
\Bowtie
\\
\inl(1)
& = &
(1, 0)
& &
\inr(1)
& = &
(1,1)
\\
& = &
\Bowtie
& &
& = &
1
\\
\inl(x\vee y)
& = &
(x\vee y, 0)
& &
\inr(x \wedge y)
& = &
(1, x\wedge y)
\\
& = &
(x\vee y, 0 \wedge 0)
& &
& = &
(1 \vee 1, x \wedge y)
\\
& = &
(x,0) \amalg (y,0)
& &
& = &
(1, x) \amalg (1, y)
\\
& = &
\inl(x) \amalg \inl(y)
& &
& = &
\inr(x) \amalg \inr(y).
\end{array}$$

Similarly,
$$\begin{array}{rclcrcl}
\pi_{1}(0)
& = &
0
& \hspace*{1em} &
\pi_{2}(0)
& = &
0
\\
\pi_{1}(1)
& = &
1
& &
\pi_{2}(1)
& = &
1
\\
\pi_{1}(\Bowtie)
& = &
1
& &
\pi_{2}(\Bowtie)
& = &
0
\\
\pi_{1}((x,y) \amalg (x',y'))
& = &
\pi_{1}(x \vee x', y\wedge y')
& &
\pi_{2}((x,y) \amalg (x',y'))
& = &
\pi_{2}(x \vee x', y\wedge y')
\\
& = &
x\vee x'
& &
& = &
y\wedge y'
\\
& = &
\pi_{1}(x,y) \vee \pi_{1}(x',y')
& &
& = &
\pi_{2}(x,y) \wedge \pi_{2}(x',y')
\end{array}$$
}

The $\ltimes$ construction yields a three-element algebra $2\ltimes 2$
that will be described more directly below,
following~\cite{Heckmann97}.  We use it as dualising object.

\begin{example}
\label{ThreeEx}
For the two-element frame $2 = \{0,1\}$ the Plotkin algebra $2\ltimes 2$
is a three-element set, which we can also describe as:
$$\begin{array}{rclcrcccl}
\three
& = &
\{0, \Bowtie, 1\}
& \qquad\mbox{where}\qquad &
0 & \leq & \Bowtie & \leq & 1.
\end{array}$$

\noindent This order is obviously both complete and cocomplete. It is
determined for $a,b\in\three$ by:
\begin{equation}
\label{ThreeOrderChar}
\begin{array}{rcl}
a\leq b
& \mbox{iff} &
\mbox{ both }\left\{\begin{array}{rcl}
a=1 & \Rightarrow & b=1 \\
b=0 & \Rightarrow & a=0.
\end{array}\right.
\end{array}
\end{equation}

\noindent The isomorphism $j = (j_{1},j_{2})\colon \three
\conglongrightarrow 2\ltimes 2$ is given by:
$$\begin{array}{rclcrccclcrcl}
j(0) & = & (0,0)
& \qquad &
j(\Bowtie) & = & (1,0) & = & \Bowtie
& \qquad &
j(1) & = & (1,1).
\end{array}$$

\noindent The two components $j_{i} = \pi_{i} \after j \colon \three
\rightarrow 2$ are monotone, and satisfy $j_{1} \geq j_{2}$.

This isomorphism $\three \cong 2\ltimes 2$ makes $\three$ into a
(directed complete) Plotkin algebra, via $\Bowtie\in\three$ and
$\amalg \colon \three\times\three \rightarrow \three$ determined by:
$$\begin{array}{rcl}
a \amalg b
& = &
\left\{\begin{array}{ll}
0 \quad & \mbox{if }a=b=0 \\
1 & \mbox{if }a=b=1 \\
\Bowtie & \mbox{otherwise.}
\end{array}\right.
\end{array}$$

\noindent Finally we notice that the two maps $j_{1}, j_{2} \colon
\three \rightarrow 2$ are maps of Plotkin algebras:
\begin{equation}
\label{ThreeTwoDiag}
\vcenter{\xymatrix@C+1pc{
(2, 0, 1, \vee, 1) & (\three, 0, 1, \amalg, \Bowtie)\ar[l]_-{j_1}\ar[r]^-{j_2} &
   (2, 0, 1, \wedge, 0)
}}
\end{equation}

\auxproof{
The following diagram commutes.
$$\xymatrix{
\three\times\three\ar[rr]^-{j}_-{\cong}\ar[d]_{\amalg} & & 
   (2\ltimes 2)\times (2\ltimes 2)\ar[d]^{\amalg}
\\
\three\ar[rr]_-{j}^-{\cong} & & 2\ltimes 2
}$$

We check this in detail, where we write $j\colon \three \rightarrow 2\times 2$
for the inclusion.
\begin{itemize}
\item If $a=b=0$ in $\three$, then:
$$\begin{array}{rcccccccccl}
j(a)\amalg j(b)
& = &
(0,0) \amalg (0,0)
& = &
(0\vee 0, 0\wedge 0)
& = &
(0,0)
& = &
j(0)
& = &
j(a\amalg b).
\end{array}$$

\item Similarly, if $a=b=1$ in $\three$, then:
$$\begin{array}{rcccccccccl}
j(a)\amalg j(b)
& = &
(1,1) \amalg (1,1)
& = &
(1\vee 1, 1\wedge 1)
& = &
(1,1)
& = &
j(1)
& = &
j(a\amalg b).
\end{array}$$

\item Otherwise, let $a=\Bowtie$, or $b=\Bowtie$, or $a=0$
  and $b=1$, or $a=1$ and $b=0$. For $a=\Bowtie$ we have:
$$\begin{array}{rcccccccccl}
j(a)\amalg j(b)
& = &
(1,0) \amalg (b_{1},b_{2})
& = &
(1\vee b_{1}, 0\wedge b_{2})
& = &
(1,0)
& = &
j(\Bowtie)
& = &
j(a\amalg b).
\end{array}$$

\noindent And if $a=0$ and $b=1$, then:
$$\begin{array}{rcccccccccl}
j(a)\amalg j(b)
& = &
(0,0) \amalg (1,1)
& = &
(0\vee 1, 0\wedge 1)
& = &
(1,0)
& = &
j(\Bowtie)
& = &
j(a\amalg b).
\end{array}$$
\end{itemize}
}
\end{example}

The following result is the analogue of Lemma~\ref{DcpoTwoLem}, but
with the dcpo $\three$ instead of $2$. The correspondence is mentioned
in~\cite{BattenfeldKS14}, just before Lemma~4.11.

\begin{lemma}
\label{DcpoThreeLem}
For a dcpo $X$ there is a bijective correspondence between:
$$\begin{prooftree}
\begin{prooftree}
\xymatrix{X\ar[r]^-{f} & \three \mbox{ in $\Dcpo$}}
\Justifies
\xymatrix{X\ar@<-.5ex>[r]_-{g_2}\ar@<+.5ex>[r]^-{g_1} & 2
   \mbox{ in $\Dcpo$ with $g_{1} \geq g_{1}$}}
\end{prooftree}
\Justifies
(U_{1}, U_{2}) \in \Open(X) \ltimes \Open(X)
\end{prooftree}$$

\noindent As a result there is an isomorphism:
$$\begin{array}{rclcrcl}
\Dcpo(X, \three)
& \cong & 
\Open(X)\ltimes\Open(X)
& \quad\mbox{given by}\quad &
f
& \longmapsto & 
(\, \set{x}{f(x)\neq 0}, \; \set{x}{f(x)=1} \,).
\end{array}$$

\noindent This is an isomorphism of Plotkin algebras, where the left
hand side carries the pointwise Plotkin algebra structure inherited
from $\three$.
\end{lemma}

The (equivalent) structures in this lemma form predicates on the dcpo
$X$. The last description tells that such a predicate is a pair of
opens $U_{1},U_{2}\in\Open(X)$ with $U_{1} \supseteq U_{2}$.  This
predicate is true if $U_{1}=U_{2}=X$ and false if
$U_{1}=U_{2}=\emptyset$. In this `logic', predicates come equipped
with a binary operation $\amalg$; its logical interpretation is not
immedidately clear.

\begin{myproof}
The second, lower correspondence is given by Lemma~\ref{DcpoTwoLem},
so we concentrate on the first one. It works as follows.
\begin{itemize}
\item Given $f\colon X \rightarrow \three$ in $\Dcpo$ we obtain
  continuous maps $\overline{f}_{i} = j_{i} \after f \colon X
  \rightarrow 2$ by compositon, with $f_{1} \geq f_{2}$, since $j_{1}
  \geq j_{2}$, see Example~\ref{ThreeEx}.

\item In the other direction, given $g = (g_{1}, g_{2})$ we define
  $\overline{g} \colon X \rightarrow \three$ as:
$$\begin{array}{rcl}
\overline{g}(x)
& = &
\left\{\begin{array}{ll}
0 \quad & \mbox{if } g_{1}(x) = 0 \\
\Bowtie & \mbox{if } g_{1}(x) = 1 \mbox{ and } g_{2}(x) = 0 \\
1 & \mbox{if } g_{2}(x) = 1.
\end{array}\right.
\end{array}$$

\noindent We first show that $\overline{g}$ is monotone. So let $x
\leq y$ in $X$.  We use the characterisation~\eqref{ThreeOrderChar}.
\begin{itemize}
\item Let $\overline{g}(x) = 1$, so that $g_{2}(x) = 1$. But then
$g_{2}(y) \geq g_{2}(x) = 1$, so that $\overline{g}(y) = 1$.

\item If $\overline{g}(y) = 0$, then $g_{1}(x) \leq g_{1}(y) = 0$, so
  that $\overline{g}(x) = 0$.
\end{itemize}

\noindent Next, let $(x_{i})$ be a directed collection in $X$.  Since
$\overline{g}$ is monotone we have $\bigvee_{i}\overline{g}(x_{i}) \leq
\overline{g}(\bigvee_{i}x_{i})$. For the reverse inequality we
use~\eqref{ThreeOrderChar} again.
\begin{itemize}
\item Let $\overline{g}(\bigvee_{i}x_{i}) = 1$, so that
  $g_{2}(\bigvee_{i}x_{i}) = \bigvee_{i}g_{2}(x_{i}) = 1$. Then
  $g_{2}(x_{i}) = 1$ for some index $i$, for which then
  $\overline{g}(x_{i}) = 1$. Hence $\bigvee_{i}\overline{g}(x_{i}) =
  1$.

\item Let $\bigvee_{i}\overline{g}(x_{i}) = 0$, so that
  $\overline{g}(x_{i}) = 0$ for all $i$, and thus $g_{1}(x_{i}) = 0$.
  But then $g_{1}(\bigvee_{i}x_{i}) = \bigvee_{i} g_{1}(x_{i}) =
  0$. Hence $\overline{g}(\bigvee_{i}x_{i}) = 0$.
\end{itemize}
\end{itemize}

\noindent It is easy to see that $\overline{\overline{f}} = f$
and $\overline{\overline{g}} = g$. \QED

\auxproof{
$$\begin{array}{rcl}
\overline{\overline{f}}(x) = 0
& \Longleftrightarrow &
\overline{f}_{1}(x) = j_{1}(f(x)) = 0 \\
& \Longleftrightarrow &
f(x) = 0 
\\
\overline{\overline{f}}(x) = 1
& \Longleftrightarrow &
\overline{f}_{2}(x) = j_{2}(f(x)) = 1 \\
& \Longleftrightarrow &
f(x) = 1.
\end{array}$$

\noindent Also:
$$\begin{array}{rcl}
\overline{\overline{g}}_{1}(x) = 0
& \Longleftrightarrow &
j_{1}(\overline{g}(x)) = 0 \\
& \Longleftrightarrow &
\overline{g}(x) = 0 \\
& \Longleftrightarrow &
g_{1}(x) = 0
\\
\overline{\overline{g}}_{2}(x) = 1
& \Longleftrightarrow &
j_{2}(\overline{g}(x)) = 1 \\
& \Longleftrightarrow &
\overline{g}(x) = 1 \\
& \Longleftrightarrow &
g_{2}(x) = 1
\end{array}$$
}

\auxproof{
We write $\varphi \colon \Dcpo(X, \three) \conglongrightarrow \Open(X)
\ltimes \Open(X)$, defined by:
$$\begin{array}{rcl}
\varphi(f)
& = &
(\, \set{x}{\overline{f}_{1}(x) = j_{1}(f(x))=1}, \; 
   \set{x}{\overline{f}_{2}(x) = j_{2}(f(x))=1} \,) \\
& = &
(\, \set{x}{f(x)\neq 0}, \; \set{x}{f(x)=1} \,).
\end{array}$$

\noindent We claim that this is a map of Plotkin algebras.
$$\begin{array}{rcl}
\varphi(\mathbf{0})
& = &
(\, \set{x}{\mathbf{0}(x)\neq 0}, \; \set{x}{\mathbf{0}(x)=1} \,) \\
& = &
(\, \set{x}{0\neq 0}, \; \set{x}{0=1} \,) \\
& = &
(\, \emptyset, \; \emptyset \,) \\
& = &
0
\\
\varphi(\mathbf{1})
& = &
(\, \set{x}{\mathbf{1}(x)\neq 0}, \; \set{x}{\mathbf{1}(x)=1} \,) \\
& = &
(\, X, \; X \,) \\
& = &
1
\\
\varphi(\mathbf{\Bowtie})
& = &
(\, \set{x}{\mathbf{\Bowtie}(x)\neq 0}, \; \set{x}{\mathbf{\Bowtie}(x)=1} \,) \\
& = &
(\, X, \; \emptyset \,) \\
& = &
\Bowtie
\\
\varphi(f \amalg g)
& = &
(\, \set{x}{f(x)\amalg g(x)\neq 0}, \; \set{x}{f(x)\amalg g(x)(x)=1} \,) \\
& = &
(\, \set{x}{f(x)\neq 0 \mbox{ or } g(x)\neq 0}, \; 
   \set{x}{f(x) = 1 \mbox{ and } g(x)=1} \,) \\
& = &
(\, \set{x}{f(x)\neq 0} \cup \set{x}{g(x)\neq 0}, \; 
   \set{x}{f(x) = 1} \cap \set{x}{g(x)=1} \,) \\
& = &
(\, \set{x}{f(x)\neq 0}, \; \set{x}{f(x) = 1} \,) 
  \amalg (\, \set{x}{g(x)\neq 0}, \; \set{x}{g(x)=1} \,) \\
& = &
\varphi(f) \amalg \varphi(g).
\end{array}$$
}
\end{myproof}

Here is another fundamental correspondence, see
also~\cite[Obs.~4.10]{BattenfeldKS14}.

\begin{lemma}
\label{FrmThreeLem}
For frames $X,Y$ there is a bijective correspondence:
$$\begin{prooftree}
\xymatrix{X\ltimes X\ar[r]^-{f} & Y\ltimes Y \quad \mbox{in $\DcPA$}}
\Justifies
\xymatrix{X\ar[r]_-{g_1} & Y\mbox{ in $\CLJO$}\quad\mbox{and}\quad
  X\ar[r]_-{g_2} & Y\mbox{ in $\PreFrmZ$} \quad \mbox{with $g_{1} \geq g_{2}$}}
\end{prooftree}$$
\end{lemma}

\begin{myproof}
The correspondence is given as follows.
\begin{itemize}
\item For $f\colon X\ltimes X \rightarrow Y\ltimes Y$ in $\DcPA$ we take
the following continuous functions.
$$\xymatrix@C-1pc{
\overline{f}_{1} = \Big(X\ar[r]^-{\inl} & X\ltimes X\ar[r]^-{f} & 
   Y\ltimes Y\ar[r]^-{\pi_1} & Y\Big)
\quad\mbox{and}\quad
\overline{f}_{2} = \Big(X\ar[r]^-{\inr} & X\ltimes X\ar[r]^-{f} & 
   Y\ltimes Y\ar[r]^-{\pi_2} & Y\Big)
}$$

\noindent They preserve $0,1,\Bowtie, \amalg$ by~\eqref{PiMapDiag}
and~\eqref{InMapDiag}. For instance,
$$\begin{array}{rcccccccl}
\overline{f}_{1}(1)
& = &
\pi_{1}\big(f(\inl(1))\big) 
& = &
\pi_{1}\big(f(\Bowtie)\big) 
& = &
\pi_{1}(\Bowtie) 
& = &
1.
\end{array}$$

\noindent And:
$$\begin{array}{rcl}
\overline{f}_{1}(x \vee y)
\hspace*{\arraycolsep}=\hspace*{\arraycolsep}
\pi_{1}\big(f(\inl(x \vee y))\big) 
& = &
\pi_{1}\big(f(\inl(x) \amalg \inl(y))\big) \\
& = &
\pi_{1}\big(f(\inl(x)) \amalg f(\inl(y))\big) \\
& = &
\pi_{1}\big(f(\inl(x))\big) \vee \pi_{1}\big(f(\inl(y))\big) \\
& = &
\overline{f}_{1}(x) \vee \overline{f}_{1}(y).
\end{array}$$

\auxproof{
We check the remaining equations.
$$\begin{array}{rcccccccl}
\overline{f}_{1}(0)
& = &
\pi_{1}\big(f(\inl(0))\big) 
& = &
\pi_{1}\big(f(0)\big) 
& = &
\pi_{1}(0) 
& = &
0
\end{array}$$

\noindent And for the second map:
$$\begin{array}{rcl}
\overline{f}_{2}(0)
& = &
\pi_{2}\big(f(\inr(0))\big) \\
& = &
\pi_{2}\big(f(\Bowtie)\big) \\
& = &
\pi_{2}(\Bowtie) \\
& = &
0
\\
\overline{f}_{2}(1)
& = &
\pi_{2}\big(f(\inr(1))\big) \\
& = &
\pi_{2}\big(f(1)\big) \\
& = &
\pi_{2}(1) \\
& = &
1
\\
\overline{f}_{2}(x \wedge y)
& = &
\pi_{2}\big(f(\inr(x \wedge y))\big) \\
& = &
\pi_{2}\big(f(\inr(x) \amalg \inr(y))\big) \\
& = &
\pi_{2}\big(f(\inr(x)) \amalg f(\inr(y))\big) \\
& = &
\pi_{2}\big(f(\inr(x))\big) \wedge \pi_{2}\big(f(\inr(y))\big) \\
& = &
\overline{f}_{2}(x) \wedge \overline{f}_{2}(y).
\end{array}$$
}

We claim that for $(x,x')\in X\ltimes X$ the following two equations
hold.
$$\begin{array}{rclcrcl}
\overline{f}_{1}(x)
& = &
\pi_{1}(f(x,x'))
& \qquad\mbox{and}\qquad &
\overline{f}_{2}(x')
& = &
\pi_{2}(f(x,x')).
\end{array}\eqno{(*)}$$

\noindent We only prove the first one, since the second one works
analogously. We have to prove $\overline{f}_{1}(x) = \pi_{1}(f(x,0)) =
\pi_{1}(f(x,x'))$. The inequality $\leq$ holds by monotonicity, so it
suffices to prove $\geq$. In $Y\ltimes Y$ we have:
$$\begin{array}{rcccccl}
f(x, 0) \amalg f(x, x')
& = &
f\big((x, 0) \amalg (x, x')\big)
& = &
f(x \vee x, 0 \wedge x')
& = &
f(x, 0)
\end{array}$$

\noindent By applying the first projection we obtain:
$$\begin{array}{rcccl}
\pi_{1}(f(x,0)) \vee \pi_{1}(f(x,x'))
& = &
\pi_{1}\big(f(x, 0) \amalg f(x, x')\big)
& = &
\pi_{1}(f(x,0)).
\end{array}$$

\noindent Hence $\pi_{1}(f(x,x')) \leq \pi_{1}(f(x,0))$. 

\auxproof{
Similarly,
$$\begin{array}{rcccccl}
f(1, x') \amalg f(x, x')
& = &
f\big((1, x') \amalg (x, x')\big)
& = &
f(1 \vee x, x' \wedge x')
& = &
f(1, x')
\end{array}$$

\noindent By applying the second projection we obtain:
$$\begin{array}{rcccl}
\pi_{2}(f(1,x')) \wedge \pi_{2}(f(x,x'))
& = &
\pi_{2}\big(f(1, x') \amalg f(x, x')\big)
& = &
\pi_{2}(f(1,x')).
\end{array}$$

\noindent Hence $\pi_{2}(f(1,x')) \leq \pi_{2}(f(x,x'))$.
}

We use these equations~$(*)$ to prove $\overline{f}_{1} \geq
\overline{f}_{2}$.  For an arbitrary $x\in X$ we have $(x,x)\in
X\ltimes X$, and so:
$$\begin{array}{rcccccl}
\overline{f}_{1}(x)
& \smash{\stackrel{(*)}{=}} &
\pi_{1}(f(x,x))
& \geq &
\pi_{2}(f(x,x))
& \smash{\stackrel{(*)}{=}} &
\overline{f}_{2}(x).
\end{array}$$

\item In the other direction, given $g_{1} \colon X \rightarrow Y$
in $\CLJO$ and $g_{2} \colon X \rightarrow Y$ in $\PreFrmZ$ we
define $\overline{g} \colon X\ltimes X \rightarrow Y\ltimes Y$ by:
$$\begin{array}{rcl}
\overline{g}(x,x')
& = &
(\, g_{1}(x), \; g_{2}(x') \, ).
\end{array}$$

\noindent This is well-defined: we have $x \geq x'$, so $g_{1}(x) \geq
g_{1}(x') \geq g_{2}(x')$. It is easy to see that $\overline{g}$ is a
continuous map of Plotkin algebras.

\auxproof{
$$\begin{array}{rcl}
\overline{g}(0,0)
& = &
(g_{1}(0), g_{2}(0)) \\
& = &
(0, 0)
\\
\overline{g}(1,1)
& = &
(g_{1}(1), g_{2}(1)) \\
& = &
(1, 1)
\\
\overline{g}(\Bowtie)
& = &
(g_{1}(1), g_{2}(0)) \\
& = &
(1, 0) \\
& = &
\Bowtie
\\
\overline{g}\big((x,x') \amalg (y,y')\big)
& = &
\overline{g}\big(x \vee y, x' \wedge y'\big) \\
& = &
(g_{1}(x \vee y), g_{2}(x' \wedge y')) \\
& = &
(g_{1}(x) \vee g_{1}(y), g_{2}(x') \wedge g_{2}(y')) \\
& = &
(g_{1}(x), g_{2}(x')) \amalg (g_{1}(y), g_{2}(y')) \\
& = &
\overline{g}(x,x') \amalg \overline{g}(y,y').
\end{array}$$
}
\end{itemize}

\noindent We prove that these operations yield a bijective
correspondence. First,
$$\begin{array}{rcccccccl}
\overline{\overline{g}}_{1}(x)
& = &
\pi_{1}\big(\overline{g}(\inl(x))\big) 
& = &
\pi_{1}\big(\overline{g}(x, 0)\big) 
& = &
\pi_{1}(g_{1}(x), g_{2}(0)) 
& = &
g_{1}(x).
\end{array}$$

\auxproof{
\noindent Similarly:
$$\begin{array}{rcccccccl}
\overline{\overline{g}}_{2}(x)
& = &
\pi_{2}\big(\overline{g}(\inr(x))\big) 
& = &
\pi_{2}\big(\overline{g}(1, x)\big) 
& = &
\pi_{2}(g_{1}(1), g_{2}(x)) 
& = &
g_{2}(x)
\end{array}$$
}

\noindent Similarly we get $\overline{\overline{g}}_{2}(x) =
g_{2}(x)$. Next, in the other direction,
$$\begin{array}[b]{rcccccl}
\overline{\overline{f}}(x,x')
& = &
(\, \overline{f}_{1}(x), \; \overline{f}_{2}(x')\,) 
& \smash{\stackrel{(*)}{=}} &
(\, \pi_{1}\big(f(x,x')\big), \; \pi_{2}\big(f(x,x')\big) \,) 
& = &
f(x,x').
\end{array}\eqno{\QEDbox}$$
\end{myproof}

As announced, we will use the dcpo $\three$ as dualising object, in:
$$\vcenter{\xymatrix@R-2pc{
\op{\DcPA}\ar@/^2ex/[dd]^{\Hom(-,\three)} \\
\dashv \\
\Dcpo\ar@/^2ex/[uu]^{\Hom(-,\three)}\ar@(dl,dr)_{\wp}
}}
\qquad
\begin{prooftree}
\xymatrix@C+1pc{Y\ar[r]^-{\DcPA} & \Hom(X, \three)}
\Justifies
\xymatrix{X\ar[r]_-{\Dcpo} & \Hom(Y,\three)}
\end{prooftree}
\qquad
\vcenter{\xymatrix@R-.5pc@C-2pc{
\op{\DcPA}\ar@/^0.7em/[rr] & \top & 
   \,\EM(\wp)\ar@/^0.6em/[ll]
\\
& \Kl(\wp)\ar[ul]^{\Pred}\ar[ur]_{\Stat} &
}}\hspace*{0em}$$

\noindent For directed complete Plotkin algebra $Y\in\DcPA$ the homset
$\Hom(Y,\three)$ of maps in $\DcPA$ is a dcpo, via the pointwise
ordering. The above adjunction is then obtained via the usual swapping
of arguments.

\auxproof{
We show that a directed join $f = \bigvee_{i}f_{i}$ is a map of (directed
complete) Plotkin algebras, if all $f_i$'s are. This works because
the sum $\amalg$ is continuous. Thus:
$$\begin{array}{rcccccccl}
f(x \amalg y)
& = &
\bigvee_{i} f_{i}(x \amalg y) 
& = &
\bigvee_{i} f_{i}(x) \amalg f_{i}(y) 
& = &
(\bigvee_{i} f_{i}(x)) \amalg (\bigvee_{i} f_{i}(y))
& = &
f(x) \amalg f(y).
\end{array}$$
}

We call the induced monad the Plotkin power domain on $\Dcpo$. It can
be described as:
$$\begin{array}{rcl}
\wp(X)
& = &
\DcPA\big(\Dcpo(X, \three), \three\big) \\
& \cong &
\DcPA\big(\Open(X)\ltimes\Open(X), 2\ltimes 2\big) \\
& \cong &
\set{(f_{1}, f_{2})}{f_{1} \in \CLJO(\Open(X), 2), 
  f_{2}\in \PreFrmZ(\Open(X), 2), \mbox{ with }f_{1} \geq f_{2}}.
\end{array}$$

\noindent The first isomorphism is based on Lemma~\ref{DcpoThreeLem}
and Example~\ref{ThreeEx}. The second one comes from
Lemma~\ref{FrmThreeLem}.

The map $f_{1} \colon \Open(X) \rightarrow 2$ in $\CLJO$ corresponds
to a non-empty closed subset of $X$, see Lemma~\ref{CLJOTwoLem}.  The
function $f_{2} \colon \Open(X) \rightarrow 2$ in $\PreFrmZ$
correponds to a proper Scott open filter, and in the sober case, to a
non-empty compact saturated subset, as discussed already in
Subsection~\ref{DcpoPreFrmSubsec}.

In~\cite{Heckmann97} `valuations' of the form $\Open(X) \rightarrow
2\ltimes 2$, for a topological space $X$, form the elements of a
monad.  In contrast, here we arrive at maps of the form
$\Open(X)\ltimes\Open(X) \rightarrow 2\ltimes 2$.

\auxproof{
\begin{itemize}
\item Let $K$ be non-empty, but $\varphi(\emtpyset) = 1$. The latter
  yields $\emptyset\in\varphi^{-1}(1)$, so that $K =
  \bigcap\varphi^{-1}(1) \subseteq \emptyset$.  By assumption, this is
  not the case.

\item If $\varphi(\emptyset) = 0$, then
  $\emptyset\not\in\varphi^{-1}(1)$. But then
  $K\not\subseteq\emptyset$, see~\cite[Lemma]{KeimelP94}, \ie,
  $K\neq\emptyset$.
\end{itemize}

Finally, we have for the maps $f_{1} \in \CLJO(\Open(X), 2)$ and
$f_{2}\in \PreFrmZ(\Open(X), 2)$,
$$\begin{array}{rcl}
f_{1} \geq f_{2}
& \Longleftrightarrow &
\bigcap\setin{V}{\Closed(X)}{f_{1}(\neg V) = 0} ??
   \bigcap \setin{U}{\Open(X)}{f_{2}(U) = 1}.
\end{array}$$
}

\section{Dualising with $[0,1]$}\label{UnitSec}

The next series of examples starts from adjunctions that are obtained
by homming into the unit interval $[0,1]$. The quantitative logic that
belongs to these examples is given in terms of effect modules. These
can be seen as ``probabilistic vector spaces'', involving scalar
multiplication with scalars from the unit interval $[0,1]$, instead of
from $\mathbb{R}$ or $\mathbb{C}$. We provide a crash course for these
structures, and refer to~\cite{JacobsM16,Jacobs15a,ChoJWW15b}
or~\cite{DvurecenskijP00} for more information. A systematic
description of the `probability' monads below can be found
in~\cite{Jacobs17a}.

A partial commutative monoid (PCM) consists of a set $M$ with a
partial binary operation $\ovee$ and a zero element $0\in M$. The
operation $\ovee$ is commutative and associative, in an appropriate
partial sense. One writes $x \orthogonal y$ if $x \ovee y$ is defined.

An \emph{effect algebra} is a PCM with an orthosupplement
$(-)^{\bot}$, so that $x \ovee x^{\bot} = 1$, where $1 = 0^{\bot}$,
and $x \orthogonal 1$ implies $x=0$. An effect algebra is
automatically a poset, via the definition $x \leq y$ iff $x\ovee z =
y$ for some $z$. The main example is the unit interval $[0,1]$, with
$x \orthogonal y$ iff $x+y \leq 1$, and in that case $x\ovee y = x+y$;
the orthosupplement is $x^{\bot} = 1-x$. A map of effect algebras
$f\colon E \rightarrow D$ is a function that preserves $1$ and
$\ovee$, if defined. We write $\EA$ for the resulting category. Each
Boolean algebra is an effect algebra, with $x \orthogonal y$ iff
$x\wedge y = 0$, and in that case $x\ovee y = x\vee y$. This yields a
functor $\BA \rightarrow \EA$, which is full and faithful.

An \emph{effect module} is an effect algebra $E$ with an action
$[0,1]\times E \rightarrow E$ that preserves $\ovee, 0$ in each
argument separately. A map of effect modules $f$ is a map of effect
algebras that preserves scalar multiplication: $f(r\cdot x) = r\cdot
f(x)$. We thus get a subcategory $\EMod \hookrightarrow \EA$. For each
set $X$, the set $[0,1]^{X}$ of fuzzy predicates on $X$ is an effect
module, with $p \orthogonal q$ iff $p(x) + q(x) \leq 1$ for all $x\in
X$, and in that case $(p\ovee q)(x) = p(x) + q(x)$. Orthosupplement is
given by $p^{\bot}(x) = 1 - p(x)$ and scalar multiplication by $r\cdot
p \in [0,1]^{X}$, for $r\in [0,1]$ and $p\in [0,1]^{X}$, by $(r\cdot
p)(x) = r\cdot p(x)$. This assignment $X \mapsto [0,1]^{X}$ yields a
functor $\Sets \rightarrow \op{\EMod}$ that will be used
below. Important examples of effect modules arise in quantum
logic. For instance, for each Hilbert space $\mathscr{H}$, the set
$\Ef(\mathscr{H}) = \set{A\colon \mathscr{H} \rightarrow \mathscr{H}}{
  0 \leq A \leq \idmap}$ of effects is an effect module.  More
generally, for a (unital) $C^*$-algebra $A$, the set of effects
$[0,1]_{A} = \setin{a}{A}{0 \leq a \leq 1}$ is an effect
module. In~\cite{FurberJ13a} it is shown that taking effects yields a
full and faithful functor:
\begin{equation}
\label{PUEModFunDiag}
\xymatrix{
\CstarPU\ar[rr]^-{[0,1]_{(-)}} & & \EMod
}
\end{equation}

\noindent Here we write $\CstarPU$ for the category of $C^*$-algebras
with positive unital maps.

An \emph{MV-algebra}~\cite{CignoliDM00} can be understood as a
`commutative' effect algebra. It is an effect algebra with a join
$\vee$, and thus also a meet $\wedge$, via De Morgan, in which the
equation $(x \vee y)^{\bot} \ovee x = y^{\bot} \ovee (x\wedge y)$
holds. There is a subcategory $\MVA \hookrightarrow \EA$ with maps
additionally preserving joins $\vee$ (and hence also $\wedge$). Within
an MV-algebra one can define (total) addition and subtraction
operations as $x + y = x \ovee (x^{\bot} \wedge y)$ and $x - y =
(x^{\bot} + y)^{\bot}$. The unit interval $[0,1]$ is an MV-algebra, in
which $+$ and $-$ are truncated (to $1$ or $0$), if needed.

There is a category $\MVMod$ of \emph{MV-modules}, which are
MV-algebras with $[0,1]$-scalar multiplication. Thus $\MVMod$ is twice
a subcategory in: $\MVA \hookleftarrow \MVMod \hookrightarrow
\EMod$. The effect module $[0,1]^{X}$ of fuzzy predicates is an
MV-module. For a commutative $C^*$-algebra $A$ the set of effects
$[0,1]_{A}$ is an MV-module. In fact there is a full and faithful
functor:
\begin{equation}
\label{MIUMVModFunDiag}
\xymatrix{
\CCstarMIU\ar[rr]^-{[0,1]_{(-)}} & & \MVMod
}
\end{equation}

\noindent where $\CCstarMIU$ is the category of commutative
$C^*$-algebras, with MIU-maps, preserving multiplication, involution
and unit (aka.\ $*$-homomorphisms).

Having seen this background information we continue our series of
examples.

\subsection{Sets and effect modules}\label{SetsEModSubsec}

As noted above, fuzzy predicates yield a functor
$\Sets\rightarrow\op{\EMod}$. This functor involves homming into
$[0,1]$, and has an adjoint that is used as starting point for several
variations.
$$\vcenter{\xymatrix@R-2pc{
\op{\EMod}\ar@/^2ex/[dd]^{\Hom(-,[0,1])} \\
\dashv \\
\Sets\ar@/^2ex/[uu]^{\Hom(-,[0,1])}\ar@(dl,dr)_{\mathcal{E}=\EMod([0,1]^{(-)}, [0,1])}
}}
\quad
\begin{prooftree}
\xymatrix@C+.5pc{Y\ar[r]^-{\EMod} & [0,1]^{X}}
\Justifies
\xymatrix{X\ar[r]_-{\Sets} & \EMod(Y, [0,1])}
\end{prooftree}
\quad
\vcenter{\xymatrix@R-.5pc@C-2.5pc{
\op{\EMod}\ar@/^0.7em/[rr] & \top & \EM(\mathcal{E})\rlap{$ = \CCHsep$}\ar@/^0.6em/[ll]
\\
& \Kl(\mathcal{E})\ar[ul]^{\Pred}\ar[ur]_{\Stat} &
}}\hspace*{6em}$$

\noindent The induced monad $\Exp$ is the \emph{expectation} monad
introduced in~\cite{JacobsM12b}. It can be understood as an extension
of the (finite probability) distribution monad $\Dst$, since $\Exp(X)
\cong \Dst(X)$ if $X$ is a finite set. The triangle corollary on the
right says in particular that Kleisli maps $X \rightarrow \Exp(Y)$ are
in bijective correspondence with effect module maps $[0,1]^{Y}
\rightarrow [0,1]^{X}$ acting as predicate transformers, on fuzzy
predicates.

The category of algebras $\EM(\Exp)$ of the expectation monad is the
category $\CCHsep$ of convex compact Hausdorff spaces, with a
separation condition (see~\cite{JacobsM12b,JacobsMF16} for details).
State spaces in quantum computing are typically such convex compact
Hausdorff spaces.

Using the full and faithfulness of the functor $[0,1]_{(-)} \colon
\CstarPU \rightarrow \EMod$ from~\eqref{PUEModFunDiag}, the expectation
monad can alternatively be described in terms of the states of the
commutative $C^*$-algebra $\ell^{\infty}(X)$ of bounded functions $X
\rightarrow \mathbb{C}$, via:
\begin{equation}
\label{StatIsoEqn}
\begin{array}{rcl}
\Stat(\ell^{\infty}(X))
\hspace*{\arraycolsep}\smash{\stackrel{\text{def}}{=}}\hspace*{\arraycolsep}
\CstarPU\big(\ell^{\infty}(X), \mathbb{C}\big)
& \smash{\stackrel{\eqref{PUEModFunDiag}}{\cong}} &
\EMod\big([0,1]_{\ell^{\infty}(X)}, [0,1]_{\mathbb{C}}\big) \\
& = &
\EMod\big([0,1]^{X}, [0,1]\big)
\hspace*{\arraycolsep} = \hspace*{\arraycolsep}
\Exp(X).
\end{array}
\end{equation}

\noindent In this way one obtains the result from~\cite{FurberJ13a} that
there is a full \& faithful functor:
\begin{equation}
\label{KlECstarFunDiag}
\xymatrix{
\Kl(\Exp)\ar[rr] & & \op{\big(\CCstarPU\big)}
}
\end{equation}

\noindent embedding the Kleisli category $\Kl(\Exp)$ of the
expectation monad into commutative $C^*$-algebras with positive unital
maps. On objects this functor~\eqref{KlECstarFunDiag} is given by $X
\mapsto \ell^{\infty}(X)$.

\auxproof{
We consider two maps $\sDst \Rightarrow \Exp$, namely:
$$\xymatrix@R-2pc@C-1pc{
\sDst(X)\ar[rr]^-{\sigma^\Diamond} & & \Exp(X)
\\
\omega\ar@{|->}[rr] & & \lamin{p}{[0,1]^{X}}{\sum_{x}p(x)\cdot\omega(x)}
}$$

\noindent and:
$$\xymatrix@R-2pc@C-1pc{
\sDst(X)\ar[rr]^-{\sigma^\Box} & & \Exp(X)
\\
\omega\ar@{|->}[rr] & & 
   \lamin{p}{[0,1]^{X}}{\sum_{x}p(x)\cdot\omega(x) + (1-\sum_{x}\omega(x))}
}$$

Naturality: for $f\colon X \rightarrow Y$ in $\Sets$ we have:
$$\begin{array}{rcl}
\big(\Exp(f) \after \sigma^{\Diamond})(\omega)(q)
& = &
\Exp(f)\big(\sigma^{\Diamond}(\omega)\big)(q) \\
& = &
\sigma^{\Diamond}(\omega)(q \after f) \\
& = &
\sum_{x} q(f(x))\cdot \omega(x) \\
& = &
\sum_{y}\sum_{x \in f^{-1}(y)} q(y) \cdot \omega(x) \\
& = &
\sum_{y} q(y) \cdot \big(\sum_{x \in f^{-1}(y)}\omega(x)\big) \\
& = &
\sum_{y} q(y) \cdot \sDst(f)(\omega)(y) \\
& = &
\sigma^{\Diamond}\big(\sDst(f)(\omega)\big)(q) \\
& = &
\big(\sigma^{\Diamond} \after \sDst(f)\big)(\omega)(q)
\end{array}$$

\noindent And similarly: 
$$\begin{array}{rcl}
\big(\Exp(f) \after \sigma^{\Box})(\omega)(q)
& = &
\Exp(f)\big(\sigma^{\Box}(\omega)\big)(q) \\
& = &
\sigma^{\Box}(\omega)(q \after f) \\
& = &
\sum_{x} q(f(x))\cdot \omega(x) + (1 - \sum_{x}\omega(x)) \\
& = &
\sum_{y}\sum_{x \in f^{-1}(y)} q(y) \cdot \omega(x) +
   (1 - \sum_{y}\sum_{x \in f^{-1}(y)} \omega(x)) \\
& = &
\sum_{y} q(y) \cdot \big(\sum_{x \in f^{-1}(y)}\omega(x)\big) +
   (1 - \sum_{y}\sDst(f)(\omega)(y)) \\
& = &
\sum_{y} q(y) \cdot \sDst(f)(\omega)(y) + (1 - \sum_{y}\sDst(f)(\omega)(y)) \\
& = &
\sigma^{\Box}\big(\sDst(f)(\omega)\big)(q) \\
& = &
\big(\sigma^{\Box} \after \sDst(f)\big)(\omega)(q)
\end{array}$$

Preservation of units:
$$\begin{array}{rcl}
\big(\sigma^{\Diamond} \after \eta^{\sDst}\big)(x)(p)
& = &
\sigma^{\Diamond}\big(\eta^{\sDst}(x)\big)(p) \\
& = &
\sum_{z} p(z) \cdot \eta^{\sDst}(x)(z) \\
& = &
p(x) \\
& = &
\eta^{\Exp}(x)(p)
\\
\big(\sigma^{\Box} \after \eta^{\sDst}\big)(x)(p)
& = &
\sigma^{\Box}\big(\eta^{\sDst}(x)\big)(p) \\
& = &
\sum_{z} p(z) \cdot \eta^{\sDst}(x)(z) + (1 - \sum_{z}\eta(x)(z)) \\
& = &
p(x) + (1 - 1) \\
& = &
\eta^{\Exp}(x)(p).
\end{array}$$

\noindent Preservation of multiplication:
$$\begin{array}{rcl}
\big(\mu^{\Exp} \after \Exp(\sigma^{\Diamond}) \after
   \sigma^{\Diamond}\big)(\Omega)(p)
& = &
\mu^{\Exp}\big(\Exp(\sigma^{\Diamond})(\sigma^{\Diamond}(\Omega))\big)(p) \\
& = &
\Exp(\sigma^{\Diamond})(\sigma^{\Diamond}(\Omega))(\lam{k}{k(p)}) \\
& = &
\sigma^{\Diamond}(\Omega)\big((\lam{k}{k(p)}) \after \sigma^{\Diamond}\big) \\
& = &
\sigma^{\Diamond}(\Omega)\big(\lam{\omega}{\sigma^{\Diamond}(\omega)(p)}\big) \\
& = &
\sum_{\omega} \sigma^{\Diamond}(\omega)(p)\cdot \Omega(\omega) \\
& = &
\sum_{\omega} \big(\sum_{x} \omega(x)\cdot p(x)\big)\cdot \Omega(\omega) \\
& = &
\sum_{x} \sum_{\omega} p(x) \cdot \Omega(\omega) \cdot \omega(x) \\
& = &
\sum_{x}  p(x) \cdot \big(\sum_{\omega}  \Omega(\omega) \cdot \omega(x)\big) \\
& = &
\sum_{x} p(x) \cdot \mu^{\sDst}(\Omega)(x) \\
& = &
\sigma^{\Diamond}\big(\mu^{\sDst}(\Omega)\big)(p) \\
& = &
\big(\sigma^{\Diamond} \after \mu^{\sDst}\big)(\Omega)(p).
\end{array}$$

\noindent Similarly,
$$\begin{array}{rcl}
\lefteqn{\big(\mu^{\Exp} \after \Exp(\sigma^{\Box}) \after
   \sigma^{\Box}\big)(\Omega)(p)} \\
& = &
\mu^{\Exp}\big(\Exp(\sigma^{\Box})(\sigma^{\Box}(\Omega))\big)(p) \\
& = &
\Exp(\sigma^{\Box})(\sigma^{\Box}(\Omega))(\lam{k}{k(p)}) \\
& = &
\sigma^{\Box}(\Omega)\big((\lam{k}{k(p)}) \after \sigma^{\Box}\big) \\
& = &
\sigma^{\Box}(\Omega)\big(\lam{\omega}{\sigma^{\Box}(\omega)(p)}\big) \\
& = &
\sum_{\omega} \sigma^{\Box}(\omega)(p)\cdot \Omega(\omega) +
   (1 - \sum_{\omega}\Omega(\omega)) \\
& = &
\sum_{\omega} \big(\sum_{x} \omega(x)\cdot p(x) + 
   (1 - \sum_{x}\omega(x))\big)\cdot \Omega(\omega) +
   (1 - \sum_{\omega}\Omega(\omega)) \\
& = &
\big(\sum_{x} \sum_{\omega} p(x) \cdot \Omega(\omega) \cdot \omega(x)\big) +
   \sum_{\omega} (\Omega(\omega) - \sum_{x}\Omega(\omega)\cdot\omega(x)) +
   (1 - \sum_{\omega}\Omega(\omega)) \\
& = &
\sum_{x}  p(x) \cdot \big(\sum_{\omega} \Omega(\omega) \cdot \omega(x)\big) +
   (1 - \sum_{\omega} \Omega(\omega) \cdot \omega(x)) \\
& = &
\sum_{x} p(x) \cdot \mu^{\sDst}(\Omega)(x) + (1 - \sum_{x}\mu^{\sDst}(\Omega)(x)) \\
& = &
\sigma^{\Box}\big(\mu^{\sDst}(\Omega)\big)(p) \\
& = &
\big(\sigma^{\Box} \after \mu^{\sDst}\big)(\Omega)(p).
\end{array}$$

For a Kleisli map $f\colon X \rightarrow \sDst(Y)$ we now get two
predicate transformers $f^{\Diamond}, f^{\Box} \colon [0,1]^{Y}
\rightarrow [0,1]^{X}$, namely:
$$\begin{array}{rcccl}
f^{\Diamond}(q)(x)
& = &
\sigma^{\Diamond}(f(x))(q)
& = &
\sum_{y} q(y)\cdot f(x)(y)
\\
f^{\Box}(q)(x)
& = &
\sigma^{\Box}(f(x))(q)
& = &
\sum_{y} q(y)\cdot f(x)(y) + (1 - \sum_{y}f(x)(y)).
\end{array}$$

\noindent They are as expected. The two associated algebras maps
$\sDst([0,1]) \rightrightarrows [0,1]$ are obtained as:
$$\begin{array}{rcccl}
\tau^{\Diamond}(\omega)
& = &
\sigma^{\Diamond}(\omega)(\idmap)
& = &
\sum_{r\in[0,1]}r\cdot\omega(r)
\\
\tau^{\Box}(\omega)
& = &
\sigma^{\Box}(\omega)(\idmap)
& = &
\sum_{r\in[0,1]}r\cdot\omega(r) + (1-\sum_{r\in[0,1]}\omega(r))
\end{array}$$

\noindent It is easy to see that $f^{\Diamond}$ preserves $0,
\ovee$.  Dually, $f^{\Box}$ preserves $1, \owedge$. For
preservation of $\owedge$ let $p^{\bot} \orthogonal q^{\bot}$, that is
$(1-p(x)) + (1-q(x)) \leq 1$, or equivalently, $1 \leq
p(x)+q(x)$. Then:
$$\begin{array}{rcccl}
(p\owedge q)(x)
& = &
1 - \big((1-p(x)) + (1-q(x))\big)
& = &
p(x) + q(x) - 1.
\end{array}$$

\noindent And thus:
$$\begin{array}{rcl}
f^{\Box}(p \owedge q)(x)
& = &
\sum_{y} (p(y) + q(y) - 1)\cdot f(x)(y) + (1 - \sum_{y}f(x)(y)) \\
& = &
\big(\sum_{y} p(y)\cdot f(x)(y)\big) - 
   \big(\sum_{y} q(y)\cdot f(x)(y)\big) - 2\sum_{y}f(x)(y)) \\
& = &
\sum_{y} p(y)\cdot f(x)(y) + (1 - \sum_{y}f(x)(y))\big) + \\
& & \qquad
   \sum_{y} q(y)\cdot f(x)(y) + (1 - \sum_{y}f(x)(y))\big) - 1 \\
& = &
f^{\Box}(p)(x) \owedge f^{\Box}(q)(x)
\end{array}$$
}

\subsection{Compact Hausdorff spaces and effect modules}\label{CHEModSubsec}

In the previous example we have used the \emph{set} $\EMod(E, [0,1])$
of effect module maps $E \rightarrow [0,1]$, for an effect module $E$.
It turns out that this homset has much more structure: it is a compact
Hausdorff space. The reason is that the unit interval $[0,1]$ is
compact Hausdorff, and so the function space $[0,1]^{E}$ too, by
Tychonoff. The homset $\EMod(E, [0,1]) \hookrightarrow [0,1]^{E}$ can
be described via a closed subset of maps satisfying the effect module
map requirements.  Hence $\EMod(E, [0,1])$ is compact Hausdorff
itself. We thus obtain the following situation.
$$\vcenter{\xymatrix@R-2pc{
\op{\EMod}\ar@/^2ex/[dd]^{\Hom(-,[0,1])} \\
\dashv \\
\CH\ar@/^2ex/[uu]^{\Hom(-,[0,1])}\ar@(dl,dr)_{\mathcal{R}=\EMod(\Cont(-,[0,1]), [0,1])}
}}
\quad
\begin{prooftree}
\xymatrix@C+.5pc{Y\ar[r]^-{\EMod} & \Cont(X,[0,1])}
\Justifies
\xymatrix{X\ar[r]_-{\CH} & \EMod(Y, [0,1])}
\end{prooftree}
\quad
\vcenter{\xymatrix@R-.5pc@C-2.5pc{
\op{\EMod}\ar@/^0.7em/[rr] & \top & \EM(\mathcal{R})\rlap{$ = \CCHsep$}\ar@/^0.6em/[ll]
\\
& \Kl(\mathcal{R})\ar[ul]^{\Pred}\ar[ur]_{\Stat} &
}}\hspace*{6em}$$

\noindent For a compact Hausdorff space $X$, the subset $\Cont(X,
          [0,1]) \hookrightarrow [0,1]^{X}$ of continuous maps $X
          \rightarrow [0,1]$ is a (sub) effect module. The induced
          monad $\Rad(X) = \EMod\big(\Cont(X, [0,1]), [0,1]\big)$ is
          the \emph{Radon} monad. Using the full \& faithful
          functor~\eqref{PUEModFunDiag} the monad can equivalently be
          described as $X \mapsto \Stat(\Cont(X))$, where $\Cont(X)$
          is the commutative $C^*$-algebra of functions $X \rightarrow
          \mathbb{C}$. The monad occurs in~\cite{Mislove12} as part of
          a topological and domain-theoretic approach to information
          theory. The main result of~\cite{FurberJ13a} is the
          equivalence of categories
$$\begin{array}{rcl}
\Kl(\Rad) 
& \simeq &
\op{\big(\CCstarPU\big)}
\end{array}$$

\noindent between the Kleisli category of this Radon monad $\Rad$ and
the category of commutative $C^*$-algebras and positive unital
maps. This shows how (commutative) $C^*$-algebras appear in
state-and-effect triangles (see also~\cite{Jacobs15a,ChoJWW15b}).

The algebras of the Radon monad are convex compact Hausdorff spaces
(with separation), like for the expectation monad $\Exp$,
see~\cite{JacobsM12b} for details.

\subsection{Compact Hausdorff spaces and MV-modules}\label{CHMVModSubsec}

The adjunction $\op{\EMod} \rightleftarrows \CH$ can be restricted to
an adjunction $\op{\MVMod} \rightleftarrows \CH$, involving MV-modules
instead of effect modules. This can be done since continuous functions
$X \rightarrow [0,1]$ are appropriately closed under joins $\vee$, and
thus form an MV-module. Additionally, for an MV-module $E$, the
MV-module maps $E \rightarrow [0,1]$ form a compact Hausdorff space
(using the same argument as in the previous subsection).

Via this restriction to an adjunction $\op{\MVMod} \rightleftarrows
\CH$ we hit a wall again.

\begin{lemma}
For a compact Hausdorff space $X$, the unit $\eta \colon X \rightarrow
\MVMod\big(\Cont(X, [0,1]), [0,1]\big)$, given by $\eta(x)(p) = p(x)$,
is an isomorphism in $\CH$.
\end{lemma}

This result can be understood as part of the Yosida duality for Riesz
spaces. It is well-known in the MV-algebra community, but possibly not
precisely in this form. For convenience, we include a proof.

\begin{myproof}
We only show that the unit $\eta$ is an isomorphism, not that it is
also a homeomorphism. Injectivity is immediate by Urysohn.  For
surjectivity, we first establish the following two auxiliary results.
\begin{enumerate}
\item For each $p\in\Cont(X, [0,1])$ and $\omega\in\MVMod\big(\Cont(X,
  [0,1]), [0,1]\big)$, if $\omega(p) = 0$, then there is an $x\in X$ with
  $p(x) = 0$.

If not, then $p(x) > 0$ for all $x\in X$. Hence there is an inclusion
$X \subseteq \bigcup_{r > 0} p^{-1}\big((r,1]\big)$. By compactness
  there are finitely many $r_{i}$ with $X \subseteq \bigcup_{i}
  p^{-1}\big((r_{i}, 1]\big)$.  Thus for $r = \bigwedge_{i} r_{i} > 0$
    we have $p(x) > r$ for all $x\in X$. Find an $n\in\NNO$ with
    $n\cdot r \geq 1$. The $n$-fold sum $n\cdot p$ in the MV-module
    $\Cont(X, [0,1])$ then satisfies $p(x) = 1$ for all $x$, so that
    $n\cdot p = 1$ in $\Cont(X, [0,1])$. But now we get a
    contradiction: $1 = \omega(1) = \omega(n\cdot p) = n\cdot
    \omega(p) = 0$.

\item For each finite collection of maps $p_{1}, \ldots, p_{n} \in
  \Cont(X, [0,1])$ and for each function $\omega\in\MVMod\big(\Cont(X,
       [0,1]), [0,1]\big)$ there is an $x\in X$ with $\omega(p_{i}) =
       p_{i}(x)$ for all $1 \leq i \leq n$.

For the proof, define $p\in\Cont(X, [0,1])$ using the MV-structure
of $\Cont(X, [0,1])$ as:
$$\begin{array}{rcl}
p
& = &
{\displaystyle\bigvee}_{\!i} \big(p_{i} - \omega(p_{i})\cdot 1\big) \vee 
   \big(\omega(p_{i})\cdot 1 - p_{i}\big).
\end{array}$$

\noindent Since the state $\omega \colon \Cont(X,[0,1]) \rightarrow
          [0,1]$ preserves the MV-structure we get in $[0,1]$:
$$\begin{array}{rcccl}
\omega(p)
& = &
{\displaystyle\bigvee}_{\!i} \big(\omega(p_{i}) - \omega(p_{i})\cdot 1\big)
   \vee \big(\omega(p_{i})\cdot 1 - \omega(p_{i})\big)
& = &
0.
\end{array}$$

\noindent Hence by the previous point there is an $x\in X$ with
$p(x) = 0$. But then $p_{i}(x) = \omega(p_{i})$, as required.
\end{enumerate}

\noindent Now we can prove surjectivity of the unit map $\eta \colon X
\rightarrow \MVMod\big(\Cont(X, [0,1]), [0,1]\big)$. Let $\omega\colon
\Cont(X, [0,1]) \rightarrow [0,1]$ be an MV-module map. Define for
each $p\in\Cont(X, [0,1])$ the subset $U_{p} = \setin{x}{X}{\omega(p)
  \neq p(x)}$. This subset $U_{p} \subseteq X$ is open since it can be
written as $f^{-1}(\R-\{0\})$, for the continuous function $f(x) =
p(x) - \omega(p)$.

Suppose towards a contradiction that $\omega \neq \eta(x)$ for all
$x\in X$.  Thus, for each $x\in X$ there is a $p\in\Cont(X, [0,1])$
with $\omega(p) \neq \eta(x)(p) = p(x)$. This means $X \subseteq
\bigcup_{p}U_{p}$. By compactness of $X$ there are finitely many
$p_{i}\in\Cont(X, [0,1])$ with $X \subseteq \bigcup_{i} U_{p_i}$. The
above second point however gives an $x\in X$ with $\omega(p_{i}) =
p_{i}(x)$ for all $i$. But then $x\not\in \bigcup_{i} U_{p_i}$. \QED
\end{myproof}

\subsection{Sets and directed complete effect modules}\label{SetsDcEModSubsec}

In the remainder of this paper we shall consider effect modules with
additional completeness properties (w.r.t.\ its standard order), as
in~\cite{JacobsW15a}. Specifically, we consider $\omega$-complete, and
directed-complete effect modules. In the first case each ascending
$\omega$-chain $x_{0} \leq x_{1} \leq \cdots$ has a least upperbound
$\bigvee_{n}x_{n}$; and in the second case each directed subset $D$
has a join $\bigvee D$. We write the resulting subcategories as:
$$\xymatrix{
\DcEMod\;\ar@{^(->}[r] & \wEMod\;\ar@{^(->}[r] & \EMod
}$$

\noindent where maps are required to preserve the relevant joins
$\bigvee$.

We start with the directed-complete case. The adjunction $\op{\EMod}
\rightleftarrows \Sets$ from Subsection~\ref{SetsEModSubsec} can be
restricted to an adjunction as on the left below.
$$\hspace*{-0.5em}\vcenter{\xymatrix@R-2pc{
\op{\DcEMod}\ar@/^2ex/[dd]^{\Hom(-,[0,1])} \\
\dashv \\
\Sets\ar@/^2ex/[uu]^{\Hom(-,[0,1])}\ar@(dl,dr)_{\mathcal{E}_{\infty}=\DcEMod([0,1]^{(-)}, [0,1])}
}}
\begin{prooftree}
\xymatrix@C+1pc{Y\ar[r]^-{\DcEMod} & [0,1]^{X}}
\Justifies
\xymatrix@C-.5pc{X\ar[r]_-{\Sets} & \DcEMod(Y, [0,1])}
\end{prooftree}
\vcenter{\xymatrix@R-.5pc@C-2.5pc{
\op{\DcEMod}\ar@/^0.7em/[rr] & \top & \EM(\mathcal{E}_{\infty})\rlap{$=\!\!\Conv_{\infty}$}\ar@/^0.6em/[ll]
\\
& \Kl(\mathcal{E}_{\infty})\ar[ul]^{\Pred}\ar[ur]_{\Stat} &
}}\hspace*{6em}$$

\auxproof{
We briefly check the adjunction.
\begin{itemize}
\item $\overline{f}(x) = \lam{y}{f(y)(x)}$ is in $\EMod(Y, [0,1])$,
  since for each directed collection $y_{i}$ we have:
$$\begin{array}{rcl}
\overline{f}(x)(\bigvee_{i}y_{i})
& = &
f(\bigvee_{i}y_{i})(x) \\
& = &
\big(\bigvee_{i} f(y_{i})\big)(x) \\
& = &
\bigvee_{i} f(y_{i})(x) \qquad \mbox{since joins are pointwise in } [0,1]^X \\
& = &
\bigvee_{i} \overline{f}(x)(y_{i})
\end{array}$$

\item $\overline{g} = \lam{y}{\lam{x}{g(x)(y)}}$ preserves joins, since:
$$\begin{array}{rcl}
\overline{g}(\bigvee_{i}y_{i})
& = &
\lam{x}{g(x)(\bigvee_{i} y_{i})} \\
& = &
\lam{x}{\bigvee_{i} g(x)(y_{i})} \\
& = &
\bigvee_{i} \lam{x}{\overline{g}(y_{i})(x)} 
   \qquad \mbox{since joins are pointwise in } [0,1]^X \\
& = &
\bigvee_{i} \overline{g}(y_{i}).
\end{array}$$
\end{itemize}
}

\noindent The resulting monad $\Exp_{\infty} =
\DcEMod\big([0,1]^{(-)}, [0,1]\big)$ on $\Sets$ is in fact
isomorphic\footnote{This isomorphism $\Exp_{\infty} \cong
  \Dst_{\infty}$ in Proposition~\ref{DcEModMonadProp} is inspired by
  work of Robert Furber (PhD Thesis, forthcoming): he noticed the
  isomorphism $\NStat(\ell^{\infty}(X)) \cong \Dst_{\infty}(X)$
  in~\eqref{NStatIsoEqn}, which is obtained here as a corollary to
  Proposition~\ref{DcEModMonadProp}.} to the infinite (discrete
probability) distribution monad $\Dst_{\infty}$,
see~\cite{Jacobs16g}. We recall, for a set $X$,
$$\begin{array}{rcl}
\Dst_{\infty}(X)
& = &
\set{\omega \colon X \rightarrow [0,1]}{\supp(\omega)\mbox{ is countable,
   and }\sum_{x}\omega(x) = 1}.
\end{array}$$

\noindent The subset $\supp(\omega) \subseteq X$ contains the elements
$x\in X$ with $\omega(x) \neq 0$. The requirement in the definition of
$\Dst_{\infty}(X)$ that $\supp(\omega)$ be countable is superfluous,
since it follows from the requirement $\sum_{x} \omega(x) =
1$. Briefly, $\supp(\omega) \subseteq \bigcup_{n>0} X_{n}$, where
$X_{n} = \setin{x}{X}{\omega(x) > \frac{1}{n}}$ contains at most $n-1$
elements (see \textit{e.g.}~\cite[Prop.~2.1.2]{Sokolova05}).

\begin{proposition}
\label{DcEModMonadProp}
There is an isomorphism of monads $\Dst_{\infty} \cong \Exp_{\infty}$, where
$\Exp_{\infty}$ is the monad induced by the above adjunction $\op{\DcEMod}
\rightleftarrows \Sets$.
\end{proposition}

\begin{myproof}
For a subset $U\subseteq X$ we write $\indic{U} \colon X \rightarrow
[0,1]$ for the `indicator' function, defined by $\indic{U}(x) = 1$ if
$x\in U$ and $\indic{U}(x) = 0$ if $x\not\in U$. We write $\indic{x}$
for $\indic{\{x\}}$. This function $\indic{(-)} \colon \Pow(X)
\rightarrow [0,1]^{X}$ is a map of effect algebras that preserves all
joins.

Let $h\in\Exp_{\infty}(X)$, so $h$ is a Scott continuous map of effect
modules $h \colon [0,1]^{X} \rightarrow [0,1]$. Define $\overline{h}
\colon X \rightarrow [0,1]$ as $\overline{h}(x) =
h(\indic{x})$. Notice that if $U\subseteq X$ is a finite subset, then:
$$\begin{array}{rcccccccccccl}
1
& = &
h(1)
& = &
h(\indic{X})
& \geq &
h(\indic{U})
& = &
h(\bigovee_{x\in U} \indic{x})
& = &
\bigovee_{x\in U} h(\indic{x})
& = &
\bigovee_{x\in U} \overline{h}(x).
\end{array}$$

\noindent We can write $X$ as directed union of its finite subsets,
and thus also $\indic{X} = \bigvee\set{\indic{U}}{U \subseteq X \mbox{
    finite}}$. But then $\overline{h} \in \Dst_{\infty}(X)$, because
$h$ preserves directed joins:
$$\begin{array}{rcccccccl}
1
& = &
h(\indic{X})
& = &
\bigvee \set{h(\indic{U})}{U \subseteq X \mbox{ finite}} 
& = &
\bigvee \set{\sum_{x\in U} \overline{h}(x)}{U \subseteq X \mbox{ finite}} 
& = &
\sum_{x\in X} \overline{h}(x).
\end{array}$$

Conversely, given $\omega \in \Dst_{\infty}(X)$ we define
$\overline{\omega} \colon [0,1]^{X} \rightarrow [0,1]$ as
$\overline{\omega}(p) = \sum_{x\in X} p(x) \cdot \omega(x)$. It is
easy to see that $\overline{\omega}$ is a map of effect modules. It is
a bit more challenging to see that it preserves directed joins
$\bigvee_{i} p_{i}$, for $p_{i}\in [0,1]^{X}$.

First we write the countable support of $\omega$ as $\supp(\omega) =
\{x_{0}, x_{1}, x_{2}, \ldots\}\subseteq X$ in such a way that
$\omega(x_{0}) \geq \omega(x_{1}) \geq \omega(x_{2}) \geq \cdots$.  We
have $1 = \sum_{x\in X}\omega(x) = \sum_{n\in\NNO} \omega(x_{n})$.
Hence, for each $N\in\NNO$ we get:
$$\begin{array}{rcl}
\sum_{n > N}\omega(x_{n})
& = &
1 - \sum_{n \leq N} \omega(x_{n}).
\end{array}$$

\noindent By taking the limit $N \rightarrow \infty$ on both sides we
get:
$$\begin{array}{rcccccccl}
\lim\limits_{N\rightarrow\infty}\sum_{n > N}\omega(x_{n})
& = &
1 - \lim\limits_{N\rightarrow\infty}\sum_{n \leq N} \omega(x_{n})
& = &
1 - \sum_{n\in\NNO} \omega(x_{n})
& = &
1 - 1
& = &
0.
\end{array}$$

\noindent We have to prove $\overline{\omega}(\bigvee_{i}p_{i}) =
\bigvee_{i}\overline{\omega}(p_{i})$. The non-trivial part is
$(\leq)$. For each $N\in\NNO$ we have:
$$\begin{array}{rcl}
\overline{\omega}(\bigvee_{i}p_{i})
& = &
\sum_{n\in\NNO} (\bigvee_{i}p_{i})(x_{n})\cdot \omega(x_{n}) \\
& = &
\sum_{n\in\NNO} (\bigvee_{i}p_{i}(x_{n}))\cdot \omega(x_{n}) \\
& = &
\sum_{n\in\NNO} \bigvee_{i}p_{i}(x_{n})\cdot \omega(x_{n}) \\
& = &
\Big(\sum_{n\leq N} \bigvee_{i}p_{i}(x_{n})\cdot \omega(x_{n})\Big) +
   \Big(\sum_{n > N} \bigvee_{i}p_{i}(x_{n})\cdot \omega(x_{n})\Big) \\
& = &
\Big(\bigvee_{i}\sum_{n\leq N} p_{i}(x_{n})\cdot \omega(x_{n})\Big) +
   \Big(\sum_{n > N} \bigvee_{i}p_{i}(x_{n})\cdot \omega(x_{n})\Big) \\
& \leq &
\Big(\bigvee_{i}\sum_{n\leq N} p_{i}(x_{n})\cdot \omega(x_{n})\Big) +
   \Big(\sum_{n > N} \omega(x_{n})\Big) \qquad \mbox{since } p_{i}(x) \in [0,1].
\end{array}$$

\noindent Hence we are done by taking the limit $N \rightarrow
\infty$.  Notice that we use that the join $\bigvee$ can be moved
outside a finite sum. This works precisely because the join is taken
over a directed set.

What remains is to show that these mappings $h\mapsto \overline{h}$
and $\omega\mapsto\overline{\omega}$ yield an isomorphism
$\Dst_{\infty}(X) \cong \Exp_{\infty}(X)$, which is natural in $X$,
and forms an isomorphism of monads.  This is left to the interested
reader. \QED

\auxproof{
$$\begin{array}{rcl}
\overline{\overline{h}}(p)
& = &
\sum_{x} p(x) \cdot \overline{h}(x) \\
& = &
\sum_{x} p(x) \cdot h(\indic{x}) \\
& = &
\sum_{x} h(p(x) \cdot \indic{x}) \\
& = &
\bigvee \set{\sum_{x\in U} h(p(x) \cdot \indic{x})}
   {U \subseteq X \mbox{ finite}} \\
& = &
\bigvee \set{h(\bigovee_{x\in U} p(x) \cdot \indic{x})}
   {U \subseteq X \mbox{ finite}} \\
& = &
h\big(\bigvee \set{\bigovee_{x\in U} p(x) \cdot \indic{x}}
   {U \subseteq X \mbox{ finite}}\big) \\
& = &
h(p) 
\\
\overline{\overline{\omega}}(x)
& = &
\overline{\omega}(\indic{x}) \\
& = &
\sum_{x'} \indic{x}(x') \cdot \omega(x') \\
& = &
1 \cdot \omega(x) \\
& = &
\omega(x).
\end{array}$$

Let's now write $\sigma_{X} \colon \Dst_{\infty}(X) \rightarrow
\Exp_{\infty}(X)$ for this map, given by $\sigma_{X}(\omega)(p) =
\sum_{x} p(x)\cdot \omega(x)$.  For $f\colon X \rightarrow Y$ we have
for $\omega\in\Dst_{\infty}(X)$ and $q \in [0,1]^{Y}$,
$$\begin{array}{rcl}
\big(\Exp_{\infty}(f) \after \sigma_{X}\big)(\omega)(q)
& = &
\Exp_{\infty}(f)\big(\sigma_{X}(\omega)\big)(q) \\
& = &
\sigma_{X}(\omega)(q \after f) \\
& = &
\sum_{x} q(f(x))\cdot \omega(x) \\
& = &
\sum_{y, x \in f^{-1}(y)} q(y) \cdot \omega(x) \\
& = &
\sum_{y} q(y) \cdot \big(\sum_{x \in f^{-1}(y)} \omega(x)\big) \\
& = &
\sum_{y} q(y) \cdot \Dst_{\infty}(f)(\omega)(y) \\
& = &
\sigma_{Y}\big(\Dst_{\infty}(f)(\omega)\big)(q) \\
& = &
\big(\sigma_{Y} \after \Dst_{\infty}(f)\big)(\omega)(q).
\end{array}$$

\noindent Also, 
$$\begin{array}{rcl}
\big(\sigma_{X} \after \eta^{\Dst_{\infty}}_{X}\big)(x)(p)
& = &
\sigma_{X}(\eta(x))(p) \\
& = &
\sum_{x'} p(x')\cdot \eta(x)(x') \\
& = &
p(x) \cdot 1 \\
& = &
\eta^{\Exp_{\infty}}_{X}(x)(p).
\end{array}$$

\noindent Further,
$$\begin{array}{rcl}
\big(\mu^{\Exp_{\infty}} \after \Exp_{\infty}(\sigma) \after \sigma\big)(\Omega)(p)
& = &
\mu^{\Exp_{\infty}}\big(\Exp_{\infty}(\sigma)(\sigma(\Omega))\big)(p) \\
& = &
\Exp_{\infty}(\sigma)(\sigma(\Omega))(\lam{k}{k(p)}) \\
& = &
\sigma(\Omega)\big((\lam{k}{k(p)}) \after \sigma\big) \\
& = &
\sigma(\Omega)\big(\lam{\omega}{\sigma(\omega)(p)}\big) \\
& = &
\sum_{\omega} \sigma(\omega)(p) \cdot \Omega(\omega) \\
& = &
\sum_{\omega} (\sum_{x} p(x) \cdot \omega(x)) \cdot \Omega(\omega) \\
& = &
\sum_{\omega, x} p(x) \cdot \omega(x) \cdot \Omega(\omega) \\
& = &
\sum_{x} p(x) \cdot (\sum_{\omega} \omega(x) \cdot \Omega(\omega)) \\
& = &
\sum_{x} p(x) \cdot \mu^{\Dst_{\infty}}(\Omega)(x) \\
& = &
\sigma\big(\mu^{\Dst_{\infty}}(\Omega)\big)(p) \\
& = &
\big(\sigma \after \mu^{\Dst_{\infty}}\big)(\Omega)(p).
\end{array}$$
}
\end{myproof}

As a result, the Eilenberg-Moore category $\EM(\Exp_{\infty})$ is
isomorphic to $\EM(\Dst_{\infty}) = \Conv_{\infty}$, where
$\Conv_{\infty}$ is the category of countably-convex sets $X$, in
which convex sums $\sum_{n\in\NNO}r_{n}x_{n}$ exist, where $x_{n}\in
X$ and $r_{n}\in [0,1]$ with $\sum_{n}r_{n} = 1$.

We briefly look at the relation with $C^*$-algebras (actually
$W^*$-algebras), like in Subsection~\ref{SetsEModSubsec}. We write
$\WstarNPU$ for the category of $W^*$-algebras with normal positive
unital maps. The term `normal' is used in the operator algebra
community for what is called `Scott continuity' (preservation of
directed joins) in the domain theory community. This means that taking
effects yields a full and faithful functor:
\begin{equation}
\label{NPUDcEModFunDiag}
\xymatrix{
\WstarNPU\ar[rr]^-{[0,1]_{(-)}} & & \DcEMod
}
\end{equation}

\noindent This is similar to the situation in~\eqref{PUEModFunDiag}
and~\eqref{MIUMVModFunDiag}. One could also use $AW^*$-algebras
here. Next, there is now a full and faithful functor to the category
of commutative $W^*$-algebras:
\begin{equation}
\label{KlDWstarFunDiag}
\xymatrix{
\Kl(\Dst_{\infty}) \cong \Kl(\Exp_{\infty})\ar[rr] & & \CWstarNPU
}
\end{equation}

\noindent On objects it is given by $X \mapsto \ell^{\infty}(X)$. This
functor is full and faithful since there is a bijective
correspondence:
$$\begin{prooftree}
\xymatrix{\ell^{\infty}(X)\ar[r] & \ell^{\infty}(Y)
   \rlap{\hspace*{12.6em}in $\CWstarNPU$}} 
\Justifies
\xymatrix{Y\ar[r] & \NStat(\ell^{\infty}(X))
   \rlap{$\;\cong\Exp_{\infty}(X)\cong\Dst_{\infty}(X)$\hspace*{3em}in $\Sets$}} 
\end{prooftree}\hspace*{12em}$$

\noindent where the isomorphism $\cong$ describing normal states is
given, like in~\eqref{StatIsoEqn}, by:
\begin{equation}
\label{NStatIsoEqn}
\begin{array}{rcl}
\NStat(\ell^{\infty}(X))
\hspace*{\arraycolsep}\smash{\stackrel{\text{def}}{=}}\hspace*{\arraycolsep}
\WstarNPU\big(\ell^{\infty}(X), \mathbb{C}\big)
& \smash{\stackrel{\eqref{NPUDcEModFunDiag}}{\cong}} &
\DcEMod\big([0,1]_{\ell^{\infty}(X)}, [0,1]_{\mathbb{C}}\big) \\
& = &
\DcEMod\big([0,1]^{X}, [0,1]\big) \\
& = &
\Exp_{\infty}(X) \\
& \cong &
\Dst_{\infty}(X).
\end{array}
\end{equation}

\subsection{Measurable spaces and $\omega$-complete effect 
   modules}\label{MeaswEModSubsec}

In our final example we use an adjunction between effect modules and
measurable spaces (instead of sets or compact Hausdorff spaces). We
write $\Meas$ for the category of measurable spaces $(X,\Sigma_{X})$,
where $\Sigma_{X} \subseteq \Pow(X)$ is the $\sigma$-algebra of
measurable subsets, with measurable functions between them (whose
inverse image maps measurable subsets to measurable subsets). We use
the unit interval $[0,1]$ with its standard Borel $\sigma$-algebra
(the least one that contains all the usual opens). A basic fact in
this situation is that for a measurable space $X$, the set $\Meas(X,
[0,1])$ of measurable functions $X \rightarrow [0,1]$ is an
$\omega$-effect module. The effect module structure is inherited via
the inclusion $\Meas(X, [0,1]) \hookrightarrow [0,1]^{X}$. Joins of
ascending $\omega$-chains $p_{0} \leq p_{1} \leq \cdots$ exists,
because the (pointwise) join $\bigvee_{n} p_{n}$ is a measurable
function again. In this way we obtain a functor $\Meas(-, [0,1])
\colon \Meas \rightarrow \op{\wEMod}$.

In the other direction there is also a hom-functor $\wEMod(-, [0,1])
\colon \op{\wEMod} \rightarrow \Meas$. For an $\omega$-effect module
$E$ we can provide the set of maps $\wEMod(E, [0,1])$ with a
$\sigma$-algebra, namely the least one that makes all the evaluation
maps $\evmap_{x} \colon \wEMod(E,[0,1]) \rightarrow [0,1]$ measurable,
for $x\in E$. This function $\evmap_{x}$ is given by $\evmap_{x}(p) =
p(x)$. This gives the following situation.

$$\vcenter{\xymatrix@R-2pc{
\op{\wEMod}\ar@/^2ex/[dd]^{\Hom(-,[0,1])} \\
\dashv \\
\Meas\ar@/^2ex/[uu]^{\Hom(-,[0,1])}\ar@(dl,dr)_{\Giry = \wEMod(\Meas(-,[0,1]), [0,1])}
}}
\quad
\begin{prooftree}
\xymatrix@C+1pc{Y\ar[r]^-{\wEMod} & \Meas(X, [0,1])}
\Justifies
\xymatrix{X\ar[r]_-{\Meas} & \wEMod(Y, [0,1])}
\end{prooftree}
\quad
\vcenter{\xymatrix@R-.5pc@C-2pc{
\op{\wEMod}\ar@/^0.7em/[rr] & \top & \EM(\Giry)\ar@/^0.6em/[ll]
\\
& \Kl(\Giry)\ar[ul]^{\Pred}\ar[ur]_{\Stat} &
}}\hspace*{6em}$$

\noindent We use the symbol $\Giry$ for the induced monad because of
the following result.

\begin{proposition}
\label{GiryProp}
The monad $\Giry = \wEMod\big(\Meas(-,[0,1]), [0,1]\big)$ on $\Meas$
in the above situation is (isomorphic to) the Giry
monad~\cite{Giry82}, given by probability measures:
$$\begin{array}{rcccl}
\mathrm{Giry}(X)
& \smash{\stackrel{\text{def}}{=}} &
\set{\phi\colon \Sigma_{X} \rightarrow [0,1]}
   {\phi \mbox{ is a probability measure}}
& = &
\wEA(\Sigma_{X}, [0,1]).
\end{array}$$
\end{proposition}

\begin{myproof}
The isomorphism involves Lebesgue integration:
$$\hspace*{5em}\xymatrix{
\llap{$\Giry(X) = \wEMod\big(\Meas(X,[0,1]),\;$} [0,1]\big)
   \ar@/^2ex/[rr]^-{I \mapsto (M\mapsto I(\indic{M}))}
   & \cong & 
\wEA\rlap{$(\Sigma_{X}, [0,1]) = \mathrm{Giry}(X)$}
   \ar@/^2ex/[ll]^-{\phi \mapsto (p\mapsto \int p \intd\phi))}
}$$

\noindent See~\cite{Jacobs13a} or~\cite{JacobsW15a} for more
details. \QED
\end{myproof}

The above triangle is further investigated in~\cite{Jacobs13a}. It
resembles the situation described in~\cite{ChaputDPP14} for Markov
kernels (the ordinary, not the abstract, ones).

\subsection{Dcpo's and directed complete effect modules}\label{DcpoDcEModSubsec}

In our final example we briefly consider another variation of the
adjunction $\op{\DcEMod} \leftrightarrows \Sets$ in
Subsection~\ref{SetsDcEModSubsec}, now with an adjunction
$\op{\DcEMod} \leftrightarrows \Dcpo$ between the categories of
directed complete effect modules and partial orders. This brings us
into the realm of probabilistic power domains, which has its own
thread of research, see
\textit{e.g.}~\cite{Heckmann94,JonesP89,Keimel08,Keimel09,KeimelP09,Saheb80,TixKP05}. Our
only aim at this stage is to show how the current approach connects to
that line of work. The most significant difference is that we use the
unit interval $[0,1]$, whereas it is custom for probabilistic power
domains to use the extended non-negative real numbers $\setin{r}{\R}{r
  \geq 0} \cup \{\infty\}$. Consequently, we use effect modules
instead of cones.

Using that the unit interval $[0,1]$, with its usual order, is a dcpo,
and that its multiplication, and also its partial addition, is
Scott continuous in each variable, we obtain:
$$\vcenter{\xymatrix@R-2pc{
\op{\DcEMod}\ar@/^2ex/[dd]^{\Hom(-,[0,1])} \\
\dashv \\
\Dcpo\ar@/^2ex/[uu]^{\Hom(-,[0,1])}\ar@(dl,dr)_{\mathcal{V}=\DcEMod([0,1]^{(-)}, [0,1])}
}}
\quad
\begin{prooftree}
\xymatrix@C+1.5pc{Y\ar[r]^-{\DcEMod} & \Dcpo(X, [0,1])}
\Justifies
\xymatrix@C-.5pc{X\ar[r]_-{\Dcpo} & \DcEMod(Y, [0,1])}
\end{prooftree}
\quad
\vcenter{\xymatrix@R-.5pc@C-2.5pc{
\op{\DcEMod}\ar@/^0.7em/[rr] & \top & \EM(\mathcal{V})\ar@/^0.6em/[ll]
\\
& \Kl(\mathcal{V})\ar[ul]^{\Pred}\ar[ur]_{\Stat} &
}}$$

\noindent The induced monad $\mathcal{V}$ is a restricted version of
the monad of valuations, that uses the extended real numbers, as
mentioned above. It is unclear what its category of Eilenberg-Moore
algebras is.

\auxproof{
We check the adjunction. First of all, for a dcpo $X$ the homset
$\Hom(X, [0,1])$ of Scott continuous functions $X \rightarrow [0,1]$
is a dcpo (since $\Dcpo$ is cartesian closed), and also an effect
module via pointwise partial sum $\ovee$ and scalar multiplication. In
the other direction, for $Y\in\DcEMod$ the homset $\Hom(Y, [0,1])$ of
Scott continuous effect module maps is a dcpo, via the pointwise
join. This join is again a map of effect modules via continuity of the
operations on $[0,1]$. For instance, for a directed collection
$(h_{i})$ in $\Hom(Y,[0,1])$, define $h(y) =
\bigvee_{i}h_{i}(y)$. This $h$ is continous, and preserves $\ovee$
since:
$$\begin{array}{rcl}
h(y\ovee z)
& = &
\bigvee_{i} h_{i}(y\ovee z) \\
& = &
\bigvee_{i} h_{i}(y)\ovee h_{i}(z) \\
& = &
\bigvee_{i} h_{i}(y)\ovee \bigvee_{i}h_{i}(z) \qquad
   \mbox{by directedness} \\
& = &
h(y) \ovee h(z).
\end{array}$$

\noindent The adjunction involves the usual argument swapping and
verifications.
\begin{itemize}
\item For $f\colon Y \rightarrow \Dcpo(X,[0,1])$ in $\DcEMod$ we
  obtain $\overline{f} \colon X \rightarrow \DcEMod(Y,[0,1])$ by
  $\overline{f}(x)(y) = f(y)(x)$. This is well-defined.  Each
  $\overline{f}(x) \colon Y \rightarrow [0,1]$ is a map of effect
  modules, for instance because:
$$\begin{array}{rcl}
\overline{f}(x)(r\cdot y)
& = &
f(r\cdot y)(x) \\
& = &
\big(r\cdot f(y)\big)(x) \\
& = &
r \cdot f(y)(x) \\
& = &
r \cdot \overline{f}(x)(y).
\end{array}$$

\item In the other direction, for $g\colon X \rightarrow \Hom(Y,
  [0,1])$ in $\Dcpo$ we have $\overline{g} \colon Y \rightarrow
  \Hom(X, [0,1])$ in $\DcEMod$, given by $\overline{g}(y)(x) =
  g(x)(y)$. This $\overline{g}$ is a map of effect modules, for
  instance because:
$$\begin{array}{rcl}
\overline{g}(r\cdot y)
& = &
\lam{x}{\overline{g}(r\cdot y)(x)} \\
& = &
\lam{x}{g(x)(r\cdot y)} \\
& = &
\lam{x}{r\cdot g(x)(y)} \\
& = &
r\cdot \big(\lam{x}{g(x)(y)}\big) \\
& = &
r\cdot \big(\lam{x}{\overline{g}(y)(x)}\big) \\
& = &
r \cdot \overline{g}(y).
\end{array}$$
\end{itemize}
}

\subparagraph*{\textbf{Acknowledgements}} Several people have contributed to
the ideas and examples presented here, including, in alphabetical
order: Kenta Cho, Robert Furber, Helle Hansen, Klaus Keimel, Bas and
Bram Westerbaan.  Thanks to all of them!

\bibliographystyle{plain}


\end{document}